\documentclass[11pt,a4paper]{article}
\usepackage[a4paper, total={6in, 8in}]{geometry}

\usepackage{tikz}
\usetikzlibrary{calc,shapes,arrows,intersections,snakes,automata}
\usepackage{pgfplots}
\usepackage{arydshln}
\usepackage{textcomp}
\usepackage{comment}
\usepackage[linesnumbered,algoruled,boxed,lined]{algorithm2e}
\usepackage{setspace}
\usepackage{graphicx}

\usepackage{bm}
\usepackage{amsmath}
\usepackage{amssymb}

\newtheorem{definition}{Definition}
\newtheorem{example}{Example}[section]

\newcommand{\Eb}{E_{\text{b}}}
\newcommand{\Es}{E_{\text{s}}}
\newcommand{\No}{N_0}

\makeatletter
\newcommand{\pushright}[1]{\ifmeasuring@#1\else\omit\hfill$\displaystyle#1$\fi\ignorespaces}
\newcommand{\pushleft}[1]{\ifmeasuring@#1\else\omit$\displaystyle#1$\hfill\fi\ignorespaces}
\makeatother

\makeatletter \g@addto@macro\@floatboxreset\centering \makeatother

\newcommand{\maxs}{\max{}^{\star}}
\DeclareMathOperator*{\maxso}{\maxs}

\title{Advances in Detection and Error Correction for Coherent Optical Communications: Regular, Irregular, and Spatially Coupled LDPC Code Designs}
\author{Laurent Schmalen${}^\star$, Stephan ten Brink${}^\dagger$, and Andreas Leven${}^\star$}
\date{}

\begin{document}
\maketitle

\newcommand{\IE}[1]{I_E^{[{#1}]}}
\newcommand{\IEtilde}[1]{\tilde{I}_E^{[{#1}]}}
\newcommand{\IA}[1]{I_A^{[{#1}]}}

\newcommand{\field}{\mathbb{F}}
\newcommand\blfootnote[1]{%
  \begingroup
  \renewcommand\thefootnote{}\footnote{#1}%
  \addtocounter{footnote}{-1}%
  \endgroup
}

\section{Introduction}

\emph{Forward error correction} (FEC) in optical communications has been first demonstrated in 1988~\cite{Grover1988}. Since then, coding technology has evolved significantly. This pertains not only to the codes but also to encoder and decoder architectures. 
Modern high-speed optical communication systems require high-performing FEC engines that support throughputs of 100\,GBit/s or multiples thereof, that have low power consumption, that realize  \emph{net coding gains} (NCGs) close to the theoretical limits at a target \emph{bit error rate} (BER) of below $10^{-15}$, and that are preferably adapted to the peculiarities of the optical channel. \blfootnote{${}^\star$ L. Schmalen and A. Leven are with Nokia Bell Labs, Lorenzstr. 10, 70435 Stuttgart, Germany. E-mail: \texttt{\{first.last\}@nokia-bell-labs.com}}
\blfootnote{${}^\dagger$ S. ten Brink is with the University of Stuttgart, Institute of Telecommunications, Pfaffenwaldring 47, 70569 Stuttgart, Germany.}
\blfootnote{This is the version of the following article: ``Advances in Detection and Error Correction for Coherent Optical Communications: Regular, Irregular, and Spatially Coupled LDPC Code Designs'', which appeared as Chapter 3 in the book \emph{Enabling Technologies for High Spectral-efficiency Coherent Optical Communication Networks} edited by X. Zhou and C. Xie, which has been published in final form at DOI:10.1002/9781119078289 (ISBN 9781118714768 (print) and ISBN  9781119078289 (online)). This article may be used for non-commercial purposes in accordance with Wiley Terms and Conditions for Self-Archiving.}

Forward error correction coding is based on deterministically adding redundant bits to a source information bit sequence. After transmission over a noisy channel, a decoding system tries to exploit the redundant information for fully recovering the source information. Several methods for generating the redundant bit sequence from the source information bits are known. Transmission systems with 100\,GBit/s and 400\,GBit/s today typically use one of two coding schemes to generate the redundant information: \emph{Block-Turbo Codes} (BTCs) or \emph{Low-Density Parity-Check} (LDPC) codes. In coherent systems, so-called soft information is usually ready available and can be used in high performing systems within a soft-decision decoder architecture. Soft-decision information means that no binary 0/1 decision is made before entering the forward error correction decoder. Instead, the (quantized) samples are used together with their statistics to get improved estimates of the original bit sequence. This chapter will focus on soft-decision decoding of LDPC codes and the evolving spatially coupled LDPC codes. 

In coherent optical communications, the signal received after carrier recovery may 
be affected by different distortions than those that commonly occur in wireless communications. For instance, the signal at the input of the signal space demapper may be affected by phase slips (also called \emph{cycle slips}~\cite{FludgerOFC2012}), with a probability depending on the non-linear phase noise introduced by the optical transmission link~\cite{LeongJLTSlips}. The phase slips are \emph{not} an effect of the physical waveform channel but, rather, an artifact of 
coarse blind phase recovery algorithms with massive parallelization at the initial digital signal processing (DSP) receiver steps~\cite{PfauOFC2014}.
If such a phase slip is ignored, error propagation will occur at the receiver and all data following the phase slip cannot be properly recovered. Several approaches to mitigate phase slips have been proposed. Of these, the most common is differential coding, rendering a phase slip into a single error event. In order to alleviate the penalty caused by differential coding, iterative decoding between an FEC decoder and a differential decoder can be beneficial~\cite{HoeherTurboDQPSK}. This solution leads however to an increased receiver complexity, as several executions of a soft-input soft-output differential decoder (usually based on the BCJR algorithm\footnote{termed after the initial letters of its inventors Bahl, Cocke, Jelinek and Raviv~\cite{Bahl0374}.}) have to be carried out.

In this chapter, we first show how the use of differential coding and the presence of phase slips in the transmission channel affect the total achievable information rates and capacity of a system. By means of the commonly used \emph{Quadrature Phase-Shift Keying} (QPSK) modulation, we show that the use of differential coding \emph{does not decrease the capacity}, i.e., the total amount of reliably conveyable information over the channel remains the same. It is a common misconception that the use of differential coding introduces an unavoidable ``differential loss''. This \emph{perceived} differential loss is rather a consequence of simplified differential detection and decoding at the receiver. Afterwards, we show how  capacity-approaching coding schemes based on LDPC and spatially coupled LDPC codes can be constructed by combining iterative demodulation and decoding. For this, we first show how to modify the differential decoder to account for phase slips and then how to use this modified differential decoder to construct good LDPC codes. This construction method can serve as a blueprint to construct good and practical LDPC codes for other applications with iterative detection, such as higher order modulation formats with non-square constellations~\cite{SchmalenOFC14}, multi-dimensional optimized modulation formats~\cite{BuelowACP}, turbo equalization to mitigate ISI (e.g., due to nonlinearities)~\cite{Koetter0104,FujimoriTE} and many more. Finally, we introduce the class of \emph{spatially coupled} (SC)-LDPC codes, which are a specialization of LDPC codes with some outstanding properties and which can be decoded with a very simple windowed decoder. We show that the universal behavior of spatially coupled codes makes them an ideal candidate for iterative differential demodulation/detection and decoding. 

This chapter is structured as follows: In Sec.~\ref{sec:differential_coding_optical} we formally introduce the notation, system model and differential coding. We highlight some pitfalls that one may encounter when phase slips occur on the equivalent channel. We propose a modified differential decoder that is necessary to construct a capacity-approaching system with differential coding. In Sec.~\ref{sec:ldpc_diff_mod}, we introduce LDPC codes and iterative detection. We highlight several possibilities of realizing the interface between the LDPC decoder and the detector and give design guidelines for finding good degree distributions of the LDPC code. We show that with iterative detection and LDPC codes, the differential loss can be recovered to a great extend. Finally, in Sec.~\ref{sec:sc_codes}, we introduce SC-LDPC codes and show how a very simple construction can be used to realize codes that outperform LDPC codes while having similar decoding complexity.

\vspace*{-0.3ex}
\section{Differential Coding for Optical Communications}\label{sec:differential_coding_optical}
\vspace*{-0.7ex}
In this section, we describe and study the effect of differential coding on coherent optical communication systems and especially on the maximum conveyable information rate (the so-called \emph{capacity}). We assume a simple, yet accurate channel model based on \emph{additive white Gaussian noise} (AWGN) and random phase slips. We start by giving a rigorous description of higher-order modulation schemes frequently used in coherent communications and then introduce in Sec.~\ref{sec:channel_model} the channel model taking into account phase slips which are due to imperfect phase estimation in the coherent receiver. We will then introduce differential coding and show how the differential decoder has to be modified in order to properly take into account phase slips. We show that differential coding as such does not limit the capacity of a communication system, provided that an adequate receiver is used.

\subsection{Higher-Order Modulation Formats}

In this section, the interplay of coding and modulation will be discussed in detail. We only take on an IQ-perspective of digital modulation, representing digital modulation symbols as complex numbers. The sequence of complex numbers (where I denotes the real part and Q the imaginary part) is then used to generate the actual waveform (taking into account pulse shaping and eventually electronic pre-distortion), i.e., to drive the optical modulators generating the I and Q component. For a thorough overview of coding and modulation in the context of coherent communications, we refer the interested reader to~\cite{BeygiOverview,DjordjevicRyan}.

\begin{figure}[tbh!]
\centering
\includegraphics{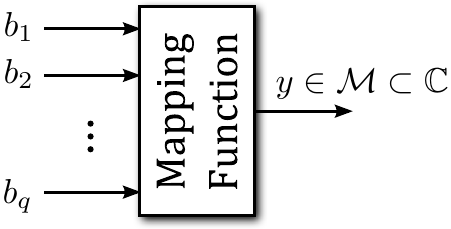}
\vspace*{-1ex}
\caption{Mapping of a group of $q$ bits $(b_1,b_2,\ldots,b_q)$ to a modulation symbol $y\in\mathbb{C}$}
\label{fig:mapping}
\end{figure}

When talking about digital modulation, especially in the context of coded modulation, we are mostly interested in the \emph{mapping function}, which is that part of the modulator that assigns (complex) modulation symbols to bit patterns. We introduce in what follows the notation necessary for describing the mapping function. Let 
$q$ denote the number of bits that are assigned to one complex modulation symbol $y\in\mathbb{C}$, and let $\bm{b} =(b_1, b_2,\ldots, b_q)\in\field_2^q$ be a binary $q$-tuple with $\field_2=\{0,1\}$ denoting the field of binary numbers.  
The one-to-one modulation mapping function $y = \phi(\bm{b})$   maps the $q$-tuple $\bm{b}$ to the (complex) modulation symbol $y\in\mathbb{C}$, where $y$ is chosen from the set of $Q=2^q$ modulation symbols $\mathcal{M}=\{M_0, M_1, \ldots, M_{Q-1}\}$. The set $\mathcal{M}$ is also commonly referred to as \emph{constellation}. The mapping function is illustrated in Fig.~\ref{fig:mapping}.
In this chapter, we only consider one-to-one mappings. One such mapping is $\phi_{\text{Nat}}(\bm{b}) = \phi_{\text{Nat}}(b_1,b_2,\ldots,b_q) = M_{(b_1b_2\ldots b_q)_{10}}$, where $(b_1b_2\ldots b_q)_{10}$ denotes the decimal expansion of the binary $q$-digit number $b_1b_2\ldots b_q$.

\begin{sloppypar}
In the context of differential coding of higher-order modulation formats, it is advantageous if the constellation $\mathcal{M}$ fulfills certain properties. One such property is the \emph{rotational invariance} of the constellation.
\begin{definition}[Rotational Invariance of Constellation]
We say that a constellation $\mathcal{M} = \{M_0, M_1, \ldots, M_{Q-1}\}$ exhibits a $V$-fold rotational invariance if we recover the original constellation $\mathcal{M}$ after rotating each modulation symbol $M_i$ by an amount $\frac{2\pi}{V}k$, $\forall k\in\{1,\ldots,V\}$ in the complex plane. Formally, we say that a constellation exhibits a $V$-fold rotational invariance if (with $\imath = \sqrt{-1}$)
\[
\{M_i\cdot \mathrm{e}^{\imath\frac{2\pi}{V}k} : M_i \in\mathcal{M} \} = \mathcal{M}\qquad \text{for all }k\in\{1,\ldots, V\}\,.
\]
\end{definition}
\begin{example}
Consider the two constellations with 8 and 16 points shown in Fig.~\ref{fig:qam8_and_qam16}.
The rectangular 8-QAM (\emph{quadrature amplitude modulation}) constellation of Fig.~\ref{fig:qam8_and_qam16}-(a) has a $V=2$ two-fold rotational invariance as any rotation of the constellation by $\pi$ leads again to the same constellation. The 16-QAM constellation shown in Fig.~\ref{fig:qam8_and_qam16}-(b) exhibits a $V=4$ four-fold rotational invariance as any rotation of the constellation by $\frac{\pi}{2}$ leads again to the same constellation.
\begin{figure}[h!]
\centering
\includegraphics{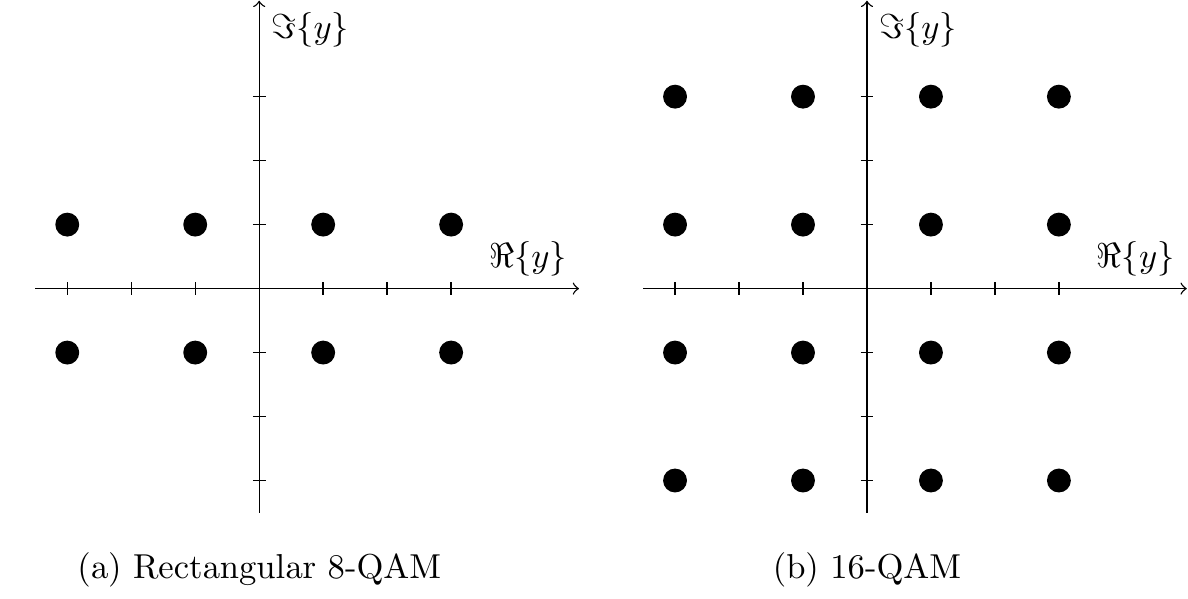}
\caption{Two common higher order constellations: (a) Rectangular 8-QAM with $Q=8$ and $V=2$, and (b) 16-QAM with $Q=16$ and $V=4$}
\label{fig:qam8_and_qam16}
\end{figure}
\hrule
\end{example}
\end{sloppypar}
Before introducing differential coding and modulation, we first describe the channel model including phase slips.

\subsection{The Phase Slip Channel Model}\label{sec:channel_model}

In coherent receivers for high-speed optical communications, it is usually not feasible to employ decision-directed blind phase recovery~\cite{PfauOFC2014} so that usually, feed-forward phase recovery algorithms have to be employed. Feed-Forward carrier recovery algorithms exploit the rotational invariance of the constellation to remove the modulation prior to estimating the phase. However, due to the necessary phase unwrapping algorithm in the feed-forward phase estimator, a phenomenon called \emph{phase slip} occurs\footnote{Sometimes, phase slips are also denotes as \emph{cycle slips}, however, we employ the term phase slip in this chapter.}. These are mostly due to coarse blind phase recovery algorithms with massive parallelization including preliminary hard decisions and phase unwrapping at the initial \emph{digital signal processing} (DSP) receiver steps~\cite{PfauOFC2014}.

\begin{figure}[t!]
\centering
\includegraphics{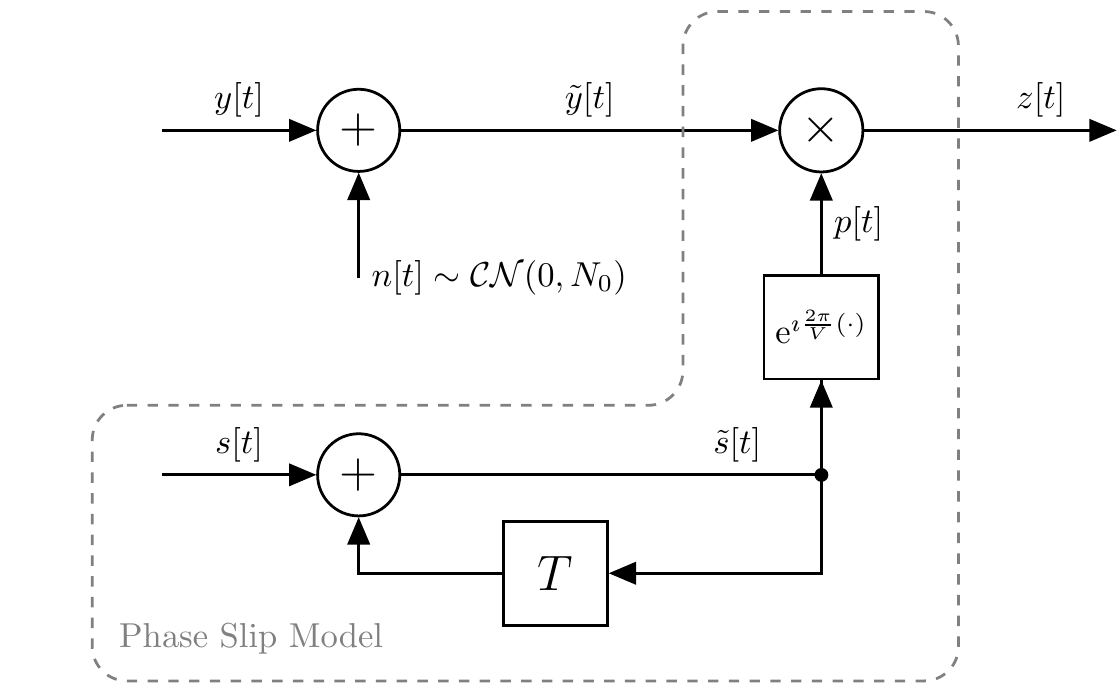}
\caption{AWGN channel model with phase slips}\label{fig:channel_model}
\end{figure}

Figure~\ref{fig:channel_model} displays the phase-slip channel model we employ in the following. The channel input is a complex modulation symbol $y[t]\in\mathcal{M}\subset\mathcal{C}$. The first noise contribution is complex-valued AWGN.
In the field of coding and in the broad body of literature on forward error correction, the terms $\Es/\No$ and $\Eb/\No$ are frequently used to characterize AWGN channels. Therein, $\Es = \mathsf{E}\{|y|^2\}$ denotes the energy per modulation symbol\footnote{Note that in this chapter we use lower case letters to denote random variables as well as their realizations to avoid confusion, unless it is not clear from the context.}. The noise $n[t] = n_I[t] + \imath n_Q[t]$ (where $\imath=\sqrt{-1}$) is characterized by the two-sided noise power spectral density $\No=2\sigma_n^2$ where $\sigma_n^2$ is the variance of both noise components $n_I[t]$ and $n_Q[t]$, i.e., $\sigma_n^2 = \mathsf{var}(n_I[t]) = \mathsf{var}(n_Q[t])$. The received symbol $z[t]$ in our model is obtained by $z[t] = (y[t] + n[t])\cdot p[t]$, where $p[t]$ describes the phase slips. Phase slips and $p[t]$ will be discussed in detail below.

Frequently, especially for comparing different coding schemes, $\Eb/\No$ is used instead of $\Es/\No$. Herein, $\Eb$ denotes the energy per \emph{information bit} whereas $\Es$ denotes the energy per \emph{transmit symbol}. For example, if a code of rate $r=4/5$, corresponding to an overhead of $\Omega=\frac{1}{r}-1\doteq 25$\%, is used, the ratio of $n$ code bits versus $k$ information bits amounts to $n/k = 5/4= 1.25$, i.e., $1.25=1/r$ code bits are transmitted for each information bit. Thereof, $q$ code bits are assigned to one modulation symbol $y[t]$. This means that if the modulation symbols will be transmitted each with energy $\Es$, the amount of energy conveyed by each information bit amounts to 
\[
\Es = \Eb\cdot q\cdot r \quad \Leftrightarrow\quad \Eb = \frac{\Es}{q\cdot r}\,.
\]
As $\Eb$ is normalized to the information bits of the transmission system, it allows us to immediately evaluate the \emph{net coding gain} (NCG).  The NCG is frequently used to assess the performance of a coding scheme and is defined as
the difference (in dB) of required $\Eb/\No$ values between coded and uncoded transmission for a given output BER. Note that the NCG takes into account
the coding rate $r$ and the number of bits assigned to each modulation symbol, which are included in $\Eb$.

In optical communications, the \emph{optical signal-to-noise ratio} (OSNR) is also frequently employed. The OSNR is the signal-to-noise ratio measured in a reference optical bandwidth, where frequently a bandwidth $B_{\rm ref}$ of 12.5\,GHz is used corresponding to $0.1$\,nm wavelength. The OSNR relates to the $\Es/\No$ and $\Eb/\No$ as 
\begin{align*}
{\rm OSNR}\bigg\vert_{\rm dB} &=\frac{\Es}{\No}\bigg\vert_{\rm dB} + 10\log_{10}\frac{R_S}{B_{\rm ref}} =\frac{\Eb}{\No}\bigg\vert_{\rm dB} + 10\log_{10}\frac{q\cdot r\cdot R_S}{B_{\rm ref}} 
\end{align*} 
where $B_{\rm ref}$ is the previously introduced reference bandwidth, $R_S$ corresponds to the symbol rate of the transmission, $r$ is the aforementioned rate of the code with $r=k/n$ and $q$ corresponds to the number of bits mapped to each modulation symbol.

Returning to the description of the channel model of Fig.~\ref{fig:channel_model}, we see that the noisy signal $\tilde{y}[t]$ additionally undergoes a potential phase rotation yielding $z[t]$. If the constellation shows a $V$-fold rotational invariance with $V$ even (which is the case for most of the practically relevant constellations), we introduce the following probabilistic phase slip model 
\begin{align*}
P(s[t] = \pm 1) &= \xi \\
P(s[t] = \pm 2) &= \xi^2 \\
\vdots &\\
P\left(s[t] = \pm \frac{V}{2}\right) &= \xi^{V/2}
\end{align*}
The probability that a phase slip occurs is thus
\begin{align}
P_{\text{slip}} &= 2\sum_{i=1}^{V/2}\xi^i = 2\left(\frac{1-\xi^{\frac{V}{2}+1}}{1-\xi}-1\right) = \frac{2\xi\left(1-\xi^{V/2}\right)}{1-\xi}\,.\label{eq:slip_versus_xi}
\end{align}
For a given phase slip probability, which may be obtained from measurements~\cite{FludgerOFC2012}, and which depends on the non-linear phase
noise introduced by the optical transmission link and variance of the additive Gaussian noise due to amplification, we obtain the value $\xi$ by solving \eqref{eq:slip_versus_xi} for $\xi$. For the practically most important cases with $V=2$, and $V=4$, we get
\begin{align}
\xi = \left\{\begin{array}{ll}
\frac{P_{\text{slip}}}{2} & \text{if}\ V=2 \\[0.8ex]
\frac{\sqrt{2P_{\text{slip}}+1}}{2}-\frac{1}{2} & \text{if}\ V=4\,.
\end{array}\right.\label{eq:xi_from_Pslip}
\end{align}
Experimental measurements~\cite{KoikeAkino} suggest that the phase slip probability depends on the equivalent bit error rate before the FEC decoder. Such a dependency was also suggested in~\cite{LeongJLTSlips}. We may thus model $P_{\text{slip}}$ empirically as
\begin{align}
P_{\text{slip}} = \left\{\begin{array}{l@{\qquad}l}
\min\left(1,\frac{\gamma}{2}\mathop{\text{erfc}}\left(\sqrt{\frac{E_s}{N_0}}\right)\right), & \text{for BPSK} \\[0.8ex]
\min\left(1,\frac{\gamma}{2}\mathop{\text{erfc}}\left(\sqrt{\frac{E_s}{2N_0}}\right)\right), & \text{for QPSK} 
\end{array}\right.\label{eq:slipprob_factor}
\end{align}
where $\gamma$ is the factor between slip rate and pre-FEC bit error rate for the equivalent BPSK channel. Given $\Es/\No$ and $\gamma$, we can compute $P_{\text{slip}}$ from~\eqref{eq:slipprob_factor} and subsequently $\xi$ from~\eqref{eq:xi_from_Pslip} or~\eqref{eq:slip_versus_xi}. Using $\xi$, we can use a pseudo-random number generator to generate a sequence of $s[t]$ with the probability mass function defined above.

\newcommand{\stat}{\mathsf{S}}

\subsection{Differential Coding and Decoding}\label{sec:differential}
Several approaches to mitigate
phase slips have been proposed in the literature. Probably the
most common is differential coding, rendering a
phase slip into a single error event. 
In this section, we restrict ourselves for simplicity to constellations with a $V$-fold rotational invariance where $V=2^v$, $v\in\mathbb{N}$, i.e., $V\in\{2,4,8,16,\ldots\}$.

We consider two different cases:
\begin{enumerate}
\item In the first case, we have $V=Q$. To each constellation point, we assign a state $\stat_{i}$, $i\in\{1,\ldots, V\}$. An example of such a constellation is the widely used QPSK constellation with $V=4$, which is shown in Fig.~\ref{fig:qpsk_state} together with its state assignment.
\begin{figure}[h!]
\centering
\vspace*{-1ex}
\includegraphics[scale=1]{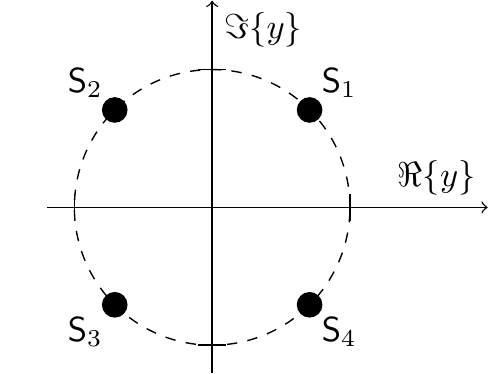}
\vspace*{-1ex}
\caption{Example of a QPSK constellation with state assignment in rotational order}
\label{fig:qpsk_state}
\end{figure}
\item In the second case, we have $Q > V$. We restrict ourselves to the practical case with $Q = J\cdot V$, where $J$ is an integer number.
In this case, we employ differential coding as described in~\cite{Weber0378}: The constellation is divided into $V$ disjoint regions such that these regions are preserved when rotating the constellation by $\pm\frac{2\pi}{V}$. 
We assign a state label $\stat_i$ to each disjoint region. The regions are selected such that each region contains exactly $J = \frac{Q}{V} = 2^j$ constellation points and such that a rotation of the constellation by an angle $\kappa\cdot \frac{2\pi}{V}$, $\kappa\in\{0,\pm 1, \pm 2, \ldots\}$ does neither change the regions nor the assignment of points to a region. For the constellation points within each region we employ a \emph{rotationally invariant} bit mapping, which means that the bit mapping of points inside a region is not changed by a rotation of the constellation by an angle $\kappa\cdot \frac{2\pi}{V}$. The popular 16-QAM constellation is an example of such a constellation with $Q=16$, $V=4$ and $J=4$. The state assignment and rotationally invariant mapping are exemplarily discussed in Example~\ref{ex:qam16ri} and shown in Fig.~\ref{fig:qam16_diff}.
\end{enumerate}

\begin{example}\label{ex:qam16ri}
We consider the transmission of the popular 16-QAM constellation~\cite{PfauJLT2009}. It can be easily verified that the 16-QAM constellation shows a $V=4$-fold rotational invariance. As shown in Fig.~\ref{fig:qam16_diff}, we label the four quadrants of the complex plane by \emph{states} $\stat_1$, $\stat_2$, $\stat_3$, and $\stat_4$. Inside the first quadrant $\stat_1$, we employ a Gray labeling (also denoted by \emph{mapping}) to assign the bits $b_3$ and $b_4$ to the four points. The mapping of the bits $b_3$ and $b_4$ in the three remaining quadrants is obtained by applying a rotational invariant mapping, i.e., by rotating the Gray mapping of $\stat_1$ by multiples of $\frac{\pi}{2}$. In this case, even by rotating the constellation by multiples of $\frac{\pi}{2}$, the bits $b_3$ and $b_4$ can always be recovered unambiguously. 
\begin{figure}[h!]
\centering
\includegraphics[scale=1]{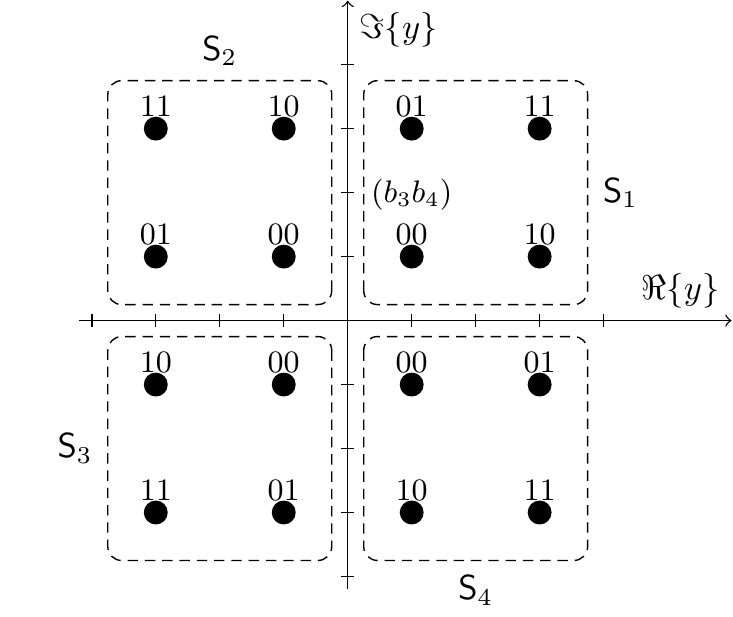}
\caption{Differential coding for the 16-QAM constellation $(V=4)$ with rotational invariant bit mapping in each quadrant}\label{fig:qam16_diff}
\end{figure}
\hrule
\end{example}

We employ differential coding with $v=\log_2(V)$ bits to encode and reliably transmit the region, i.e., the state. Within each of these regions, exactly $Q/V$ constellation points are placed, to which  a \emph{rotationally invariant} bit mapping is assigned. This means that whenever the constellation is rotated by an angle that is a multiple of $\frac{2\pi}{V}$, the bit patterns assigned to constellation points within the region can still be uniquely identified. Note that we restrict ourselves to state-region assignments such that the rotation of a complete region gives another valid region, i.e., $\forall i \in\{1,\ldots, V\}$, there exists a $j \in\{1,\ldots, V\}$, such that
\[
\left\{z\cdot \mathrm{e}^{\imath \kappa \frac{2\pi}{V}} : z \in\stat_i\right\} = \stat_{j},\qquad \forall\kappa\in\{0,\pm 1, \pm 2, \ldots\}\,.
\]
Note that this restriction does not impose any problems for practical systems as most of the practically relevant constellations can be described in this form.
In what follows, we impose another, slightly more stringent condition on the states. We assume that the states $\stat_i$ are assigned in what we denote as \emph{rotational order}. Formally,
\begin{definition}
We define a sequence of states $\stat_i$, $i\in\{1,\ldots, V\}$ that are assigned to a region of the complex plane, to be in \emph{rotational order}, if and only if the following condition
\[
\left\{z\cdot \mathrm{e}^{\imath \kappa \frac{2\pi}{V}} : z \in\stat_i\right\} = \stat_{((i+\kappa-1)\!\!\!\!\mod V)+1},\qquad \forall\kappa\in\mathbb{N}
\]
is fulfilled.
\end{definition}
We can easily verify that the state assignments of the constellations given in Fig.~\ref{fig:qpsk_state} and Fig.~\ref{fig:qam16_diff} are in rotational order. Again, note that the restriction of the states to be in \emph{rotational order} does not yet impose any major constraint, as we have not yet defined an encoding map. We group the $V$ states into the set $\mathcal{S}:=\{\stat_1, \stat_2, \ldots, \stat_V\}$.

\newcommand{\diffmem}[1]{\mathsf{d}^{[\text{mem}]}_{#1}}

\begin{figure}[tbh!]
\centering
\includegraphics[scale=1]{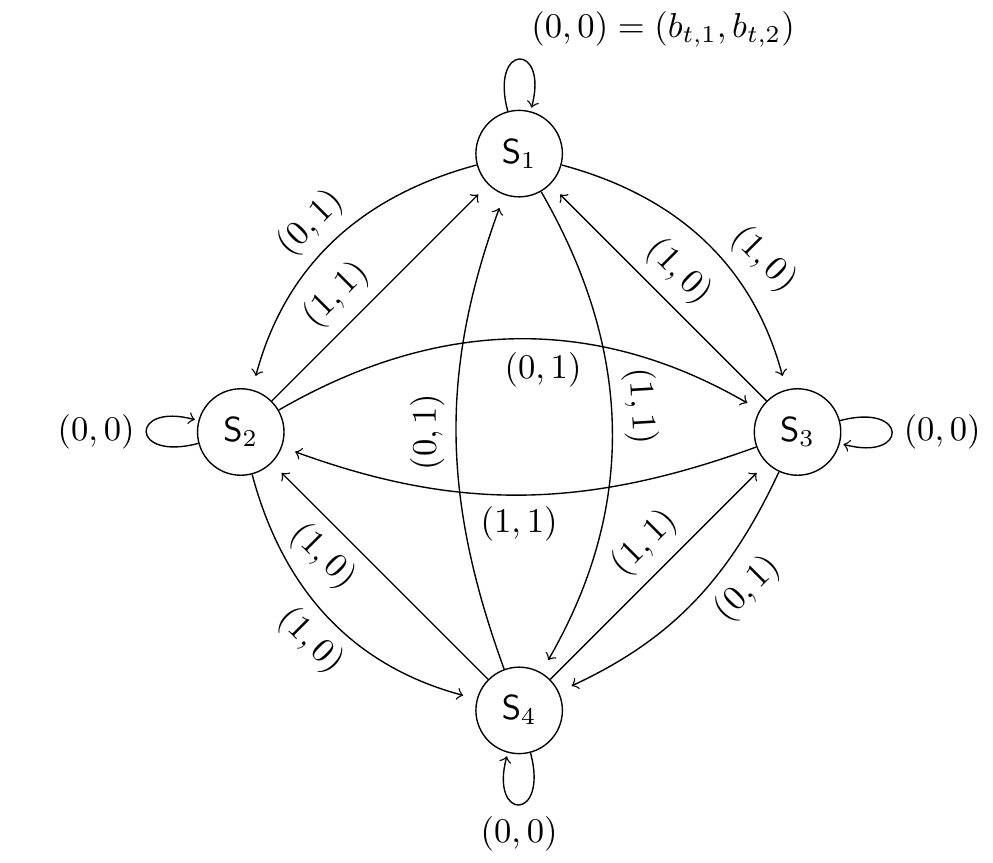}
\caption{Differential encoding state transition diagram for the  natural differential code. Arrow annotations are binary labels ($b_{t,1},b_{t,2}$)}
\label{fig:natural_state_transition}
\end{figure}

The main step in differential coding is to impose memory on the modulation. We assume that the transmission starts at time instant $t=1$. We introduce the differential memory $\diffmem{t}\in\mathcal{S}$ and set $\diffmem{0}=\stat_1$. The differential encoder can be considered to be the function 
\begin{align*}
f_{\text{diff}} : & \mathcal{S}\times\field_{2}^v \rightarrow \mathcal{S} \\
& (\diffmem{t-1},b_{t,1},b_{t,2},\ldots,b_{t,v}) \mapsto f_{\text{diff}}(\diffmem{t-1},b_{t,1},b_{t,2},\ldots, b_{t,v}) = \diffmem{t},
\end{align*}
which takes as input the bits $b_{t,1},\ldots, b_{t,v}$ and the differential memory $\diffmem{t-1}$ and generates a new state that is saved in the differential memory $\diffmem{t}$. This new state $\diffmem{t}$ selects the symbol to be transmitted (if $V=Q$) or the region from which the symbol is selected using the bits $b_{t,v+1},\ldots,b_{t,q}$. Note that the differential function is not unique but depends on the assignment of bit patterns to state transitions. 
Consider the example of the QPSK constellation shown in Fig.~\ref{fig:qpsk_state}. We can give two distinct differential encoding maps. The first differential encoding function is the \emph{natural differential code}. The state transition diagram of the natural differential code is visualized in Fig.~\ref{fig:natural_state_transition} and is also given in Tab.~\ref{tab:qpsk_natdiff}. The second encoding function, baptized \emph{Gray differential code} is given in Tab.~\ref{tab:qpsk_graydiff}. Note that all other differential coding maps for the QPSK constellation can be transformed into one of 
these two forms by elementary transformations of the constellation and the state assignment.

\begin{table}
\caption{Differential encoding map $f_{\text{diff}}$ for the natural differential code}
\label{tab:qpsk_natdiff}
\centering
\begin{tabular}{@{\quad}ccccc@{\quad}}
\hline
& & & & \\[-1.2em]
$(b_{t,1},b_{t,2})$ & $\diffmem{t-1} = \stat_1$ & $\diffmem{t-1} = \stat_2 $ & $\diffmem{t-1} = \stat_3$ & $\diffmem{t-1} = \stat_4$ \\
\hline
$(0,0)$ & $\stat_1$ & $\stat_2$ & $\stat_3$ & $\stat_4$ \\
$(0,1)$ & $\stat_2$ & $\stat_3$ & $\stat_4$ & $\stat_1$ \\
$(1,0)$ & $\stat_3$ & $\stat_4$ & $\stat_1$ & $\stat_2$ \\
$(1,1)$ & $\stat_4$ & $\stat_1$ & $\stat_2$ & $\stat_3$ \\
\hline
\end{tabular}
\end{table}

\begin{table}
\caption{Differential encoding map $f_{\text{diff}}$ for the Gray differential code}
\label{tab:qpsk_graydiff}
\centering
\begin{tabular}{@{\quad}ccccc@{\quad}}
\hline
& & & & \\[-1.2em]
$(b_{t,1},b_{t,2})$ & $\diffmem{t-1} = \stat_1$ & $\diffmem{t-1} = \stat_2 $ & $\diffmem{t-1} = \stat_3$ & $\diffmem{t-1} = \stat_4$ \\
\hline
$(0,0)$ & $\stat_1$ & $\stat_2$ & $\stat_3$ & $\stat_4$ \\
$(0,1)$ & $\stat_2$ & $\stat_3$ & $\stat_4$ & $\stat_1$ \\
$(1,0)$ & $\stat_4$ & $\stat_1$ & $\stat_2$ & $\stat_3$ \\
$(1,1)$ & $\stat_3$ & $\stat_4$ & $\stat_1$ & $\stat_2$ \\
\hline
\end{tabular}
\end{table}

\begin{figure}[htb!]
\centering\includegraphics[scale=1]{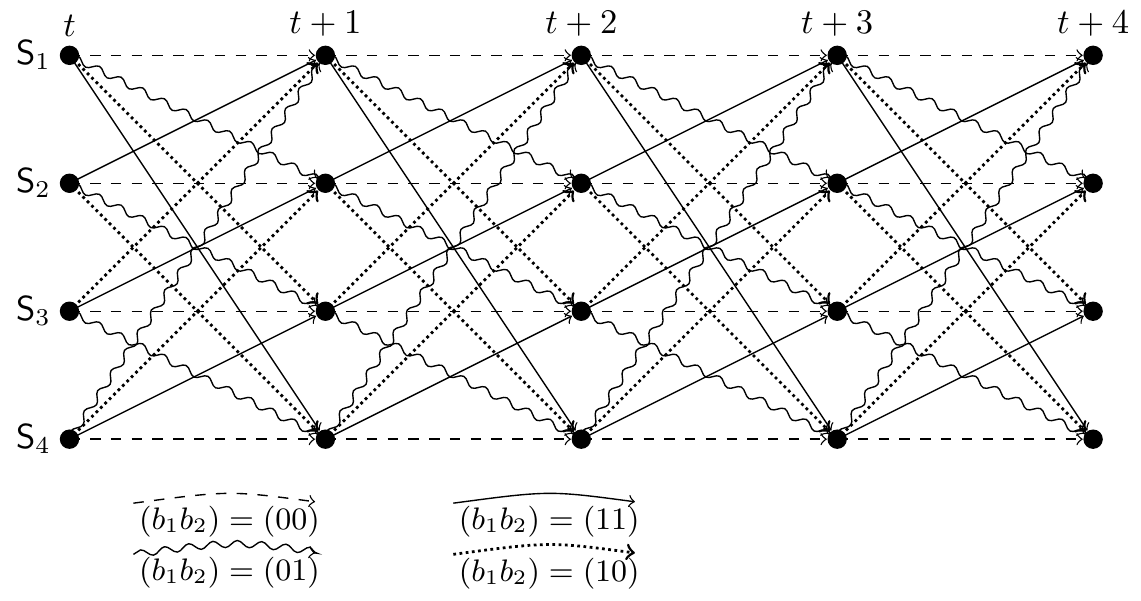}
\caption{Trellis diagram for \emph{natural} differential encoding of a constellation with $V=4$ as given in Tab.~\ref{tab:qpsk_natdiff}}
\label{fig:trellisV4nat}
\end{figure}

\begin{figure}[htb!]
\centering\includegraphics[scale=1]{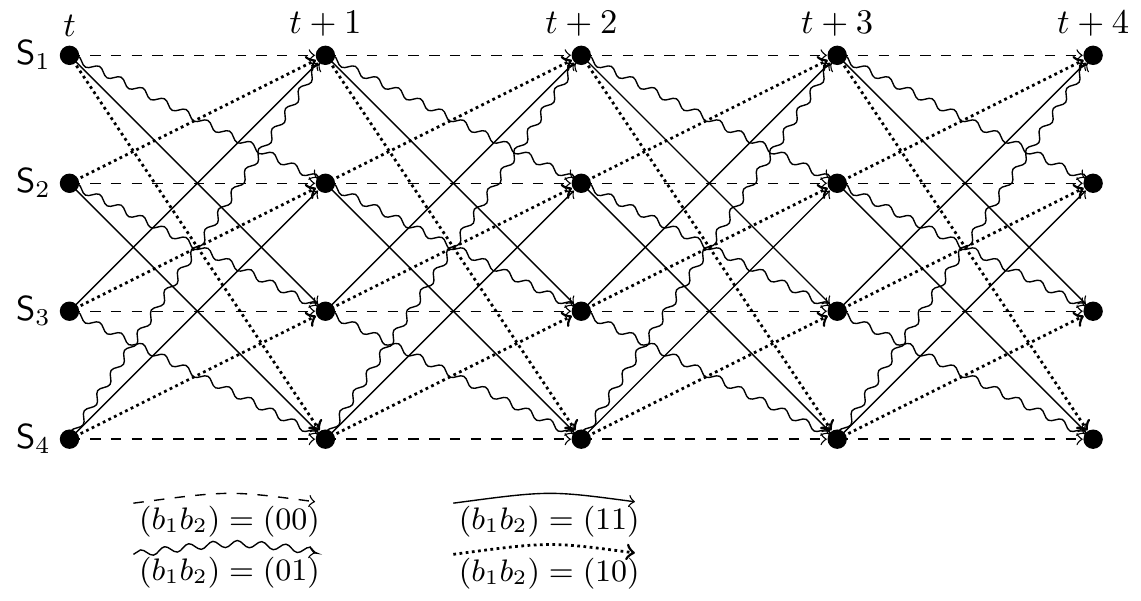}
\caption{Trellis diagram for \emph{Gray} differential encoding of a constellation with $V=4$  as given in tab.~\ref{tab:qpsk_graydiff}}
\label{fig:trellisV4gray}
\end{figure}

As the differential code can be understood as a Markov process, we can  employ the BCJR algorithm~\cite{Bahl0374} to carry out \emph{bit-wise Maximum A Posteriori (MAP)} decoding of the differential code. For this, we may represent the differential code using a so-called trellis diagram. The trellis diagram is an ``unrolled'' version of the state diagram of Fig.~\ref{fig:natural_state_transition}. Figure~\ref{fig:trellisV4nat} shows four segments of a trellis diagram for the natural differential encoding map. Four segments of the trellis diagram of the Gray differential encoding map are given in Fig.~\ref{fig:trellisV4gray}. The different input bit patterns $(b_1,b_2)$ can be distinguished by different line styles (dashed, dotted, solid and ``waved'').

\begin{figure}[htb!]
\centering\includegraphics[scale=1.05]{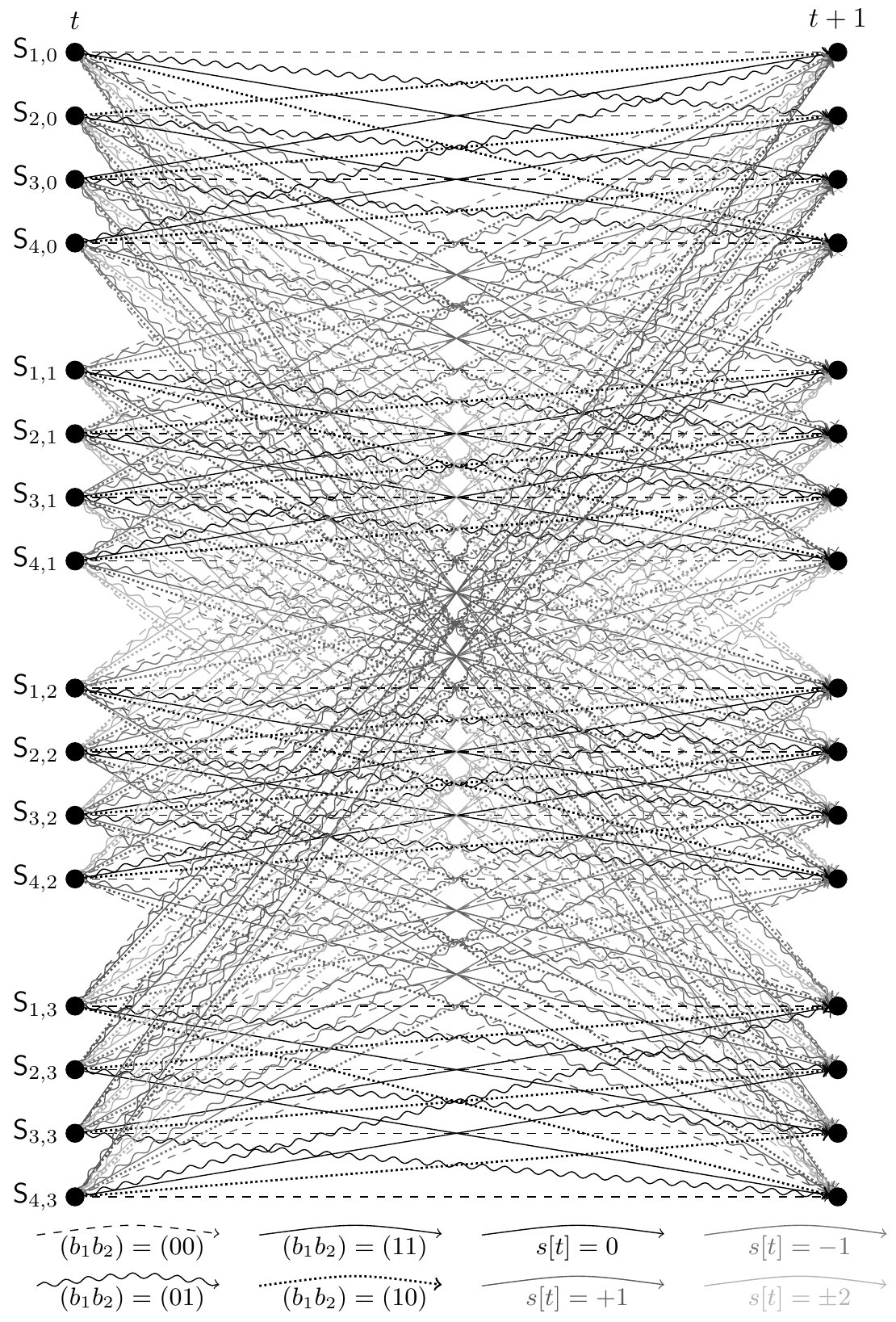}\caption{Trellis diagram for \emph{Gray} differential encoding of a constellation with $V=4$ taking into account the possible phase slips and tracking the phase slip state (indicated by four distinct line types and gray scales)}
\label{fig:trellisV4slipsnoncoll}
\end{figure}

If phase slips occur on the channel, memory is imposed on the channel as well. If this additional memory is not properly accounted for in the BCJR decoder of the differential code, the performance of the decoder will rapidly decrease, due to the decoder not being properly adapted to the channel model, as has been observed in~\cite{BisplinghoffECOC2012}. 
We therefore need to extend the trellis to properly \emph{take into account the phase slips}. One such extension introduces additional states that correspond to the memory of the phase slip channel~\cite{PatentLtB}. We introduce states $\stat_{i,\tilde{s}}$ where the second index $\tilde{s}$ tracks the current phase slip state $\tilde{s}[t]\text{ mod } 4$ (see Fig.~\ref{fig:channel_model}), while the first index $i$ is still responsible for describing the differential code. The occurrence of a phase slips ($s[t] \neq 0$)  leads to a different $\tilde{s}[t]$. For the running example of a differential code for $V=4$, we have no longer a trellis diagram (or a state transition diagram) with 4 states and $4\cdot 4=16$ state transitions, but instead a trellis diagram with $4\cdot 4 = 16$ states and $16\cdot 16 = 256$ state transitions. One segment of this extended trellis diagram is shown in Fig.~\ref{fig:trellisV4slipsnoncoll} for the Gray differential encoding map. In order to distinguish the additional state transitions corresponding to phase slips, we use grey scales. The original trellis is obtained by utilizing only those state transitions that correspond to $s[t]=0$, which correspond to the black lines. The state transitions corresponding to $s[t]=1$ and $s[t]=3$ are given by \textcolor{black!60!white}{grey} lines while the state transitions corresponding to $s[t]=2$ are given by \textcolor{black!30!white}{light grey} lines, as these have the lowest probability of occurrence.

\begin{figure}[htb!]
\centering\includegraphics[scale=1]{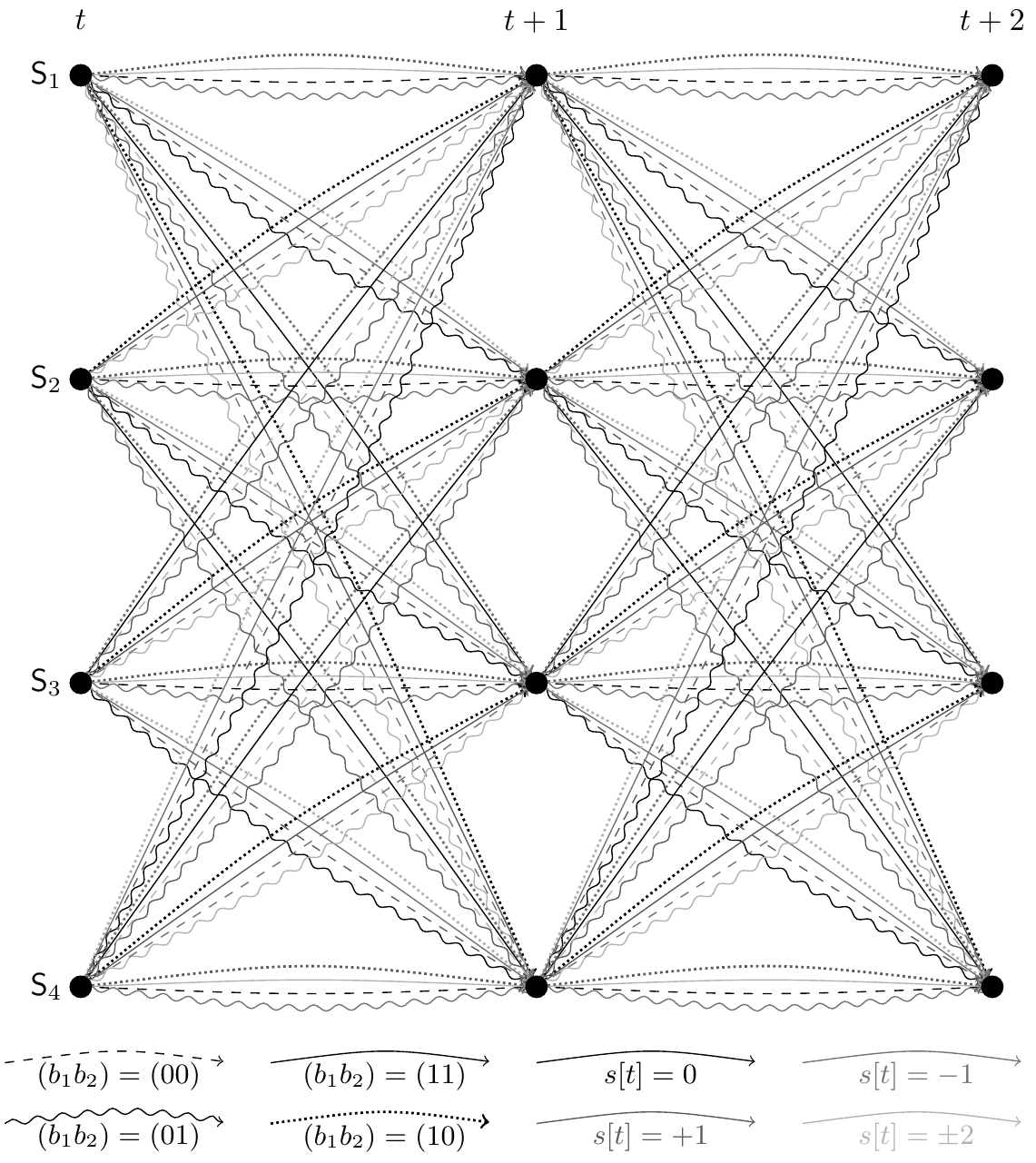}\caption{Trellis diagram for \emph{Gray} differential encoding of a constellation with $V=4$ taking into account the possible phase slip (indicated by four distinct line types and gray scales)}
\label{fig:trellisV4slipsnew}
\end{figure}

As the trellis diagram of Fig,~\ref{fig:trellisV4slipsnoncoll} may be challenging to implement, we seek for a way to reduce its complexity. By observing that the memory of the phase slip channel collapses with the memory of the differential encoder, we may get a more compact representation of the trellis and only need $V$ states. This is possible as a phase slip does not introduce a new state, but only to a different state transition to one of the $V$ existing states. In fact we have
\[
{\diffmem{t}}^\prime = \stat_{((i+s[t]-1)\!\!\!\mod V)+1}\qquad \text{with}\ \stat_i = \diffmem{t}
\]
The state transitions are given exemplarily for the case of the Gray differential encoder in Tab.~\ref{tab:qpsk_graydiff_ext}.
This means that we can still use a trellis diagram with $V$ states but have to insert additional state transitions taking into account all possible values of $s[t]$. Figure~\ref{fig:trellisV4slipsnew} shows the extended trellis diagram taking into account the possible slips, indicated by the slip value $s[t]\in\{0,1,2,3\}$. Again, we use differential grey scales to represent the state transitions corresponding to different values of $s[t]$. The trellis diagram of Fig.~\ref{fig:trellisV4slipsnew} is a simplification of the extended trellis diagram with only $V=4$ states (instead of 16) and $4\cdot 16 = 64$ state transitions (instead of~256). Another approach to take into account phase slips into an extended trellis has been presented in~\cite{KoikeAkino}. 

\begin{table}
\caption{Differential encoding map $f_{\text{diff}}$ for the Gray differential code taking into account the phase slip variable $s[t]$}
\label{tab:qpsk_graydiff_ext}
\centering
\begin{tabular}{@{\quad}cccccc@{\quad}}
\hline
& & & & & \\[-1.2em]
 $s[t]$ & $(b_{t,1},b_{t,2})$ & $\diffmem{t-1} =\stat_1$ & $\diffmem{t-1} =\stat_2$ & $\diffmem{t-1} =\stat_3$ & $\diffmem{t-1} =\stat_4$ \\
\hline
 0 & $(0,0)$ &$\stat_1$ & $\stat_2$ & $\stat_3$ & $\stat_4$ \\
 0 & $(0,1)$ &$\stat_2$ & $\stat_3$ & $\stat_4$ & $\stat_1$ \\
 0 & $(1,1)$ &$\stat_3$ & $\stat_4$ & $\stat_1$ & $\stat_2$ \\
 0 & $(1,0)$ &$\stat_4$ & $\stat_1$ & $\stat_2$ & $\stat_3$ \\
\hdashline[2.5pt/5pt]\\[-1.3em]
 1 & $(0,0)$ &$\stat_2$ & $\stat_3$ & $\stat_4$ & $\stat_1$ \\
 1 & $(0,1)$ &$\stat_3$ & $\stat_4$ & $\stat_1$ & $\stat_2$ \\
 1 & $(1,1)$ &$\stat_4$ & $\stat_1$ & $\stat_2$ & $\stat_3$ \\
 1 & $(1,0)$ &$\stat_1$ & $\stat_2$ & $\stat_3$ & $\stat_4$ \\
\hdashline[2.5pt/5pt]\\[-1.3em]
 2 & $(0,0)$ &$\stat_3$ & $\stat_4$ & $\stat_1$ & $\stat_2$ \\
 2 & $(0,1)$ &$\stat_4$ & $\stat_1$ & $\stat_2$ & $\stat_3$ \\
 2 & $(1,1)$ &$\stat_1$ & $\stat_2$ & $\stat_3$ & $\stat_4$ \\
 2 & $(1,0)$ &$\stat_2$ & $\stat_3$ & $\stat_4$ & $\stat_1$ \\
\hdashline[2.5pt/5pt]\\[-1.3em]
 3 & $(0,0)$ &$\stat_4$ & $\stat_1$ & $\stat_2$ & $\stat_3$ \\
 3 & $(0,1)$ &$\stat_1$ & $\stat_2$ & $\stat_3$ & $\stat_4$ \\
 3 & $(1,1)$ &$\stat_2$ & $\stat_3$ & $\stat_4$ & $\stat_1$ \\
 3 & $(1,0)$ &$\stat_3$ & $\stat_4$ & $\stat_5$ & $\stat_2$ \\
\hline
\end{tabular}
\end{table}

\subsection{Maximum a Posteriori Differential Decoding}
In what follows, we use the BCJR decoder~\cite{Bahl0374} to carry out bit-wise \emph{maximum a posteriori} differential decoding. The BCJR decoder makes a decision on the transmitted symbol (equivalent to a state) based on the maximization
\begin{align*}
\hat{\stat}[t] = \arg\max_{\mathsf{s}\in\{\stat_1,\ldots,\stat_V\}}P(\stat[t] = \mathsf{s} | \bm{z}_1^{\tilde{n}}).
\end{align*}
At each time instant $t$, the most probable state $\stat_t$ is computed given the complete received sequence $\bm{z}_1^{\tilde{n}}=\left(z[1],z[2],\ldots z[\tilde{n}]\right)$. We will not give a complete derivation of the BCJR algorithm and refer the interested reader to the literature, e.g., ~\cite{Bahl0374}, \cite{RyanBook}. We merely summarize the equations in the Appendix.

We use the technique of \emph{EXtrinsic Information Transfer} (EXIT) charts~\cite{tenBrink1001} to characterize the behavior of the differential decoder based on the BCJR algorithm. EXIT charts plot the extrinsic output mutual information as a function of the input mutual information and are a tool to characterize single components in iterative decoders. 
Bit interleavers statistically decouple the respective encoding/decoding
components such that a single parameter is sufficient to track their
input/output relations. This parameter may be the signal-to-noise
ratio at the output of a processing block, or, as is the case for
EXIT charts, the mutual information between transmitted bits and the
received and processed soft bit \emph{log-likelihood ratio} (LLR) values. For some
channels and some codes, the individual \emph{transfer characteristics}
(or EXIT curves) can be obtained analytically, while for most cases,
one has to resort to Monte Carlo simulation for computing the mutual information. EXIT
curves can be defined not only for channel encoders/decoders such
as convolutional codes or parity-check codes, but also for components
of many serially or parallel concatenated detection and decoding schemes:
For example, EXIT curves have been used for describing channel interfaces
such as mappers/demappers (detectors) for spectrally efficient modulation,
or equalizers of multipath channels; even the decoder of an LPDC code
can be viewed as a serial concatenation, with a variable node decoder
and a check node decoder that, both, can be described by EXIT curves,
respectively.

The main advantage of the EXIT chart technique is that the individual
component processing blocks can be studied and characterized separately
using EXIT curves, and that the interaction of two (or more) such
processing blocks can be graphically predicted in the EXIT chart without
performing a complex simulation of the actual fully-fletched concatenated
coding scheme itself. As it turns out, the EXIT curves must not intersect
to allow convergence to low bit error rates, and thus, code design
reduces to finding good pairs of EXIT curves that match well, or,
more constructively as in the case of LDPC codes, to apply curve-fitting
algorithms to determine variable and check node degree profiles that
match well. A decoding trajectory visualizes the iterative exchange
of information between the processing blocks, and shows the progress
of the decoding. 

While the EXIT chart is exact on the \emph{binary erasure channel} (BEC) 
for sufficiently long/infinite sequence lengths, the reduction to
single parameter tracking of the involved distributions is just an
approximation for other channels. It has been observed, however, that the
predicted and actually simulated decoding trajectories match quite well,
proving the usefulness of the method, with many successful code designs
performed in practice up to date.

\newcommand\solidrule[1][1cm]{\rule[0.5ex]{#1}{.4pt}}
\newcommand\dashedrule{\mbox{%
  \solidrule[2mm]\hspace{2mm}\solidrule[2mm]\hspace{2mm}\solidrule[2mm]}}
  
\begin{figure}[tb!]
\centering
\includegraphics[width=0.85\textwidth]{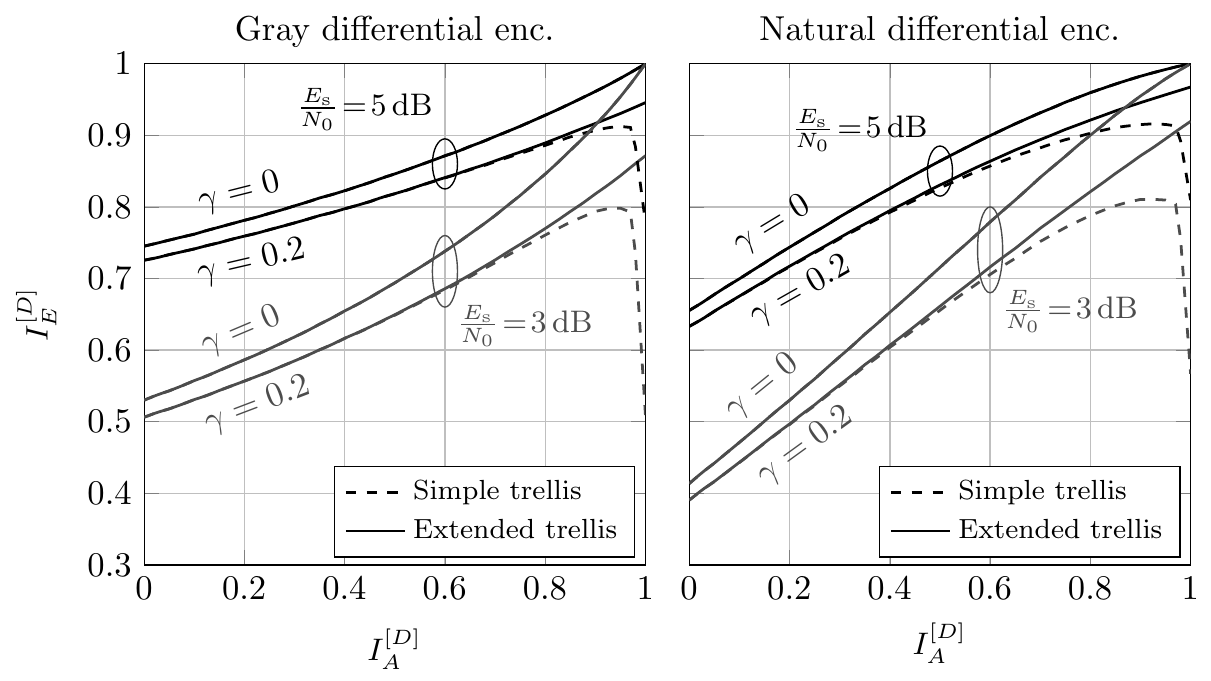}
\vspace{-4ex}
\caption{EXIT characteristics of differential detectors using the model-matched decoder with the trellis diagram of Fig.~\ref{fig:trellisV4slipsnew} (solid lines, \solidrule) and using the conventional unmatched decoder based on the trellis diagram of Fig.~\ref{fig:trellisV4gray} (or Fig.~\ref{fig:trellisV4nat}, respectively) (dashed lines, \protect\dashedrule)}
\label{fig:exit_detector}
\end{figure}

Figure~\ref{fig:exit_detector} shows the EXIT characteristics of the differential decoder for a QPSK constellation and both differential encoding maps. We can clearly see that the characteristic of the detector employing the non-matched trellis diagram has a non-increasing shape, which is an indicator of a \emph{mismatched model} used within the decoder: the decoder trellis does not leave the possibility open for phase slips to occur, but \emph{forces} the result to a simply differentially encoded target sequence, which, however, is not the case after the phase slip channel. This non-increasing shape is the reason for the error floor that has been observed in~\cite{BisplinghoffECOC2012}. The decreasing EXIT characteristic means that during iterative decoding, the overall system performance actually decreases, which can lead to a severe error floor. In~\cite{BisplinghoffECOC2014}, the authors proposed to employ \emph{hybrid turbo differential decoding} (HTDD): by a careful execution of the differential decoder only in those iterations where the extrinsic information is low enough, the operating point in the EXIT chart is in the range of an increasing characteristic. This approach allows the authors of~\cite{BisplinghoffECOC2014} to mitigate the detrimental effect of phase slips on iterative differential decoding and to realize codes with relatively low error floors which can be combated using a high-rate outer code.

If we employ the trellis diagram of Fig.~\ref{fig:trellisV4slipsnew} incorporating the phase slip model instead of the non-matched trellis diagram, we can see that the EXIT characteristics are monotonically increasing, which is a prerequisite for successful decoding with low error floors. In the next section, we use the EXIT characteristics to compute the information theoretic achievable rates of the differentially encoded system.
Further note that for $\gamma > 0$ (see Sec.~\ref{sec:channel_model}), the value of $\IE{D} < 1$, even for $\IA{D}=1$, which may entail an error floor unless the channel code is properly designed.

\subsection{Achievable Rates of the Differentially Coded Phase Slip Channel}\label{sec:capacity}

According to Shannon's information theory~\cite{Shannon1948,CoverThomas2006}, the \emph{capacity} of a communication channel is the maximum amount of information (usually expressed in terms of \emph{bits per channel use}) that can be reliably conveyed over the channel. In information theory, the capacity is usually maximized over the input distribution of the channel. In this chapter, we are only interested in the maximum achievable information rate for uniform channel inputs $y$, as we do not wish to impose any constraints on the data sequence. One possibility to achieve a non-uniform channel input is the use of \emph{constellation shaping}~\cite{ForneyShaping,SmithShaping}, which is however beyond the scope of this chapter. The comparison between the achievable rate of the channel affected by phase slips and the achievable rate of the original AWGN channel shows how much the performance may be sacrificed by the presence of phase slips. In order to compute the achievable rates of the differentially encoded channel affected by phase slips, we employ the EXIT chart technique. 

\begin{figure}[tbh!]
\includegraphics[width=0.75\textwidth]{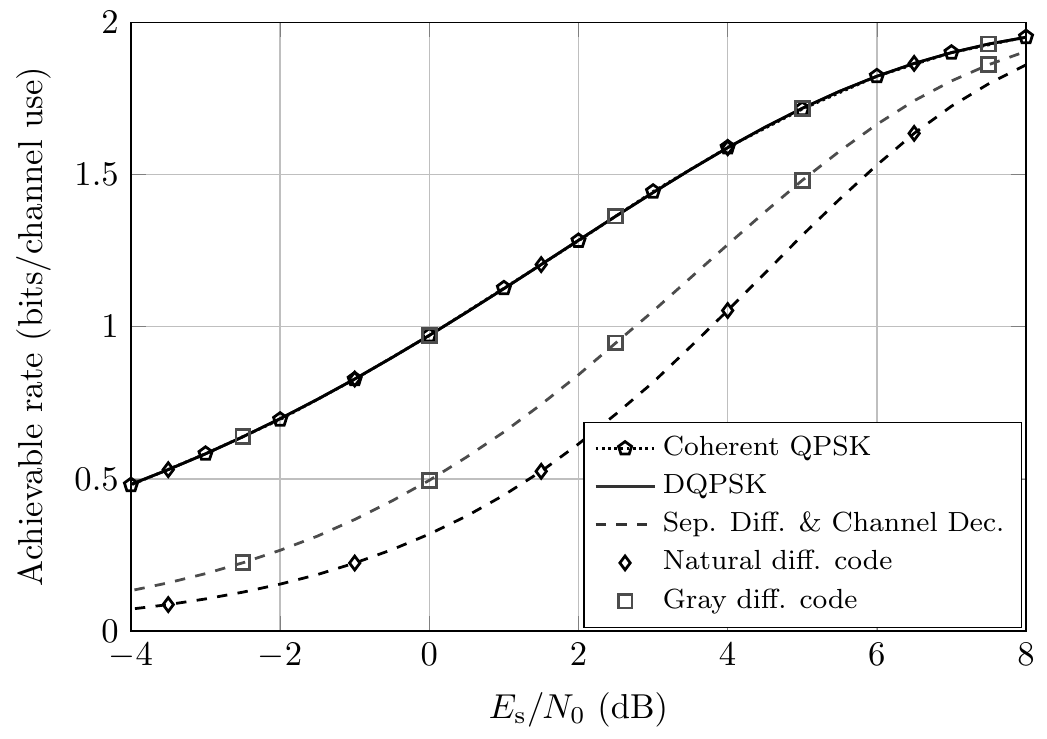}
\vspace*{-2ex}
\caption{Achievable rates of the DQPSK channel (solid lines, \solidrule) and of conventional separate differential \& channel decoding (dashed lines, \protect\dashedrule) for an AWGN channel without phase slips ($\gamma=0$)}
\label{fig:capacity_noslips}
\end{figure}

\begin{figure}[tbh!]
\centering
\includegraphics[width=0.75\textwidth]{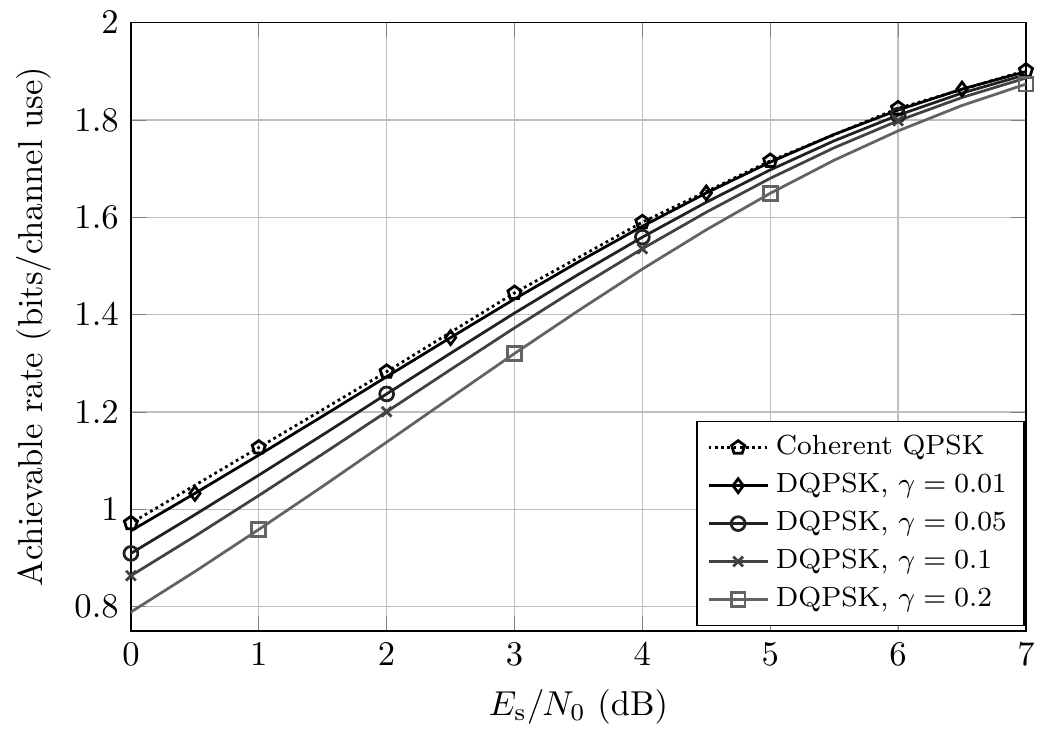}
\caption{Capacity of the differential DQPSK system for transmission over an AWGN channel affected by phase slips with probability of occurrence depending on pre-FEC bit error rate, given by \eqref{eq:slipprob_factor}}
\label{fig:capacity_factor}
\end{figure}

By utilizing a slightly modified way of computing EXIT curves of the BCJR decoder, we can also compute the achievable rates of the coded modulation schemes~\cite{GuilleniFabregas}. For this, we make use of the chain-rule of mutual information~\cite{tenBrink1001b,Ashikhmin1104}
and compute the mutual information of the equivalent bit channel experienced
by the channel decoder \emph{after} differential detection. This can
be done by (numerically, simulation-based) computing the EXIT curve $\IEtilde{D}$
of the differential detector using \emph{a priori} knowledge that is modeled
as coming from a BEC, and integrating over such curves. Specifically, EXIT curves like those depicted in Fig.~\ref{fig:exit_detector} are determined for many different $\Es/\No$-values
(and several different phase slip probabilities factors $\gamma$) but now
with \emph{a priori} knowledge based on a BEC model: By integration, we determine
the area $q\int_0^1\IEtilde{D}(I_A)\text{d}I_A$ under these curves~\cite{tenBrink1001b,Ashikhmin1104,GuilleniFabregas} and obtain the respective mutual information limits that are plotted into Figs~\ref{fig:capacity_noslips} and~\ref{fig:capacity_factor} at the corresponding $\Es/\No$-values and phase slip probabilities
factors $\gamma$, respectively. Note that this mutual information is available
to the channel decoder provided that \emph{perfect} iterative decoding
over inner differential detector and outer LDPC decoder is performed.
Thus, we still need to design an appropriate LDPC code and iterative
decoding scheme to actually approach these promised rates as closely
as possible. Indeed, the subsequent sections explain how to construct
such codes and coding schemes in more detail. The achievable rate of the non-iterative system with separate differential decoding and channel decoding is obtained from  $q\IEtilde{D}(0) = q\IE{D}(0)$.

\newcommand\mypentagon{%
  \begin{tikzpicture}
    \begin{axis}[hide axis, scale only axis, height=1.5ex]
      \addplot[mark=pentagon] coordinates {(0,0)};
    \end{axis}
  \end{tikzpicture}}
  
Figures~\ref{fig:capacity_noslips} and~\ref{fig:capacity_factor} show the numerically computed achievable rates for the QPSK constellation \emph{without} differential coding on an AWGN channel that is \emph{not affected} by phase slips (dotted lines, marker ``\mypentagon'') and additionally the achievable rates for differentially encoded QPSK for a channel affected by phase slips (solid lines) with $P_{\text{slip}} = \min\left(1,\frac{\gamma}{2}\mathop{\text{erfc}}\left(\sqrt{\frac{E_s}{2N_0}}\right)\right)$. In Fig.~\ref{fig:capacity_noslips} we set $\gamma=0$ and we observe that the achievable rate of the differential QPSK transmission equals the achievable rate of a conventional coherent QPSK transmission, \emph{independent} of the differential encoding map. Additionally, we plot the achievable rates for a simplified system that carries out differential decoding (leading to the well-known effect of error doubling) followed by error correction decoding (dashed lines). We see that at a spectral efficiency of $1.6$ (corresponding to system with $\Omega=25$\% overhead for coding), the simplified system leads to an unavoidable loss in $\Es/\No$ of $1.5$\,dB (Gray differential encoding map) or $2.5$\,dB (natural differential encoding map) respectively. This performance difference becomes even more severe if low spectral efficiencies (i.e., high coding overheads) are targeted.

If phase slips occur on the channel ($\gamma > 0$), we can observe in Fig.~\ref{fig:capacity_factor} that for high spectral efficiencies (above 1.5\,bits/channel use), the loss in information rate due to the phase slips is not severe, unless $\gamma$ becomes large. For example, for $\gamma=0.2$, the capacity loss at a spectral efficiency of $1.5$\,bit/channel use is only approximately $0.7$\,dB. 
The transmission at very low spectral efficiencies, requiring codes with very large overheads, is however seriously affected by the phase slip channel.

\section{LDPC Coded Differential Modulation}\label{sec:ldpc_diff_mod}

In the previous section, we have compared the achievable rates of various systems for an AWGN channel ($\gamma=0$) and we have found that differential coding can be used without entailing a decrease of the communication system's achievable rate. This means that at least from an information theoretic perspective, we can employ differential coding to combat phase slips without introducing any decoding penalty. Information theory however does not tell us what constructive method we may use to achieve this capacity.

One particularly promising way to approach the capacity with differential coding is the use of coded differential modulation with iterative decoding, as proposed first in~\cite{HoeherTurboDQPSK} with convolutional codes and in~\cite{FranceschiniJSAC} with LDPC codes. This scheme extends the \emph{bit-interleaved coded modulation} (BICM)~\cite{Caire0598} method to account for differential encoding and employs iterative decoding and detection~\cite{tenBrink1998,Li0699} to improve the overall system performance. The adaptation of this scheme to optical communications has been considered in~\cite{YuECOC2011} for the channel not affected by phase slips and in~\cite{PatentLtB,KoikeAkino,BisplinghoffECOC2012,BisplinghoffECOC2014} for the channel affected by phase slips. Note that other schemes have been proposed that do not rely on iterative differential decoding, including the slip resilient code presented in~\cite{SchmalenECOC14_slips,SchmalenJLT_slips} and block differential modulation~\cite{Bellini_slips}.

\begin{figure}[tb!]
\centering\includegraphics[scale=1]{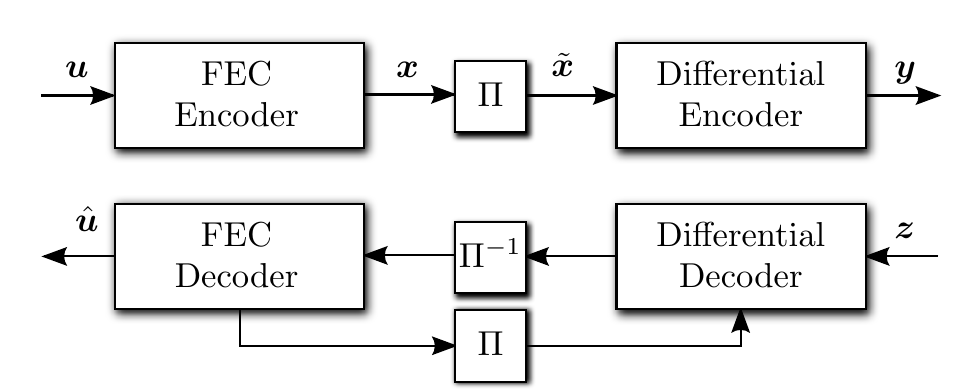}
\caption{Block diagram of LDPC coded differential modulation transmitter (top) with iterative detector (bottom)}
\label{fig:ldpc_cod_mod}
\end{figure}

Figure~\ref{fig:ldpc_cod_mod} shows the general transmitter (top) and iterative receiver (bottom) of the coded differential modulation system with iterative decoding and detection. In this general block diagram, a block FEC encoder takes as input a binary length-$k$ vector of inputs bits $\bm{u}=(u_1,u_2,\ldots, u_k)$, where $u_i\in\field_2=\{0,1\}$ and generates a binary length-$n$ vector of code bits $\bm{x} = (x_1, x_2, \ldots, x_n)$. Almost all of the popular channel codes that are used in optical communications are such block codes. The amount $n-k$ of redundant bits that are added by the FEC encoder is commonly expressed in terms of the code rate $r$ which is defined as the ratio of the information block length $k$ and the code dimension $n$, i.e.,
 \begin{align*}
 r := \frac{k}{n}\,.
 \end{align*}
In optical communications, often the overhead is  used to quantify the amount of redundant information. The overhead $\Omega$ of the code and its rate are interrelated by
 \begin{align*}
 \Omega := \frac{n}{k}-1=\frac{n-k}{k}=\frac{1}{r}-1=\frac{1-r}{r}\,.
 \end{align*}

The block $\bm{x}$ of code bits is interleaved by a permutation $\Pi$ to yield a permuted version $\tilde{\bm{x}}$. Ideally, a random permutation is employed, but sometimes, a structure in the permutation is necessary to facilitate implementation (parallelization) or to improve the error correction capabilities of the code. Note that the permutation $\Pi$ is sometimes implicitly included in the FEC encoder and does not need to be explicitly implemented. The interleaved block $\tilde{\bm{x}}$ is  differentially encoded (as discussed in Sec.~\ref{sec:differential}) yielding a block of $\tilde{n} = \lceil\frac{n}{q}\rceil$ modulation symbols (where $\lceil\tau\rceil$ denotes the smallest integer larger or equal than $\tau$).

At the receiver, the differential decoder and the FEC decoder iteratively decode the signal, where the output of the FEC decoder is used to yield an improved differential decoding result in a subsequent iteration by sharing so-called \emph{extrinsic information} between the decoder components. For a thorough description and introduction to the concept of iterative detection and decoding we refer the interested reader to~\cite{HanzoTurboBook,RyanBook}. In the remainder of this section, we assume that the employed FEC scheme is a low-density parity-check (LDPC)~\cite{GallagerLDPC,RyanBook} code. We will first give an introduction to LDPC codes and then show how irregular LDPC codes can be designed to be well-adapted to differential coding. We do not show explicitly how decoding is performed, as we intend to take on a more code design-oriented perspective. We will only give equations for performing differential decoding and LDPC decoding in the Appendix.

We restrict ourselves in the remainder of this chapter to the case where $V=Q=2^q$, i.e., every state $\stat_i$ is assigned to the modulation symbol $M_i$. We will however give hints on how to deal with the case $V < Q$ in Sec.~\ref{sec:VsmallQ}.

\subsection{Low-Density Parity-Check (LDPC) Codes}
Low-density
parity-check (LDPC) codes were
developed in the 1960s by Gallager in his landmark
Ph.D. thesis~\cite{GallagerLDPC}. These codes were not further investigated
for a long time due to the perceived complexity
of long codes. With the discovery of turbo codes in
1993~\cite{Berrou0593} and the sudden interest in iteratively decodable
codes, LDPC codes were rediscovered soon afterwards~\cite{MacKay1996,Luby1997}. In the years that followed, numerous
publications from various researchers paved the way
for a thorough understanding of this class of codes leading to numerous
applications in various communication standards,
such as, e.g., WLAN (IEEE
802.11)~\cite{IEEE2012}, DVB-S2~\cite{ETSI2009}, and 10G Ethernet (IEEE 802.3)~\cite{IEEE8023an}.
LDPC codes for soft-decision decoding in optical
communications were studied in~\cite{Miyata2009}. 
Modern high-performance FEC systems
are sometimes constructed using a soft-decision LDPC
inner code which reduces the BER to a level of $10^{-3}$
to $10^{-5}$ and a hard-decision outer code which pushes
the system BER to levels below $10^{-12}$~\cite{Miyata2009}. An outer cleanup code is used as most LDPC codes exhibit a phenomenon called \emph{error floor}: above a certain \emph{signal-to-noise ratio} (SNR), the BER does not drop rapidly anymore but follows a curve with a small slope. This effect is mainly due to the presence of \emph{trapping sets} or \emph{absorbing sets}~\cite{RichardsonFloor,DolecekFloor}. The implementation of a coding system with an outer cleanup code requires a thorough understanding of the LDPC code and a properly designed interleaver between the LDPC and outer code for avoiding that the errors at the output of the LDPC decoder---which typically occur in clusters---cause uncorrectable blocks after outer decoding. With increasing computing resources, it is now also feasible to evaluate very low target BERs of LDPC codes and optimize the codes to have very low error floors below the system's target BER~\cite{Morero}. 
A plethora
of LDPC code design methodologies exist, each with its own
advantages and disadvantages. The goal of an LDPC
code designer is to find a code that yields high coding
gains and which possesses some structure facilitating
the implementation of the encoder and decoder.
We point the interested reader to numerous
articles published on this topic, e.g.,~\cite{Bonello2011,Liva2006,Richardson2008} and references therein. An introduction to LDPC codes in the context of optical communications is given in~\cite{LevenSchmalenJLT2014}. An overview of coding schemes for optical communications is also provided in~\cite{DjordjevicRyan} and the references therein. For a thorough reference to LDPC codes together with an overview of decoding algorithms and construction methods, we refer the interested reader to~\cite{RyanBook}.

\newcommand{\bH}{\bm{H}}

An LDPC code is defined by a sparse binary parity check matrix $\bm{H}$ of dimension $m\times n$, where $n$ is the code word length (in bits) of the code and $m$ denotes the number of parity check equations defining the code. Usually\footnote{provided that the parity-check matrix has full row rank, i.e., $\mathop{\text{rank}}\bm{H} = m$. If the parity-check matrix $\bm{H}$ is rank-deficient, the number of information bits $k \geq n-m$}, the number of information bits equals $n-m$. The overhead of the code is defined as $\Omega = \frac{m}{n-m}$. A related measure is the \emph{rate} of the code, which is defined as $r = \frac{n-m}{n}$. Sparse means that the number of ``1''s in $\bm{H}$ is small compared to the number of zero entries. Practical codes usually have a fraction of ``1''s that is below 1\% by several orders of magnitude. We start by introducing some notation and terminology related to LDPC codes. Each column of the parity check matrix $\bm{H}$ corresponds to one bit of the FEC frame. The $n$ single bits of the code are also often denoted as \emph{variables}. Similarly, each row of $\bm{H}$ corresponds to a parity check equation and ideally defines a single parity bit (if $\bm{H}$ has full rank).

\newcommand{\ltw}{\mathtt{v}}
\newcommand{\ltwreg}{\mathtt{v}_{\text{reg.}}}
\newcommand{\rtw}{\mathtt{c}_{\text{reg.}}}
\newcommand{\rtwbest}{\mathtt{c}_{\text{best}}}
\newcommand{\rtwmin}{\mathtt{c}_{\min}}
\newcommand{\rtwmax}{\mathtt{c}_{\max}}
\newcommand{\ltwmax}{\mathtt{v}_{\max}}

\subsubsection{Regular and Irregular LDPC Codes}
LDPC codes are often classified into two categories: regular and irregular LDPC codes. In this chapter, we consider the latter, which also constitutes the more general, broader class of codes. The parity check matrix of regular codes has the property that the number of ``1''s in each column is constant and amounts to $\ltwreg$ (called \emph{variable degree}) and that the number of ``1''s in each row is constant and amounts to $\rtw$ (called \emph{check degree}). Clearly, $n\cdot\ltwreg = m\cdot\rtw$ has to hold and we furthermore have $r = 1-\frac{\ltwreg}{\rtw}$. Irregular LDPC codes~\cite{Richardson0201} have the property that the number of ``1''s in the different columns of $\bm{H}$ is not constant. In this chapter, we  mainly consider \emph{column-irregular} codes, which means that only the number of ``1''s in the columns is not constant but the number of ``1''s in each row remains constant. The irregularity of the parity-check matrix is often characterized by the degree profile of the parity check matrix $\bm{H}$~\cite{Richardson2008}. 

We denote the number of columns of the parity-check matrix $\bH$ with $i$ ones  by $\Lambda_i$. We say that these columns have \emph{degree} $i$. Normalizing this value to the number of total bits $n$ per codewords yields 
\[
L_i = \frac{\Lambda_i}{n}\,,
\]
which is the \emph{fraction} of columns with degree $i$, i.e., with $i$ ones  (e.g., if $L_3=\frac{1}{2}$, half the columns of $\bm{H}$ have three ``1''s). 

Similarly, we can define the \emph{check degree profile} by defining that $P_j$ denotes the number of rows of $\bH$ with exactly $j$ ``1''s. The normalized check profile is given by $R_j$, the \emph{fraction} of rows with $j$ ``1''s. We have the $R_j = \frac{P_j}{m}$. In most of the codes we consider, however, \emph{all} rows of $\bH$ have the same number of $\rtw$ ``1''s. In that case, we have $R_{\rtw} = 1$ and $R_1 = R_2 = \cdots = R_{\rtw-1} = R_{\rtw+1} = \cdots = R_{\infty} = 0$. Example~\ref{ex:irregular_H} illustrates the degree distribution of such an irregular LDPC code.

\begin{example}\label{ex:irregular_H}
Consider the following LDPC code of size $n=32$ with parity-check matrix of size $\dim{\bm{H}}=m\times n = 8\times 32$, i.e., of rate $r = \frac{32-8}{32} =  0.75$, corresponding to an overhead of $33.\bar{3}$\%. Note that the zeros in $\bm{H}$ are not shown for clarity.
\begin{align*}
\bm{H} ={\small \left(\begin{array}{*{7}{c@{\;\,}}c:*{15}{c@{\;\,}}c:*{7}{c@{\;\,}}c}
1 &   &   & 1 &   &   &   &   &   &   & 1 & 1 &   &   & 1 &   &   &   &   & 1 &   & 1 &   & 1 & 1 &   &   & 1 & 1 & 1 & 1 &   \\
1 &   &   & 1 &   &   &   &   &   &   & 1 &   &   & 1 &   & 1 &   & 1 &   &   &   &   & 1 & 1 & 1 & 1 & 1 &   & 1 &   &   & 1 \\
  & 1 &   &   &   &   &   & 1 &   & 1 &   & 1 &   &   &   & 1 & 1 &   & 1 &   &   &   &   & 1 &   & 1 &   &   & 1 & 1 & 1 & 1 \\
  &   & 1 &   & 1 &   &   &   &   & 1 &   &   & 1 &   &   & 1 &   &   & 1 &   &   & 1 & 1 &   &   & 1 & 1 &   & 1 & 1 & 1 &   \\
  & 1 &   &   &   &   &   & 1 & 1 &   &   &   & 1 & 1 &   &   &   & 1 & 1 &   & 1 &   &   &   & 1 &   &   & 1 & 1 & 1 &   & 1 \\
  &   &   &   &   & 1 & 1 &   & 1 &   &   &   & 1 &   & 1 &   & 1 & 1 &   &   &   &   & 1 &   & 1 & 1 & 1 & 1 &   &   &   & 1 \\
  &   & 1 &   & 1 &   &   &   & 1 &   &   &   &   & 1 & 1 &   & 1 &   &   & 1 & 1 &   &   &   & 1 &   & 1 & 1 &   &   & 1 & 1 \\
  &   &   &   &   & 1 & 1 &   &   & 1 & 1 & 1 &   &   &   &   &   &   &   & 1 & 1 & 1 &   &   &   & 1 & 1 & 1 &   & 1 & 1 &   
  \end{array}\right)}
\end{align*}
The first 8 columns of $\bm{H}$ have two ``1''s per column, i.e., $\Lambda_2=8$. Furthermore, the middle 16 columns each contain three ``1''s, i.e., $\Lambda_3=16$. Finally, the last 8 columns contain five ``1''s, i.e., $\Lambda_5=8$. Normalizing leads to
\[
L_2 = \frac{\Lambda_2}{n} = \frac{1}{4},\qquad L_3 = \frac{\Lambda_3}{n} = \frac{1}{2},\qquad L_5 = \frac{\Lambda_5}{n} = \frac{1}{4}\,.
\]
Note that $L_1 = L_4 = L_6 = L_7 = \cdots = 0$. 
The number of ``1''s in each row of $\bH$ is constant and amounts to $\rtw = 13$.\\[0.5ex]
\hrule
\end{example}

\subsubsection{Graph Representation of LDPC Codes}\label{sec:ldpc_graph}
LDPC codes are often represented by a so-called \emph{Tanner} graph~\cite{Richardson2008}. This graph is an undirected bipartite graph in which the nodes can be partitioned into two disjoint sets and each edge connects a node from the first set to a node from the second set. The Tanner graph allows for an easy description of the decoding algorithm of LDPC codes, which we will not detail here. We will give a summary of the iterative decoding algorithm in the Appendix. 
\begin{figure}[tb!]
\centering\includegraphics[scale=1]{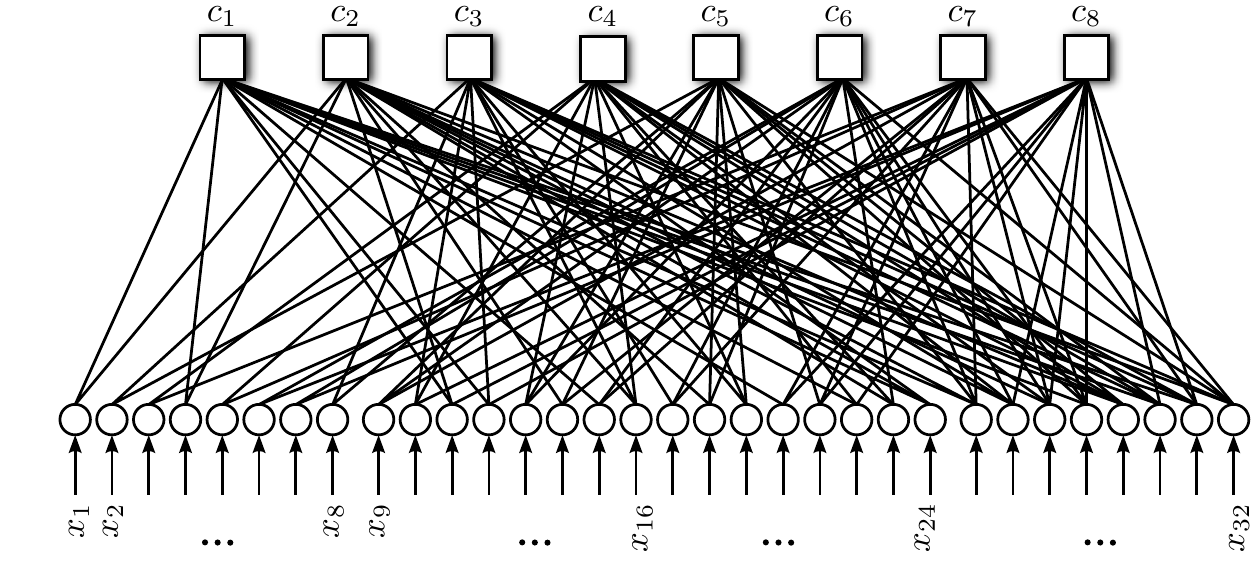}
\caption{Graph of the code defined by the parity check matrix given in Example~\ref{ex:irregular_H}}
\label{fig:ldpc_graph}
\end{figure}
Figure~\ref{fig:ldpc_graph} shows the graph representation of the toy code given in Example~\ref{ex:irregular_H}. The circular nodes on the bottom of the graph represent the \emph{variable nodes}, which correspond to the bits in the codeword. As each codeword contains $n$ bits, there are $n$ variable nodes $x_1$, $x_2, \ldots, x_n$. The variable node $x_i$ has one connection to the transmission channel (arrow from the bottom) and $j$ additional connections towards the top where $j$ equals the number of ``1''s in the $i$th column of $\bH$. For instance, the first $\Lambda_2$ variables $x_1,\ldots, x_{\Lambda_2}$ (where $\Lambda_2=8$) of the code have 2 connections towards the graph part of the code and an additional connection from the transmission channel. As in Example~\ref{ex:irregular_H}, the variable nodes can be divided into three groups, corresponding to the degree of these variables.

\newcommand{\LDPCintl}{\Pi^{[\text{LDPC}]}}

The rectangular nodes on the top of the graph are the so called \emph{check nodes}. Each check node $c_i$ corresponds to one of the $m$ rows of the parity-check matrix $\bH$ of the code and defines a code \emph{constraint}. The number of connections of the check nodes with the graph corresponds to the number of ``1''s in the respective row of $\bH$. In the above example, every row has $\rtw=13$ ``1''s, so that each of the check nodes has exactly $\rtw=13$ connected edges. If $\bH$ has a non-zero entry at row $i$ and column $j$, i.e, $H_{i,j}=1$, then an edge connects variable node $x_j$ to check node $c_i$.

\begin{figure}[tb!]
\centering\includegraphics[scale=1]{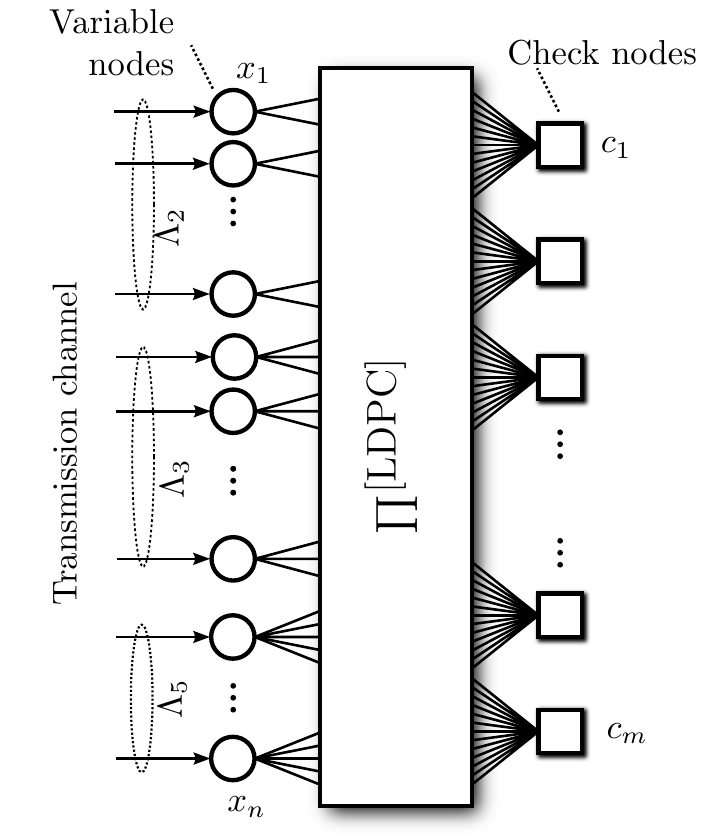}
\caption{Simplified graph representation of an irregular LDPC code with $\ltw\in\{2,3,8\}$ and $\rtw=8$}
\label{fig:ldpc_only}
\end{figure}

As drawing the graph of the code in this way quickly becomes cumbersome and confusing due to the large number of edges, we resort to a simplified (and rotated) representation shown in Fig.~\ref{fig:ldpc_only}. In this figure, we do not draw all the edges, but only the beginning and end of each edge and assume that the permutation of the edges is managed by an interleaver $\LDPCintl$. The interleaver $\LDPCintl$ thus ensures that the connections between the different nodes corresponds to the one given by the parity-check matrix~$\bH$.

\subsubsection{Design of Irregular LDPC Codes}
The design of irregular LDPC codes consists of finding good degree distributions, i.e, good values $\Lambda_i$ and $P_i$ (or $\rtw$) such that the rate of the code has the desired value (given by the system designer) and such that the NCG achievable by this code is maximized, i.e., the code is able to successfully recover the bit stream at the lowest possible $\Es/\No$ value. A comprehensive body of literature on the design of irregular codes exists~(see~\cite{RyanBook} and references therein) and we only introduce the basics to describe the optimization of codes tailored to slip-tolerant differential decoding in Sec.~\ref{sec:code_design}.

The optimization of irregular LDPC codes requires the use of \emph{edge-perspective} degree distributions~\cite{Richardson2008}. 

\begin{definition}[Edge-perspective degree distribution]
In the Tanner graph representation of the code, we denote by $\lambda_i$ the fraction of \emph{edges} that are connected to variable nodes of degree $i$. We have
\begin{align}
\lambda_i = \frac{i\cdot L_i}{\sum_{j=1}^{\infty}j\cdot L_j}\,.\label{eq:L_to_lambda}
\end{align}
Similarly, $\rho_i$ denotes the fraction of edges that are connected to check nodes of degree $i$. Again, we have
\[
\rho_i = \frac{i\cdot R_i}{\sum_{j=1}^{\infty}j\cdot R_j}\,.
\]
\end{definition}
Using the technique of EXIT charts~\cite{tenBrink1001,Ashikhmin1104,tenBrink2004}, good values of $\lambda_i$ and potentially $\rho_i$ may be found that can then be used to design a parity-check matrix $\bH$ fulfilling these degree distributions. We constrain the maximum possible variable node degree to be $\ltwmax$ and the maximum possible check node degree to be $\rtwmax$.

The inverse relationship between $\lambda_i$ and $L_i$, or between $\rho_i$ and $R_i$, respectively, reads
\begin{align}
L_i = \frac{\frac{\lambda_i}{i}}{\sum_{j=1}^{\ltwmax}\frac{\lambda_j}{j}}\qquad\text{and}\qquad R_i = \frac{\frac{\rho_i}{i}}{\sum_{j=1}^{\rtwmax}\frac{\rho_j}{j}}\,. \label{eq:inverse_lambda_L}
\end{align}
The (iterative) LDPC decoding process may be understood as a process where two decoders pass information between each other. The first decoder is the \emph{variable node decoder} (VND) which processes each of the $n$ variable nodes of the code. The second decoder is the \emph{check node decoder} (CND), which processes each of the $m$ check nodes. Each of these decoders has a certain \emph{information transfer} (EXIT) characteristic. Before describing the transfer characteristics, we introduce the $J$-function that interrelates mean $\mu$ (and variance, which amounts $2\mu$ in the case of symmetric messages, for details, see~\cite{tenBrink1001} and~\cite{Richardson2008}) and mutual information for the Gaussian random variable describing the messages that are exchanged in the iterative decoder, with
\[
J(\mu) = 1 - \int\limits_{-\infty}^{\infty} \frac{e^{-(\tau-\mu)^2/(4\mu)}}{\sqrt{4\pi\mu}} \log_2\left(1+e^{-\tau}\right)\text{d}\tau
\] 
which can be conveniently approximated~\cite{Schreckenbach2007} by
\begin{align*}
I = J(\mu) &\approx \left(1 - 2^{-H_1(2\mu)^{H_2}}\right)^{H_3}\\
\mu = J^{-1}(I) &\approx \frac{1}{2}\left(-\frac{1}{H_1}\log_2\left(1-I^{\frac{1}{H_3}}\right)\right)^{\frac{1}{H_2}}
\end{align*}
with $H_1 = 0.3073$, $H_2 = 0.8935$, and $H_3 = 1.1064$.

In the case of LDPC codes and transmission over an AWGN channel, the information transfer characteristics are obtained as~\cite{SharonTCOM2006}
\begin{align}
\IE{V} = f_V\left(\IA{V},\frac{\Es}{\No}\right) &:= \sum_{i=1}^{\ltwmax}\lambda_iJ\left(4\frac{\Es}{\No} + (i-1)J^{-1}(\IA{V})\right) \label{eq:ldpc_vnd_char}\\
\IE{C} = f_C(\IA{C}) &:= \sum_{i=1}^{\rtwmax} \frac{\rho_i}{\log(2)}\sum_{j=1}^{\infty} \frac{\left(\Phi_j\left(J^{-1}(\IA{C})\right)\right)^{i-1}}{2j(2j-1)}\label{eq:ldpc_cnd_char}
\end{align}
where
\[
\Phi_i(\mu) = \int\limits_{-1}^1\frac{2\tau^{2i}}{(1-\tau^2)\sqrt{4\pi\mu}}\exp\left(-\frac{\left(\mu-\log\frac{1+\tau}{1-\tau}\right)^2}{4\mu}\right)\text{d}\tau\,.
\]
Equation~\eqref{eq:ldpc_vnd_char} describes the characteristic of the VND while~\eqref{eq:ldpc_cnd_char} describes the characteristic of the CND. For codes with regular check node degree,~\eqref{eq:ldpc_cnd_char} can be simplified to
\begin{align*}
\IE{C} = f_C(\IA{C}) &:=  \frac{1}{\log(2)}\sum_{j=1}^{\infty} \frac{\left(\Phi_j\left(J^{-1}(\IA{C})\right)\right)^{\rtw-1}}{2j(2j-1)}
\end{align*}
As $\IA{V} = \IE{C}$ holds in the context of iterative decoding, a condition for successful decoding is that
\begin{align}
f_V\left(I, \frac{\Es}{\No}\right) > f_C^{-1}(I),\quad \forall I \in [0;1)\label{eq:condition}
\end{align}
where the inverse function $f_C^{-1}(I)$ of the strictly monotonically increasing function $f_C$ given in~\eqref{eq:ldpc_cnd_char} can be found using numerical methods. The task of the code designer is to find a degree distribution minimizing $\Es/\No$ such that~\eqref{eq:condition} is fulfilled. Usually, the condition~\eqref{eq:condition} is evaluated at discrete values of $I$ only, simplifying the implementation.

Some more conditions usually apply to the degree distributions. One of these is the so-called \emph{stability condition}~\cite{Richardson2008}, which, in the case of an AWGN channel ensures that
\[
\lambda_2 \leq \frac{\exp\left(\frac{\Es}{\No}\right)}{\sum_{i=1}^{\rtwmax}\rho_i(i-1)}\,.
\]

\subsection{Code Design for Iterative Differential Decoding}\label{sec:code_design}

As described in Sec.~\ref{sec:capacity}, the differential decoder based on the BCJR algorithm can be characterized by an EXIT characteristic $\IE{D} = f_D(\IA{D}, \Es/\No)$. Before optimizing the LDPC code towards the interworking with the differential decoding, we first have to define the \emph{decoder scheduling} as we are concerned with a three-fold iterative decoder loop: decoding iterations are carried out within the LDPC decoder and between LDPC decoder and differential decoder. In this chapter, we restrict ourselves to the following scheduling:
\begin{itemize}
\item[a)] In a first initial step, the differential decoder is executed and generates initial channel-related information.
\item[b)] Using this initial channel-related information, a \emph{single} LDPC iteration is carried out, i.e., a single execution of the check node and variable node computing processors.
\item[c)] Using the accumulated variable node information from the LDPC  graph, \emph{excluding} the intrinsic channel-related information from the initial differential decoding execution (step a)), the differential decoder is executed again, yielding improved channel-related information.
\item[d)] With the improved information from step c), another \emph{single} LDPC iteration is carried out. If the maximum number of allowed iterations is not yet reached, 
we continue with step c).
\item[e)] If the maximum number of iterations is reached, the accumulated variable node information is used to get an \emph{a posteriori} estimate of each bit.
\end{itemize}
In what follows, we now describe in detail how to find good degree distributions for iterative differential decoding. In~\cite{PfluegerSCC13} and~\cite{PfluegerVTC13}, conditions for degree distributions were derived and it was analyzed if it is possible to construct codes that work equally well for differential coding and conventional non-differential transmission. In this work, we solely consider the case of differential coding and we aim at showing different possibilities of degree distribution optimization with the goal to show the best possibility for LDPC coded differential modulation with the above mentioned decoder scheduling.

We only consider \emph{column irregular} codes in the remainder of this chapter, i.e., the number of ``1''s in each row of the parity-check matrix $\bH$ is constant and amounts to $\rtw$. Such a constraint is often imposed as it simplifies the hardware that is needed to implement the check node decoding operation, which is the most difficult operation in the LDPC decoder. The complexity of this operation scales roughly linearly with the check node degree (i.e., the number of ``1''s per row) and having a constant degree allows the hardware designer to implement a fixed and optimized check node computation engine.
The second constraint that we impose is that we only have three different variable node degrees, namely $\Lambda_2$ variable nodes of degree 2, $\Lambda_3$ variable nodes of degree 3, and $\Lambda_{\ltwmax}$ variable nodes of degree $\ltwmax$. This is in line with the findings given in~\cite{ShokrollahiDEBook} that show that the degree distributions are often sparse and that only a few different values are often sufficient. Having only three different variable node degrees simplifies the hardware implementation, especially the design of the required bit widths in a fixed point implementation.

Contrary to many degree distribution approaches proposed in the literature~\cite{Richardson2008,PfluegerSCC13,Lechner2006} we first fix the rate $r$ of the final code as the rate is usually constrained by the system design parameters (e.g., speed of analog-to-digital and digital-to-analog converters, pulse shape, channel bandwidth, framing overhead, etc.). With fixed rate $r$, we remove the dependencies~\cite{ShokrollahiDEBook} of the degree distribution. We further assume that no nodes of degree 1 are present in the code, i.e., $\Lambda_1=0$ and thus $\lambda_1 = 0$. As $\sum_i\lambda=1$, we can uniquely determine $\lambda_2$ as
\begin{align}
\lambda_2 &= 1 - \sum_{i=3}^{\ltwmax}\lambda_i\,. \label{eq:condition_l2}
\end{align}
As the rate of the code is given by~\cite{Richardson2008}
\begin{align}
r = 1 - \frac{\sum_{i=1}^{\rtwmax}\frac{\rho_i}{i}}{\sum_{i=1}^{\ltwmax}\frac{\lambda_i}{i}}\label{eq:condition_rate}
\end{align}
we can eliminate another dependency and by combining~\eqref{eq:condition_rate} with~\eqref{eq:condition_l2}, we get
\begin{align}
\lambda_3 = 3 + 6\sum_{i=4}^{\ltwmax}\lambda_i\left(\frac{1}{i}-\frac{1}{2}\right) - \frac{6}{1-r} \sum_{i=1}^{\rtwmax}\frac{\rho_i}{i}\,.\label{eq:condition_l3_gen}
\end{align}
For check-regular codes with regular check node degree $\rtw$ (i.e., $\rho_{\rtw} = 1$),~\eqref{eq:condition_l3_gen} can be simplified to
\begin{align}
\lambda_3 &= 3 - 6\left(\frac{1}{\rtw (1-r)}-\sum_{i=4}^{\ltwmax}\lambda_i\left(\frac{1}{i}-\frac{1}{2}\right)\right)\,.\label{eq:condition_l3}
\end{align}
This means that $\lambda_2$ and $\lambda_3$ are uniquely determined by $\lambda_4, \lambda_5, \ldots, \lambda_{\ltwmax}$. If we only allow $\lambda_2$, $\lambda_3$ and $\lambda_{\ltwmax}$ to be nonzero, then $\lambda_2$ and $\lambda_3$ are uniquely determined by $\lambda_{\ltwmax}$ and we have
\begin{align}
\lambda_3 &= 3 - 6\left(\frac{1}{\rtw (1-r)}-\lambda_{\ltwmax}\left(\frac{1}{\ltwmax}-\frac{1}{2}\right)\right)\label{eq:lambda3_unique}\\
\lambda_2 &= -2 - \lambda_{\ltwmax} +  6\left(\frac{1}{\rtw (1-r)}-\lambda_{\ltwmax}\left(\frac{1}{\ltwmax}-\frac{1}{2}\right)\right)\label{eq:lambda2_unique}.
\end{align}
For determining the degree distribution, the choice of the interleaving scheme between LDPC code and differential encoder/decoder is crucial. In fact, this choice determines how to select the degree distribution and finally has an influence on the overall system performance.

\subsubsection{Design of LDPC Codes -- Full Interleaving}

\newcommand{\intldiff}{\Pi^{[\text{diff}]}}

\begin{figure}[tb!]
\centering\includegraphics[scale=1]{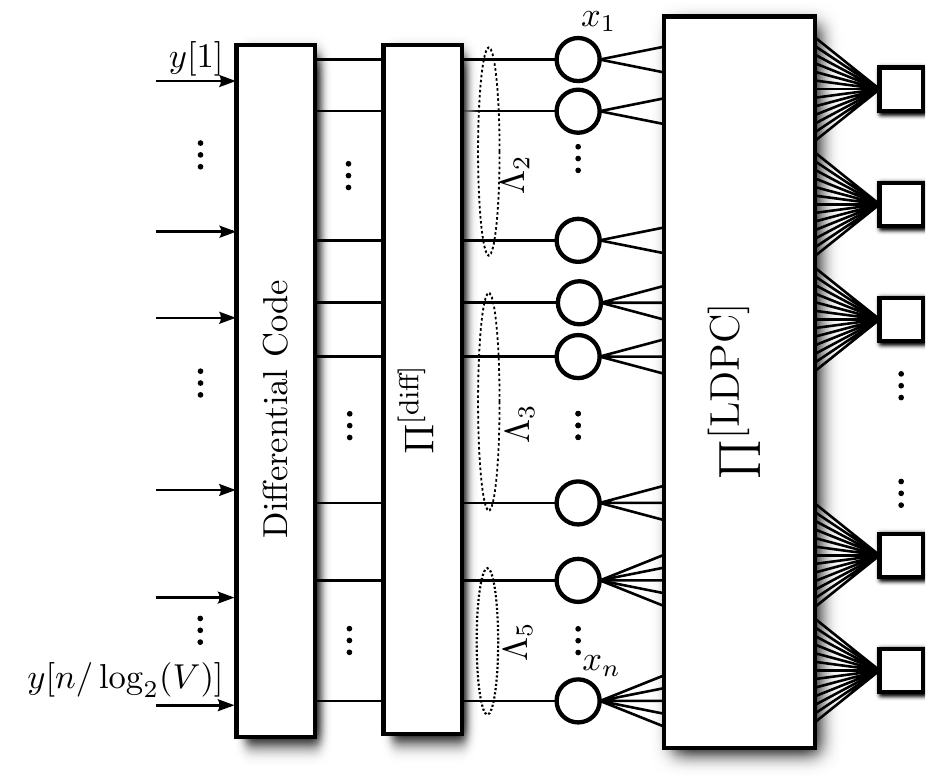}
\caption{Schematic of the LDPC code with full interleaving between LDPC code and differential decoder}
\label{fig:ldpc_full_block}
\end{figure}

The first way of interleaving consists in placing a full interleaver $\intldiff$ of size $n$ between differential code and LDPC code, as depicted in Fig.~\ref{fig:ldpc_full_block}. The interleaver $\intldiff$ is placed between the differential decoder and the variable nodes of the LDPC code, such that the interleaved output of the differential decoder mimics the transmission channel output. This is the approach that has been followed in~\cite{Lechner2006} and~\cite{FranceschiniJSAC}.

As the transmission channel is in this case the combination of differential decoder and interleaver, we need to modify the convergence condition~\eqref{eq:condition}. Instead of having a function describing the information transfer of the VND, we introduce a function $f_{V,D}$ that describes the information transfer of the combined differential decoder and VND. This combined information transfer function is given by
\begin{align*}
\IE{V,D} = \sum_{i=1}^{\ltwmax}\lambda_iJ\left(\mu_c + (i-1)J^{-1}(\IA{V,D})\right)
\end{align*}
The value $\mu_c$ is the mean of the message that is sent from the differential decoder towards the LDPC code. Using the EXIT characteristic of the differential decoder $f_D(I, \Es/\No)$, which can be prerecorded and potentially represented by a polynomial~\cite{tenBrink2004}, we can express $\mu_c$ as~\cite{tenBrink2004}
\begin{align*}
\mu_c = J^{-1}\left(f_D\left(\sum_{i=1}^{\ltwmax}L_iJ\left(i\cdot J^{-1}(\IA{V,D})\right), \frac{\Es}{\No} \right)\right)
\end{align*}
which leads to the overall EXIT characteristic of the combined VND and differential decoder
\begin{align}
f_{V,D}\left(I,\frac{\Es}{\No}\right) &= \sum_{i=2}^{\ltwmax}\lambda_iJ\!\left(\!  J^{-1}\!\left(\!f_D\!\left(\sum_{j=2}^{\ltwmax}L_jJ\left(j\cdot J^{-1}(\IA{V,D})\right),\frac{\Es}{\No} \right)\right)+ (i\!-\!1)J^{-1}(\IA{V,D})\right) \nonumber\\
&=\sum_{i=2}^{\ltwmax}\lambda_iJ\!\left(\!  J^{-1}\!\left(\!f_D\!\left(\sum_{j=2}^{\ltwmax}\frac{\lambda_jJ\left(j\cdot J^{-1}(\IA{V,D})\right)}{j\sum_{\kappa=2}^{\ltwmax}\frac{\lambda_\kappa}{\kappa}},\frac{\Es}{\No} \right)\right)+ (i\!-\!1)J^{-1}(\IA{V,D})\right)\label{eq:combined_vnd_det_full}
\end{align}
where we have used~\eqref{eq:inverse_lambda_L} in the second line of the equation. This leads to the condition for successful decoding
\begin{align}
f_{V,D}\left(I, \frac{\Es}{\No}\right) > f_C^{-1}(I),\quad \forall I \in [0;1)\,.\label{eq:condition_CMfull}
\end{align}
In this case, the stability condition reads~\cite{Lechner2006}
\begin{align}
\lambda_2 < \frac{1}{\rtw-1}\exp\left(\frac{J^{-1}\left(f_D\left(1,\frac{\Es}{\No}\right)\right)}{4}\right)\label{eq:stability_CM}
\end{align}
As the function $f_{V,D}(\cdot,\cdot)$ is not linear in $\lambda_i$ (and even not necessarily convex), the elegant linear programming based optimization~\cite{Richardson2008} cannot be applied.  We have to resort to heuristic optimization methods such as \emph{differential evolution}~\cite{ShokrollahiDEBook} or \emph{simulated annealing}. If we assume however that the degree distribution only consists of three degrees 2, 3 and $\ltwmax$, then we have seen before that by fixing $\lambda_{\ltwmax}$, the values of $\lambda_2$ and $\lambda_3$ are immediately given. Thus the problem of finding the optimal degree distribution reduces to a one-dimensional problem. By sweeping $\lambda_{\ltwmax}$ between the extremes of the admissible interval $[0,1]$, we can find the best possible degree distribution. 

We use the following binary search to find the best possible degree distribution. We first assume that the differential decoder EXIT characteristic is available for any $\Es/\No$ value between $\frac{\Es}{\No}\big\vert_{\min}$ and $\frac{\Es}{\No}\big\vert_{\max}$. We fix a minimum step size $\Delta_{\min}$ and use Algorithm~\ref{alg:binsearch} which outputs the optimum $\rtw$, $\lambda_2$, $\lambda_3$ and $\lambda_{\ltwmax}$.\SetArgSty{textnormal}
\begin{algorithm}[tp!]\label{alg:binsearch}
\KwIn{Maximum variable node degree $\ltwmax$}
\KwIn{Minimum and maximum check node degrees $\rtwmin$ and $\rtwmax$}
\KwIn{Discretization steps $D$ (mutual information) and $D_{\lambda}$}
\KwIn{Minimum stepsize $\Delta_{\min}$}
\KwOut{Optimum variable node degree distribution $\bm{\lambda}_{\text{best}} = (\lambda_2, \lambda_3, \lambda_{\ltwmax})$}
\KwOut{Optimum check node degree $\rtwbest$}
\KwOut{SNR threshold $E_\text{best}$}
\setstretch{1.198}
\Begin{$E \leftarrow \frac{1}{2}\left(\frac{\Es}{\No}\big\vert_{\min}+\frac{\Es}{\No}\big\vert_{\max}\right)$, $E_{\text{best}} \leftarrow \infty$,  $\Delta \leftarrow \frac{1}{2}\left(\frac{\Es}{\No}\big\vert_{\max}-\frac{\Es}{\No}\big\vert_{\min}\right)$

\While{$\Delta > \Delta_{\min}$}{
$\text{Success}\leftarrow \text{false}$

\For{$\rtw = \rtwmin \ldots \rtwmax$}{
compute $f_C^{-1}\left(\frac{i}{D}, \rtw\right)$, $\forall i \in\{0,1,\ldots,D-1\}$

\For{$j \in \{0,1,\ldots, D_{\lambda}$}{
$\lambda_{\ltwmax} \leftarrow \frac{j}{D_{\lambda}}$

determine $\lambda_2$ and $\lambda_3$ via~\eqref{eq:lambda2_unique} and \eqref{eq:lambda3_unique}

compute $f_{V,D}\left(\frac{i}{D}, E\right)$, $\forall i \in\{0,\ldots, D-1\}$ using~\eqref{eq:combined_vnd_det_full}

\If{$f_{V,D}\left(\frac{i}{D},E\right) > f_C^{-1}\left(\frac{i}{D}\right)$, $\forall i$ and \eqref{eq:stability_CM} fulfilled}{
\If{$E < E_{\text{best}}$}{
$E_{\text{best}} \leftarrow E$

$\bm{\lambda}_{\text{best}} = \left(\lambda_2,\lambda_3,\lambda_{\ltwmax}\right)$

$\rtwbest \leftarrow \rtw$
}
$\text{Success}\leftarrow\text{true}$
} % if
} % for j in 
} % for \rtw 
\uIf{\text{Success} = \text{True}}{
$E \leftarrow E - \frac{\Delta}{2}$
}\Else{
$E \leftarrow E + \frac{\Delta}{2}$
}
$\Delta \leftarrow \frac{\Delta}{2}$
} % while
} %\begin
\caption{Binary search for finding good LDPC code degree distribution}
\end{algorithm}
We have found that using only three different variable node degrees and only a fixed check node degree does not impose a noteworthy limitations and that the performance of the obtained codes is very close to the performance of codes designed with less constraints, provided that $\ltwmax$ is chosen large enough.

The full interleaving scheme has several limitations. For instance, the EXIT chart based optimization assumes that the messages exchanged between the different decoder components are Gaussian distributed. This is however not the case when interleaving the outputs of all different variables; in this case, the messages show rather a distribution that can be described by a \emph{Gaussian mixture}, leading to inaccuracies of the model. Even though the messages may be conveniently approximated by Gaussian distributions (if the variances of the different parts of the mixture do not vary much), the codes designed according to this model may not yield the best possible performance.

\subsubsection{Design of LDPC Codes -- Partial Interleaving}
\begin{figure}[tb!]
\centering\includegraphics[scale=1]{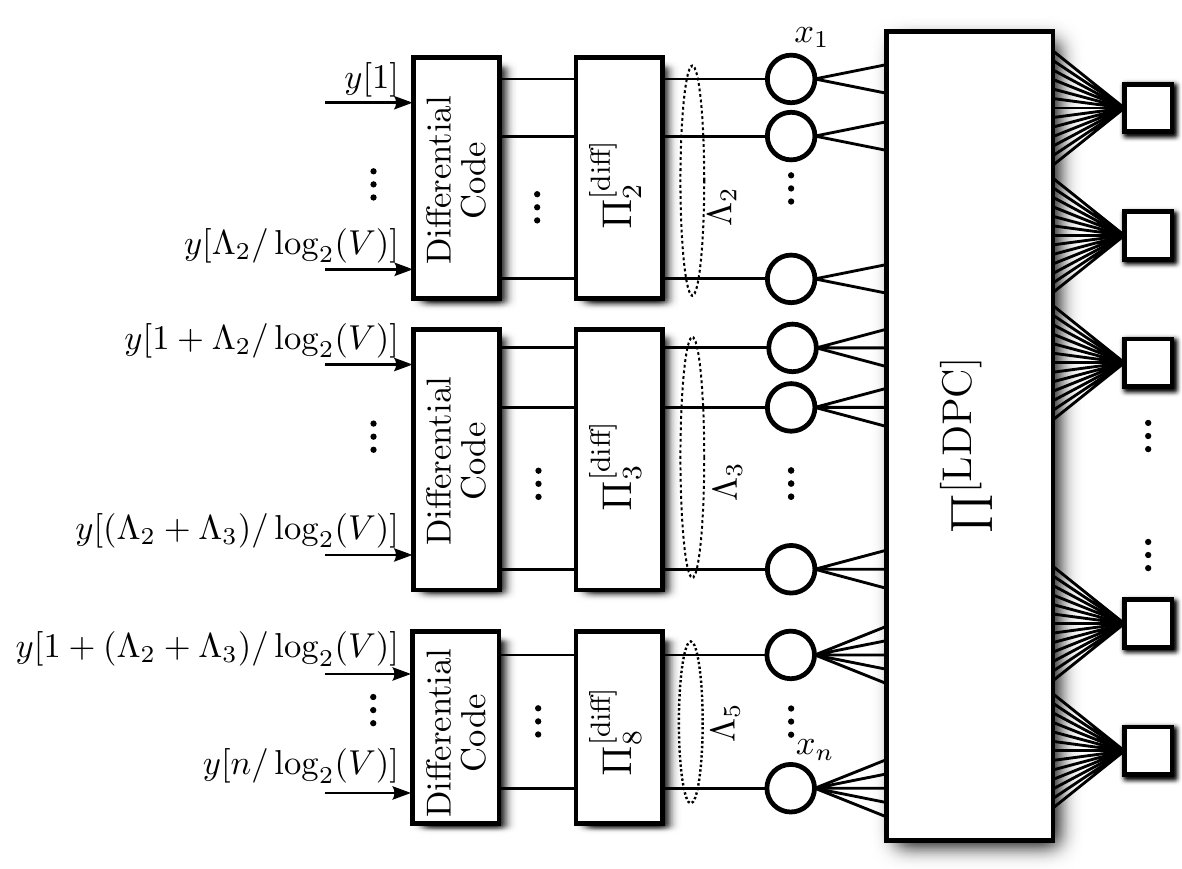}
\caption{Schematic of the LDPC code with partial interleaving between LDPC code and differential decoder}
\label{fig:ldpc_partial_block}
\end{figure}

In order to mitigate the limitations of the full interleaving approach of the previous paragraph, we  replace the single interleaver $\intldiff$ by multiple \emph{partial} interleavers, as described in~\cite{BenammarISTC2014} and inspired by the analysis for BICM-ID with convolutional codes in~\cite{Alvarado0910}. In the partial interleaving case, we group all $\Lambda_i$ variable nodes of degree $i$, assign an interleaver of size $\Lambda_i$ to these nodes and employ a separate differential decoder/encoder for this group of variable nodes.
The graph-based model with partial interleaving is shown in Fig.~\ref{fig:ldpc_partial_block} (where we assume that each $\Lambda_i$ is a multiple of $V$). 

If partial interleaving is used, the equations for analyzing the convergence and finding good degree distributions have to be modified as well. In this case, every variable node group (of degree $i$) has to be treated separately and is assigned its own differential decoder output $\mu_{c,i}$ and we can write
\begin{align}
\IE{V,D} = \sum_{i=1}^{\ltwmax}\lambda_iJ\left( \mu_{c,i} + (i-1)J^{-1}(\IA{V,D})\right)\label{eq:ievd_for_VQ}
\end{align}
where $\mu_{c,i}$ can be computed as
\begin{align*}
\mu_{c,i} = J^{-1}\left(f_D\left(J\left(i\cdot J^{-1}(\IA{V})\right) , \frac{\Es}{\No}\right) \right)\,.
\end{align*}
This leads to the overall EXIT characteristic of the combined variable node differential decoder
\begin{align*}
f_{V,D}\left(I,\frac{\Es}{\No}\right) := \sum_{i=1}^{\ltwmax}\lambda_iJ\left(  J^{-1}\left(f_D\left(J\left(i\cdot J^{-1}(I)\right),\frac{\Es}{\No} \right)\right)+ (i-1)J^{-1}(I)\right)
\end{align*}
which is a linear function in $\lambda$. Due to the linearity of $f_{V,D}(\cdot,\cdot)$, we can employ a simple linear programming optimization to find good values of $\lambda_i$ for a given check node degree distribution, as described in~\cite{Richardson2008}. However, if we only allow three different variable node degrees (and thus three partial interleavers), the problem reduces to a one-dimensional problem again, which we can solve in a similar way (using Algorithm~\ref{alg:binsearch}) as for the case with full interleaving. However, we would like to point out that due to the linearity of $f_{V,D}(\cdot,\cdot)$, it is much easier to find \emph{optimal} degree distributions. Numerical methods such as, e.g., differential evolution cannot guarantee to find the global optimum of the problem.

\subsubsection{Comparison of Interleaving Schemes -- Results}
Using both interleaving schemes and differential QPSK transmission, we design degree distributions. We impose the constraint that the maximum variable node degree shall be $\ltwmax=12$. For each interleaving scheme, we either use Gray differential encoding or natural differential encoding. For both options we have optimized codes using Algorithm~\ref{alg:binsearch} with variable degrees $\in\{2,3,12\}$ and regular check node degree. We additionally design codes for the constraint where $L_2 \leq 1-r$. This constraint is necessary to avoid a high number of degree-2 nodes. It can be shown~\cite{TillichISIT06} that a potential error floor can occur if the fraction of degree-2 nodes is larger than $1-r$. If we impose the constraint $L_2\leq 1-r$, then we can design a code that avoids---in the graph description---cycles containing only degree-2 nodes and with no information bits assigned to degree-2 nodes~\cite{SugiharaOFC2013}. The condition $L_2 \leq 1-r$ translates in the general case into 
\[
\lambda_2 \leq 2\left(\frac{1}{r}-1\right)\sum_{j=3}^{\ltwmax}\frac{\lambda_j}{j} ,
\]
which can be added to Algorithm~\ref{alg:binsearch}.

We generate codes of target rate $r=\frac{4}{5}=0.8$, a rate typically used in optical communications. A rate of $r=0.8$ is a viable selection for current and future $100$\,Gbit/s (with QPSK) or $200$\,Gbit/s systems operating in a \emph{dense wavelength division multiplex} (DWDM) setting with $50$\,Ghz channel spacing and an exploitable bandwidth of roughly $37.5$\,Ghz due to frequent filtering with non-flat frequency characteristic. The best possible achievable values of $\Es/\No$ (corresponding to $E_{\text{best}}$ in Algorithm~\ref{alg:binsearch}) for the different code designs are shown in Tab.~\ref{tab:denevl_results} and the degree distributions of the resulting codes are summarized in Tab.~\ref{tab:degree_distributions}. We can see that Gray differential coding with partial
interleaving leads to the best coded transmission schemes operating at the lowest possible $\Es/\No$ values.

\begin{table}
\caption{Theoretical thresholds values of $\Es/\No$ (in dB) for the designed codes}
\begin{tabular}{ccccc}
& \multicolumn{2}{c}{No constraint on $L_2$} & \multicolumn{2}{c}{$L_2 \leq 1-r$} \\
& Full Interl. & Partial Interl. & Full Interl. & Partial Interl. \\
\hline
Gray Diff. & $4.536$ & $4.43$ & $4.946$ & $4.907$ \\
Natural Diff. & $4.905$ & $4.839$ & $5.382$ & $5.33$
\end{tabular}
\label{tab:denevl_results}
\end{table}

\begin{table}[b!]
\vspace*{-2.6ex}
\caption{Degree distributions of all considered codes}\label{tab:degree_distributions}
\begin{tabular}{rcccc}
Code & $L_2$ & $L_3$ & $L_{12}$ & $\rtw$ \\
\hline
Gray diff., full intl. & 0.959 & 0.002 & 0.039 & 12 \\
Gray diff., partial intl. & 0.919 & 0.002 & 0.079 & 14\\
Natural diff., full intl. & 0.979 & 0.002 & 0.019 & 11\\
Natural diff., partial intl. & 0.959 & 0.002 & 0.039 & 12 \\
\hline
Gray diff., full intl., $L_2 \leq 1-r$ & 0.198 & 0.78 & 0.022 & 15\\
Gray diff., partial intl., $L_2 \leq 1-r$ & 0.198 & 0.78 & 0.022 & 15\\
Natural diff., full intl., $L_2 \leq 1-r$ & 0.198 & 0.78 & 0.022 & 15 \\
Natural diff., partial intl., $L_2 \leq 1-r$ & 0.198 & 0.78 & 0.022 & 15 \\
\end{tabular}
\end{table}

\begin{figure}[tbh!]
\centering\includegraphics[scale=1.1]{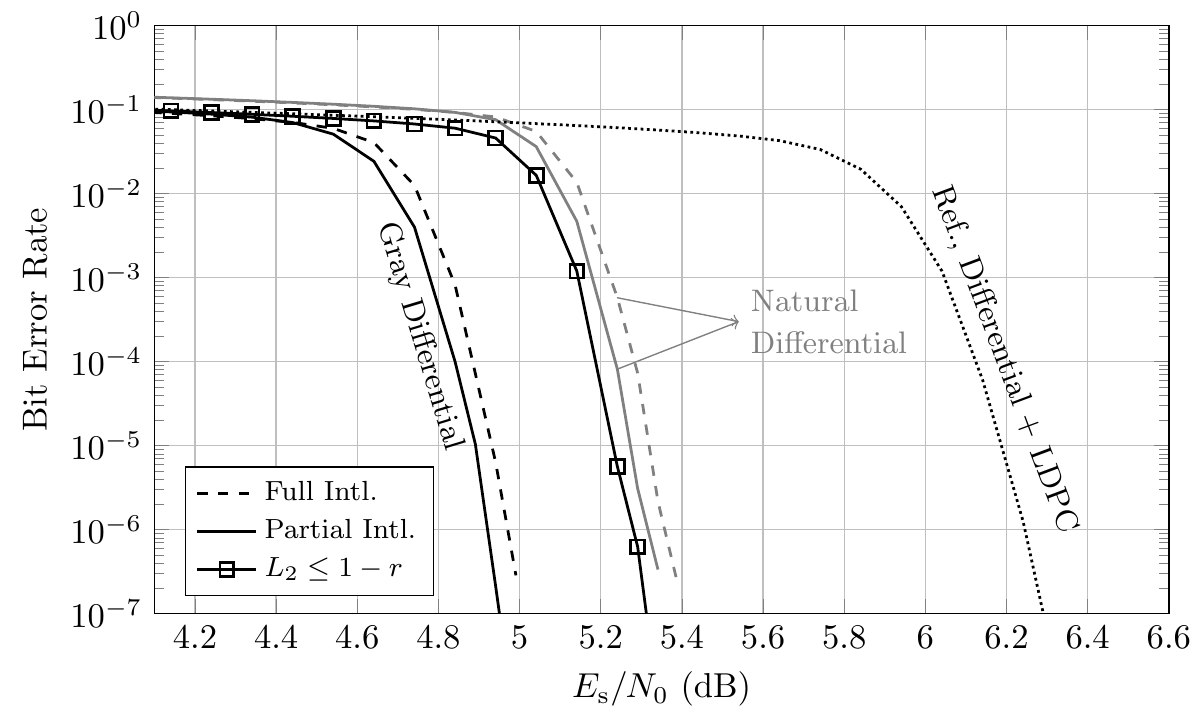}
\caption{Simulation example, QPSK with $\gamma=0$}
\label{fig:simex_1}
\end{figure}

Figure~\ref{fig:simex_1} shows a first simulation results for the case of a QPSK modulation. We have constructed codes of size $n=32000$ having the degree distributions from Tab.~\ref{tab:degree_distributions}. The utilized channel model is the phase-slip channel model from Fig.~\ref{fig:channel_model} with $\gamma = 0$. In this example, we wish to confirm the results of Tab.~\ref{tab:denevl_results}. 
As a reference scheme, we optimized an LDPC code for a simple AWGN channel with $\ltwmax = 12$ and regular check node degree $\rtw$, which is used with \emph{non-iterative} differential decoding.
We can see that the results of Tab.~\ref{tab:denevl_results} are indeed confirmed and the code with Gray differential coding and partial interleaving yields the best performance. As decoder, we use a conventional decoder as described in the Appendix of this chapter with $18$ decoding iterations, where in each iteration, we invoke the differential decoder. Note that with the use of a layered decoder~\cite{Hocevar2004}, the convergence speed can be increased and the same performance can be obtained by using only approximately 12 layered iterations. For this reason, the choice of 18 iterations is practical, as 12 layered LDPC iterations are deemed to be implementable~\cite{BisplinghoffECOC2014}.

\begin{figure}[tb!]
\centering\includegraphics[scale=1.1]{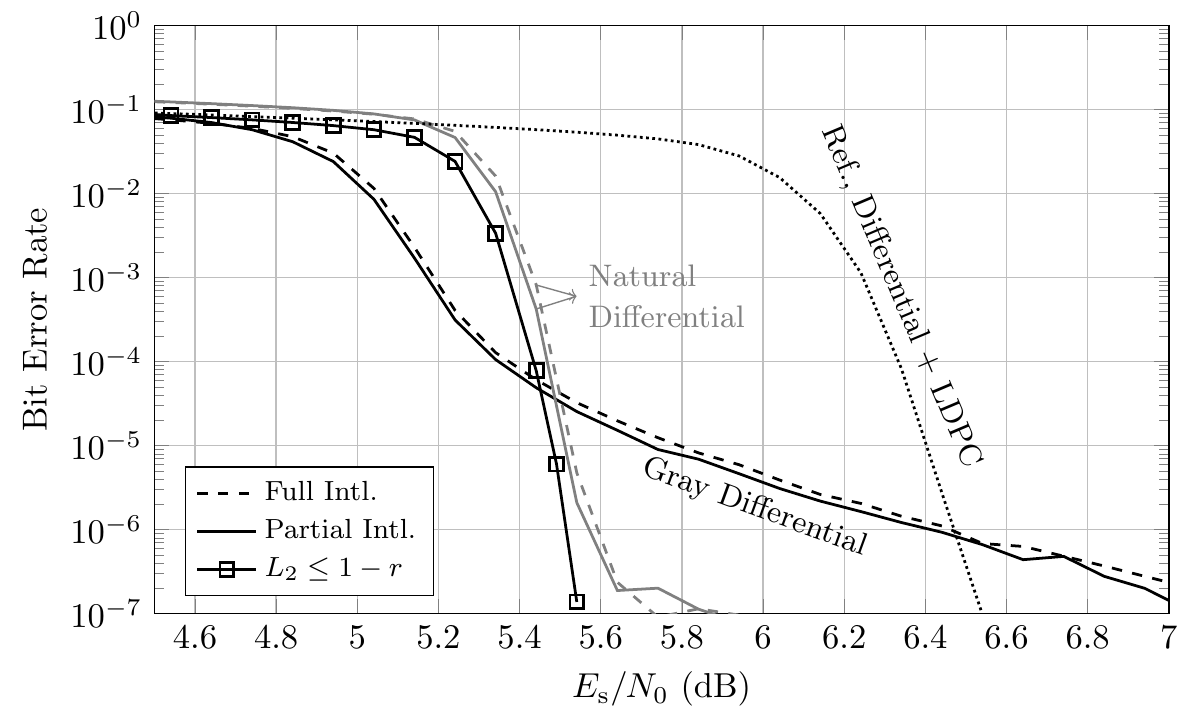}
\caption{Simulation example, QPSK with $\gamma=0.2$}
\label{fig:simex_2}
\end{figure}

In the second example, we increase the phase slip probability on the channel by choosing $\gamma=0.2$. The simulation results for this case are shown in Fig.~\ref{fig:simex_2}. We can observe that the formerly best case with Gray differential coding now shows a significant error floor.  With natural differential coding, an error floor is observed as well, however, at several orders of magnitude smaller. This floor is mainly due to the large number of degree-2 nodes and the fact that $\IE{D}(\IA{D}=1) < 1$ if $\gamma > 0$. Indeed, for this optimization, most of the variable nodes are of degree-2, i.e., $\lambda_2$ and consequently $L_2$ becomes very large. It has also been observed~\cite{PfluegerSCC13} that LDPC codes designed for differentially coded modulation require many degree-2 nodes. Degree-2 variable nodes are however a non-neglible contributor to the error floor, especially if there are cycles in the graph that connect only degree-2 variable nodes.  It has been shown that cycles containing only degree-2 variable nodes can be avoided~\cite{TillichISIT06} if $\Lambda_2 \leq m = n(1-r)$, i.e., if $L_2 \leq 1-r$, which is why we have included that constraint into the optimization. In this case, we may design a systematic code and assign only parity bits to the degree-2 variable nodes. This further reduces the error floor as the bit error rate is calculated purely based on the systematic bits and the higher the variable node degree, the more reliable a bit is after decoding.

We have added the constraint $L_2 \leq 1-r$ to the optimization routine and have found according degree distribution (summarized in Tab.~\ref{tab:degree_distributions}). In the simulation results shown in Fig.~\ref{fig:simex_1}, the performance of the code obtained with the $L_2\leq 1-r$ condition is shown by the curve with square markers for Gray differential coding. We see that in this case, we get a slightly better performance than natural differential coding, but have advantages regarding the residual bit error rate, at the expense of an increased required $\Es/\No$ to allow for successful decoding in the case where $\gamma=0$ (see Fig.~\ref{fig:simex_1}). Thus, depending on the required target bit error rate, we may either use the condition $L_2\leq 1-r$ or not. If an outer code is used that can correct up to an input bit error rate of $4.2\cdot 10^{-3}$ (e.g., the staircase code~\cite{Smithxx12}) we may use the code designed without the constraint on $L_2$, but if we use a higher rate outer code as in~\cite{BisplinghoffECOC2014,SugiharaOFC2013} that requires a very low input bit error rate, we may select the code designed with $L_2 \leq 1-r$.

We can thus summarize that there are somewhat conflicting code design strategies. If $\gamma=0$, we can use the code optimized for Gray differential coding with partial interleaving. This code will however lead to an elevated error floor if phase slips occur on the channel. This error floor has to be combated either with a properly designed outer code\footnote{Note that the implementation of a coding system with an outer cleanup code requires a thorough understanding of the LDPC code and a properly designed interleaver between the LDPC code and the outer code.} or by using only the code with the constraint $L_2 \leq 1-r$, which however leads to a suboptimal performance in the phase-slip-free case ($\gamma=0$). Another solution is the implementation of two codes, one for each case ($\gamma=0$ and large $\gamma$). This latter method can guarantee best performance depending on the channel, but requires a feedback loop from the receiver to the transmitter, which may not be available in the network and of course it requires the implementation of two different codes, which may be too complex on an \emph{application specific integrated circuit} (ASIC). In Sec.~\ref{sec:sc_codes}, we show a solution which requires only a single code and shows a more universal, channel-agnostic behavior.

\subsection{Higher Order Modulation Formats with $V < Q$}\label{sec:VsmallQ}

\begin{figure}[t!]
\centering\includegraphics[scale=1]{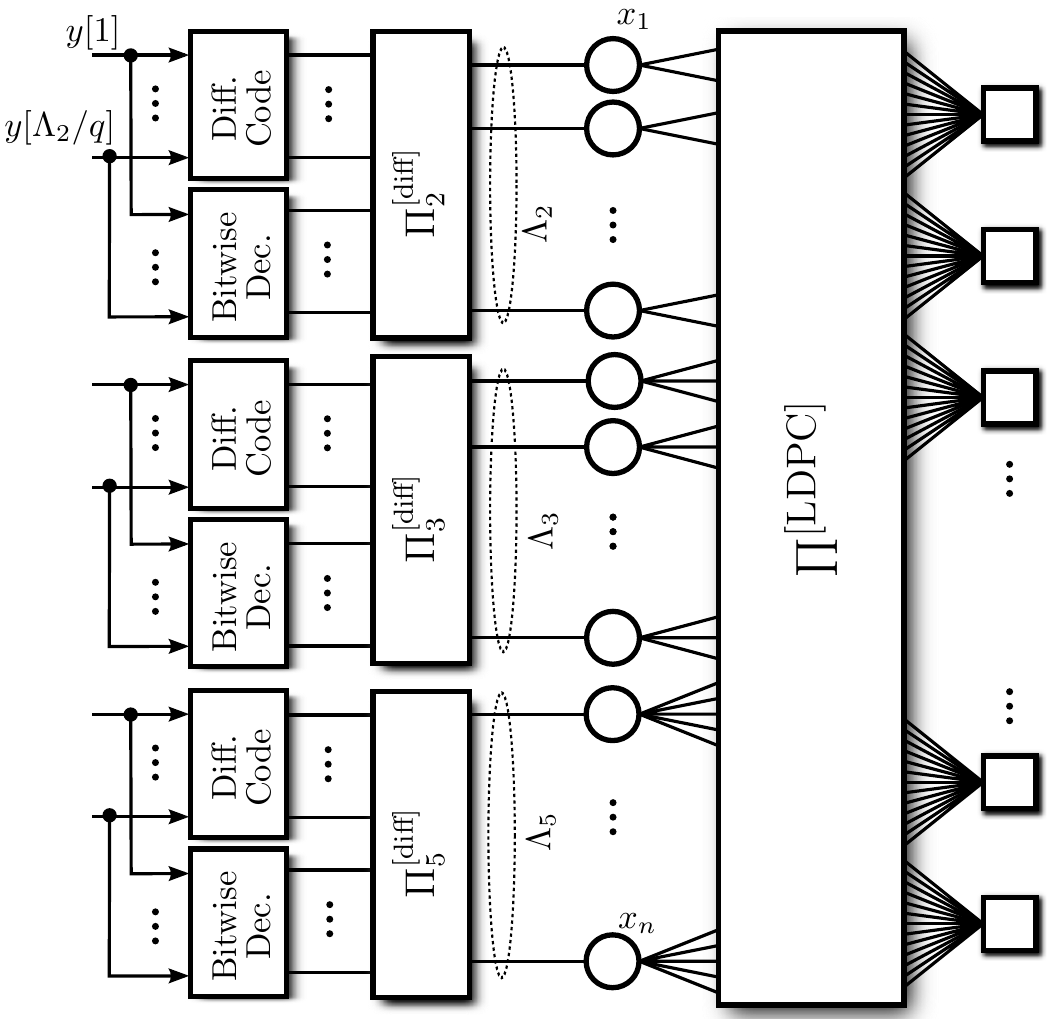}
\caption{Schematic of the LDPC code with partial interleaving between LDPC code and differential decoder for the case where $V < Q$}
\label{fig:ldpc_partial_block_VleQ}
\end{figure}

In practical systems, we often have to deal with the case where $V < Q$, e.g., if 16-QAM is used, where we have $V=4$ and $Q=16$. In this case, we may use different techniques to optimize the code. We propose to refine the method of partial interleaving and to use only differential coding on a fraction of $\frac{\log_2V}{\log_2Q} = \frac{v}{q}$ of the bits. This is shown in Fig.~\ref{fig:ldpc_partial_block_VleQ}. In this case, the value $\mu_{c,i}$ required in \eqref{eq:ievd_for_VQ} is computed as
\begin{align*}
\mu_{c,i} = \frac{v}{q}J^{-1}\left(f_D\left(J\left(i\cdot J^{-1}(\IA{V})\right) \right) \right) + \left(1-\frac{v}{q}\right)\bar{\mu}_c\,.
\end{align*}
where $\bar{\mu}_c$ denotes the mean of the log-likelihood ratios of the bits that are not differentially encoded, obtained using a conventional bit-wise decoder~\cite{GuilleniFabregas} (see~\eqref{eq:bitwise}) and averaged over all these bits. These bits correspond to the part of the constellation encoded with the rotationally invariant mapping (symbols within a region associated to a state~$\stat_i$).

Besides this simple and direct approach, we can also use more involved methods, e.g., using the technique of \emph{multi-edge-type} (MET) codes, as described in~\cite[Ch. 7]{Richardson2008,ZhangMET}, but this is outside the scope of this chapter.

\section{Coded Differential Modulation with Spatially Coupled\\ LDPC Codes}\label{sec:sc_codes}

In the 1960s and the following decades, most
coding research focused on block coding techniques,
but many practical coding schemes were based upon
convolutional codes~\cite{Costelloxx07}. With the advent of turbo
codes~\cite{Berrou0593}, the rediscovery of LDPC codes, and advances
in semiconductor technology, this suddenly changed
so that today most new coding schemes are, again,
block codes. The trend is, however, to return to
convolutional-like structures~\cite{TavaresPhD} that can be efficiently encoded and decoded using sliding-window
techniques.

In the last few years, the class of spatially coupled (SC)
code ensembles has emerged~\cite{Kudekar0211,Kudekarxx13}. 
Spatially coupled codes were originally introduced more than a decode ago~\cite{Felstroem0699} and were then called \emph{LDPC convolutional codes}. The appealing properties of SC codes were only recently noticed, when it was found that the performance  of terminated SC-LDPC codes with simple belief propagation decoding approaches the \emph{maximum a posteriori} (MAP)
thresholds of the underlying ensemble~\cite{LentmaierITAW09,Lentmaier_ISTC10}. Thus, contrary to a common belief, introducing
structure into LDPC codes leads to a class of (degenerated)
realizations of LDPC codes that demonstrate
superior performance under belief propagation
decoding. This effect of threshold saturation has been analyzed for the
binary erasure channel (BEC) in~\cite{Kudekar0211} and it has been
shown that spatially coupled LDPC codes can asymptotically
achieve the MAP threshold of the underlying
ensemble under belief propagation
decoding.
Recently, this result has been extended to
more general channels and it has been shown in~\cite{Kudekarxx13} that spatially coupled regular LDPC ensembles \emph{universally}
achieve capacity over binary-input memoryless output-symmetric channels: most codes in this ensemble
are good for each channel realization in this class of
channels.

Spatially coupled codes are now emerging in various
applications. Two examples in the context of optical communications 
are the staircase code~\cite{Smithxx12} and the braided BCH codes of~\cite{Jian1213}, which are both rate $R = 239/255$
codes targeted for 100\,GBit/s applications with hard-decision
decoding. Both codes are spatially coupled BCH product codes that allow for a natural
windowed decoder implementation. These codes can be interpreted as being generalized spatially coupled LDPC codes with variable node degree $d_v=2$ and every bit participating in two BCH component codes, where the component BCH codes are able to correct up to 4 errors each. Another example
is the IEEE 1901 power line communications
standard, where an LDPC convolutional code is specified for the wavelet physical layer~\cite{IEEE1901}. For a basic introduction to spatially coupled codes, we refer the interested reader to~\cite{LevenSchmalenJLT2014},\cite{CostelloMAGAZINE}.

\subsection{Protograph-Based Spatially Coupled LDPC Codes}
In this chapter, we restrict ourselves to the class of 
protograph-based construction of SC-LDPC codes as introduced in, e.g.,
\cite{LentmaierITAW09,Lentmaier_ISTC10,SchmalenSCC13}. A protograph~\cite{Thorpe_IPN} is a convenient
way of describing LDPC codes. Protograph codes are constructed from
the $P$-cover of a relatively small graph which conveys the main
properties of the code. In contrast to the graph representation of the
LDPC code, the protograph may contain multiple edges. The code itself
is constructed by placing $P$ copies of the protograph next to each
other (note that these have no interconnecting edges) and permuting
the edges between the different copies of the protograph, such that
the relation between the group of edges is respected. The construction of a small toy code is illustrated in Example~\ref{ex:protograph}.

\begin{example}\label{ex:protograph}
We illustrate the construction of larger codes based on protographs
using a simple toy example. Starting with a prototype matrix, also called \emph{protomatrix}
\[
\bm{B} = \left(\begin{array}{cccc}
1 & 2 & 3 & 0\\
1 & 2 & 0 & 2
\end{array}\right)
\]
we show how a $P$-cover is constructed. First, we construct an equivalent graph of the base matrix in the same way as we constructed the graph of the LDPC code in Sec.~\ref{sec:ldpc_graph}. The difference is that the non-zero entries in $\bm{B}$ indicate the number of \emph{parallel edges} connecting the variables with the checks. The graph representation of the protomatrix $\bm{B}$ is given by:
\begin{center}
\centering\includegraphics[width=0.22\textwidth]{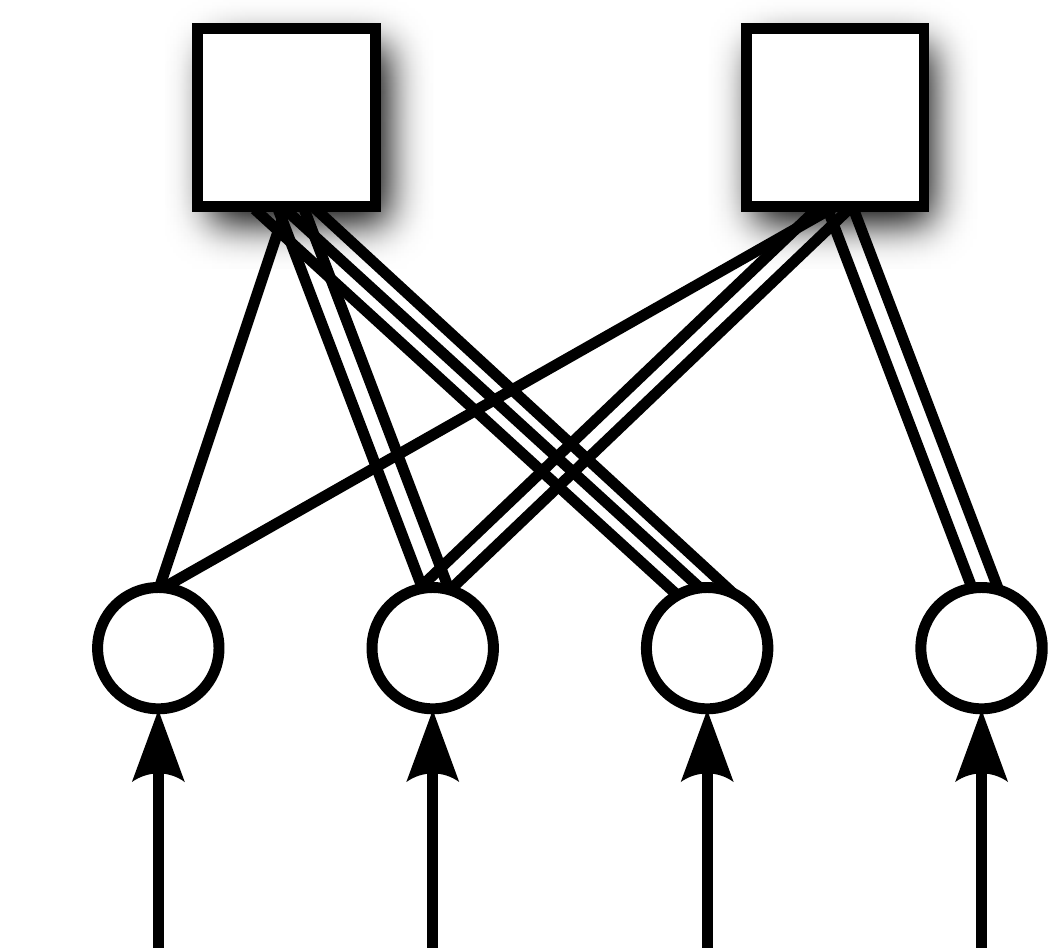}
\end{center}
In the next step, we construct the $P$-cover of this graph, which means that we simply place $P$ copies of this graph next to each other:
\begin{center}
\centering\includegraphics[width=0.72\textwidth]{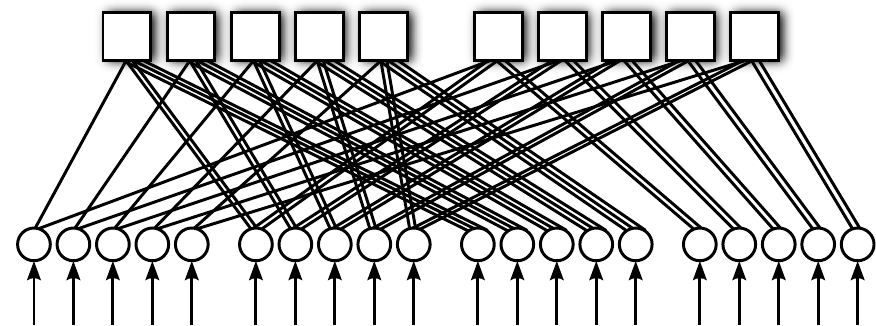}
\end{center}
This graph is still not a valid LDPC code as it contains parallel edges and the $P$ sub-graphs are not connected. In order to remove the parallel edges and to construct a more randomized code, we permute in a next step all edges that are within one \emph{edge group}, i.e., that correspond to a single entry $B_{i,j}$ of $\bm{B}$. This permutation is performed in such a way that no parallel edges persist. The final code graph is then obtained by
\begin{center}
\centering\includegraphics[width=0.72\textwidth]{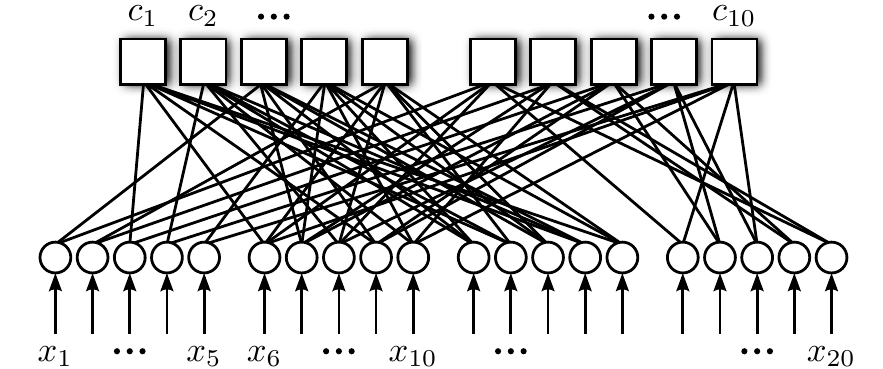}
\end{center}
Note that it is \emph{not} possible to draw this code with a single interleaver $\LDPCintl$ as the code in Fig.~\ref{fig:ldpc_only}, because it is actually a \emph{multi-edge-type} ensemble~\cite[Sec. 7.1]{Richardson2008} and for every single entry $B_{i,j}$ of $\bm{B}$, an individual interleaver is required. The parity-check matrix $\bH$ can be constructed from $\bm{B}$ by replacing each entry $B_{i,j}$ (row $i$, column $j$) of $\bm{B}$ by the superposition of $B_{i,j}$ permutation matrices\footnote{A permutation matrix $\bm{P}$ is a square binary matrix (i.e., a matrix containing only ``0'' and ``1'') where each row contains exactly one ``1'' \emph{and} where each column contains exactly one ``1''.}, chosen such that no two ``1''s are in the same position (avoiding parallel edges in the final graph). For example, a parity-check matrix corresponding to the above $\bm{B}$ with $P=5$ is given by
\begin{align*}
\bm{H} ={\small \left(\begin{array}{*{4}{c@{\;\;\;}}c:*{4}{c@{\;\;\;}}c:*{4}{c@{\;\;\;}}c:*{4}{c@{\;\;\;}}c}
   &   & 1 &   &   & 1 &   &   & 1 &   &   & 1 &   & 1 & 1 &   &   &   &   &   \\
   &   &   & 1 &   &   &   & 1 &   & 1 & 1 & 1 &   & 1 &   &   &   &   &   &   \\
 1 &   &   &   &   &   & 1 &   & 1 &   & 1 &   & 1 & 1 &   &   &   &   &   &   \\
   &   &   &   & 1 &   & 1 &   &   & 1 & 1 &   & 1 &   & 1 &   &   &   &   &   \\
   & 1 &   &   &   & 1 &   & 1 &   &   &   & 1 & 1 &   & 1 &   &   &   &   &   \\\hdashline
 1 &   &   &   &   & 1 &   & 1 &   &   &   &   &   &   &   & 1 &   &   &   & 1 \\
   & 1 &   &   &   &   & 1 &   &   & 1 &   &   &   &   &   &   &   &   & 1 & 1 \\
   &   & 1 &   &   &   & 1 &   & 1 &   &   &   &   &   &   &   & 1 &   & 1 &   \\
   &   &   & 1 &   &   &   & 1 & 1 &   &   &   &   &   &   &   & 1 & 1 &   &   \\
   &   &   &   & 1 & 1 &   &   &   & 1 &   &   &   &   &   & 1 &   & 1 &   &   \\
   \end{array}\right).}  
\end{align*}
\hrule
\end{example}

\newcommand{\bB}{\bm{B}}

We follow the approach given in~\cite{Mitchell_ITA11} to describe
protograph-based SC-LDPC codes. The protograph of a \emph{time-invariant, terminated} spatially coupled LDPC with syndrome former memory $m_s$ and replication factor $L$ is obtained from a collection of $(m_s+1)$ distinct protomatrices $\bB_i$, $i\in\{0,1,\ldots, m_s\}$ each of size $\dim\bB_i = m^\prime \times n^\prime$. The protomatrix of the spatially coupled code is then given by
\begin{align}
\bm{B}^{[\text{conv}]}(L) = \left(\begin{array}{cccc}
\bB_0 & & & \\
\bB_1 & \bB_0 & & \\
\vdots & \bB_1 & \ddots & \\
\bB_{m_s} & \vdots & \ddots & \bB_0 \\
 & \bB_{m_s} & \ddots & \bB_1 \\
 & & \ddots & \vdots \\
 & & & \bB_{m_s}
\end{array}\right)_{(L+m_s)m^\prime\times Ln^\prime}
\label{eq:protomatrix}
\end{align}
$\mathbf{B}^{[\text{conv}]}(L)$ can also be viewed as being composed by a stack of $L+m_s$ shifted (by $n^\prime$) and
overlapping versions of
\begin{align*}
  \bB_r = \left(\begin{array}{ccc} \bB_{m_s} & \cdots & \bB_0 
  \end{array}\right)\,.
\end{align*}
Note that the termination, which cuts the boundaries of the stacked matrix, leads to an
inevitable rate loss, which becomes however negligible for increasing
$L$. The rate of the code amounts to~\cite{Mitchell_ITA11}
\begin{align*}
  r_L = 1 - \left(\frac{L+m_s}{L}\right)\frac{m^\prime}{n^\prime}\,.
\end{align*}
If we are allowed to choose $L$ large enough, we immediately see that $\lim_{L\to\infty}r_L = 1 - \frac{m^\prime}{n^\prime}$, which corresponds to the rate of the original protomatrices $\bB_i$.

One can say that the convergence of spatially coupled codes is well
understood meanwhile. A thorough analysis for the binary erasure
channel is given in~\cite{Kudekar0211} and extended to general
binary input memoryless channels in~\cite{Kudekarxx13}. The convergence behavior
of the iterative decoder can be subdivided into two convergence regions
\begin{itemize}
  \item the region of \emph{macro-convergence}, where convergence
    is dominated by the code's degree distribution and
    convergence prediction by conventional EXIT charts is possible.
  \item the region of \emph{micro-convergence}, where the convergence
    is dominated by spatial coupling and termination effects.
\end{itemize}
The region of \emph{macro-convergence} is observed in the first decoding iterations and at high channel SNRs. On the other hand, the region of \emph{micro-convergence} is observed at low SNRs close to the thresholds of the code. In the region of micro-convergence, the decoding process can be visualized as a decoding wave~\cite{Kudekar0812} that slowly progresses through the graph from the boundaries onward. This wave-like behavior allows the efficient design of windowed decoders~\cite{Iyengar0711} where the decoding window follows the decoding wave. The understanding of the dynamics of the decoding wave, especially its speed~\cite{Aref1013}, are essential for designing high-performing codes and effective windowed decoders.

\subsection{Spatially Coupled LDPC Codes with Iterative Demodulation}
 In this section, we combine spatially
coupled codes with a differential decoder and use common analysis
techniques to show how the detector front-end influences the
performance of the codes. As we have seen, in conventional LDPC code design, usually
the code needs to be ``matched'' to the transfer curve of the
detection front-end. If the code is not well matched to the front-end,
a performance loss occurs. If the detector front-end has highly
varying characteristics, due to, e.g., varying channels or varying phase slip probabilities, several codes need to be implemented and always the right
code needs to be chosen for maximum performance, which appears
impractical in optical networks where a feedback channel from the receiver to the transmitter can potentially be difficult to realize.

On the other hand, in contrast to a random ensemble of the same degree
profile, spatially coupled LDPC codes can converge below the pinch-off
in the EXIT chart, so even if the code is not well matched to the
differential decoder we can hope to successfully decode due to the
micro-convergence effect. So, even with a varying channel and detector
characteristics, we can use a single code which is \emph{universally}
good in all scenarios. This means that the code design can stay
\emph{agnostic} to the channel/detector behavior. 

\renewcommand{\IE}[2]{I_{E,#2}^{[{#1}]}}
\renewcommand{\IA}[2]{I_{A,#2}^{[{#1}]}}

We determine the thresholds of the protograph-based spatially coupled codes combined
with demodulation and detection by an extension of the PEXIT technique~\cite{Liva_ProtographEXIT} with the Gaussian approximation of the check node operation~\cite{Chung0201b}. A refined version of the PEXIT technique taking into account the windowed decoder of spatially coupled codes has been presented in~\cite{HaegerOptEx14}.
The mutual information analysis used for the design of degree distributions in LDPC codes has to be modified  slightly to account for the protograph structure. Instead of analyzing and tracking a single mutual information value $I$, we now have to track an individual mutual information value for each non-zero entry at row $i$ and column $j$ of the protomatrix $\bm{B}^{[\text{conv}]}$. We denote the respective
outgoing (incoming) edge mutual information by $\IE{V,D}{i,j}$ ($\IA{V,D}{i,j}$) or by $\IE{C}{i,j}$ ($\IA{C}{i,j}$), depending if the message is computed by the combined variable node detector engine (``$V,D$'') or by the check node engine (``$C$''). Note that we assume that the messages are Gaussian distributed and can described by a single parameter,
their mean $\mu$ (with the variance $2\mu$, see~\cite{tenBrink1001} and~\cite{Richardson2008} for details).

As in the previous section, we assume that the demodulator/detector properties can be described by
means of an EXIT characteristic~\cite{tenBrink2004,Ashikhmin1104} which we 
denote by $f_D(\cdot,\cdot)$.
If the message at the input of the detector is
Gaussian distributed with mean $\mu$, then the detector output mean
for (protograph) variable $j$ is obtained by 
\begin{align}
\mu_{c,j} = J^{-1}\left(f_D\left( J\left( \sum_{i=1}^{(L+m_s)m^\prime}   B^{[\text{conv}]}_{i,j} J^{-1}(\IA{V,D}{i,j}) \right), \frac{\Es}{\No}\right)\right)\label{eq:pexit_muupdate}
\end{align}
leading to the combined variable node and detector update characteristic
\begin{align}
\IE{V,D}{i,j} = J\left(\mu_{c,j} + (B^{[\text{conv}]}_{i,j}-1)J^{-1}\left(\IA{V,D}{i,j}\right) + \sum_{\substack{k=1 \\ k\neq i}}^{(L+m_s)m^{\prime}}B^{[\text{conv}]}_{k,j}J^{-1}\left(\IA{V,D}{k,j}\right)   \right)\,,\label{eq:pexit_vnuupdate}
\end{align}
which has to be evaluated for all $(i,j)\in [1,\ldots,(L+m_s)m^\prime]\times [1,\ldots,Ln^\prime]$ where $B_{i,j}^{[\text{conv}]}\neq 0$.
The check node information is computed according to~\cite{Chung0201b}
\begin{align}
\IE{C}{i,j} = J\left(\phi^{-1}\left( 1 - \left[1\!-\!\phi\left(J^{-1}(\IA{C}{i,j})\right)\right]^{B^{[\text{conv}]}_{i,j}-1} \prod_{\substack{k=1\\ k \neq j}}^{Ln^\prime}\left[1\!-\!\phi\left(J^{-1}(\IA{C}{i,k})\right)\right]^{B^{[\text{conv}]}_{i,k}}  \right)\right).\label{eq:pexit_cnupdate}
\end{align}
which again has to be evaluated for all combinations of $(i,j)$ such that $B_{i,j}^{[\text{conv}]}\neq 0$.
The function
$\phi(\mu)$, which is used to compute the evolution of the mean of the
Gaussian messages in the check node update is given by
\begin{align}
\phi(x) = \left\{\begin{array}{ll}
    1\! -\! \frac{1}{\sqrt{4\pi x}}\!\int\limits_{-\infty}^{\infty}\tanh\!\left(\frac{u}{2}\right)\exp\left({-\frac{(u-x)^2}{4x}}\right)\text{d}u, & \!\!\text{if}\ x > 0 \\
    1, & \!\!\text{if}\ x = 0
  \end{array}\right.\label{eq:phifun}
\end{align}
Numerical approximations for (\ref{eq:phifun}) and its inverse
function are given in \cite{Chung0201b}.

\begin{figure}
\centering\includegraphics[scale=1.1]{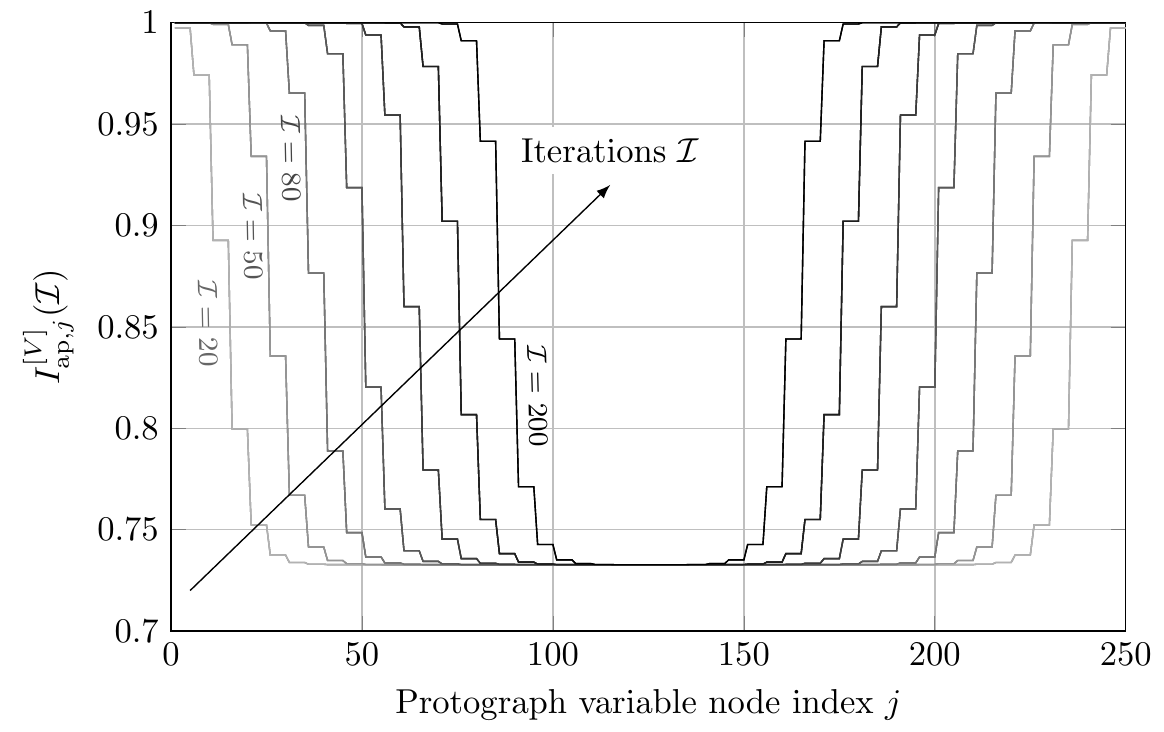}
\caption{Wave-like decoding behavior of the spatially coupled, protograph-based LDPC code ($L=50$) with Gray differential coding and $\gamma=0.2$ for $\Es/\No = 4.8$\,dB}
\label{fig:wave}
\end{figure}

The evaluation of the information is  carried out in an iterative way: First, we initialize the 
process by setting $\IA{V,D}{i,j}(1)=0$ for all possible $(i,j)$ where the ``$(1)$'' denotes the first iteration. Using~\eqref{eq:pexit_muupdate}, we first compute $\mu_{c,j}(1)$, $\forall j\in[1,Ln^\prime]$ and use $\mu_{c,j}(1)$ to compute $\IE{V,D}{i,j}(1)$ by evaluating~\eqref{eq:pexit_vnuupdate} for all $(i,j)\in[1,\ldots,(L+m_s)m^\prime]\times[1,\ldots,Ln^\prime]$. We then set $\IA{C}{i,j}(1)=\IE{V,D}{i,j}(1)$ and evaluate~\eqref{eq:pexit_cnupdate} yielding $\IE{C}{i,j}(1)$. By setting $\IA{V,D}{i,j}(2) = \IE{C}{i,j}(1)$ we may proceed to the second iteration and 
compute $\mu_{c,j}(2)$, $\IE{V,D}{i,j}(2)$ and $\IE{C}{i,j}(2)$ in this sequence. Finally, after $\mathcal{I}$ iterations, we may---for each variable node in the protograph---determine the \emph{a posteriori} reliability by
\begin{align*}
I^{[V]}_{\text{ap},j} = J\left(\mu_{c,j} +  \sum_{k=1}^{(L+m_s)m^{\prime}}B^{[\text{conv}]}_{k,j}J^{-1}\left(\IA{V,D}{k,j}(\mathcal{I})\right)   \right)\,.
\end{align*}
We illustrate the behavior of $I^{[V]}_{\text{ap},j}$, which gives an indication of the reliability of the $P$ bits that will be assigned to position $j$ in the protograph by means of an example. We consider a spatially coupled code of rate $r=0.8$ with $m_s=2$ and with $\bm{B}_0 = \bm{B}_1 = \bm{B}_2 = \left(1\ \ 1\ \ 1\ \ 1\ \ 1\right)$. We use QPSK with Gray differential coding and set $\gamma=0.2$ at $\Es/\No=4.8$\,dB, i.e., according to~\eqref{eq:slipprob_factor}, phase slips occur on the channel with a probability $P_{\text{slip}}\approx 0.0082$. Figure~\ref{fig:wave} shows the behavior of the \emph{a posteriori} mutual information $I^{[V]}_{\text{ap},j}(\mathcal{I})$ as a function of the decoding iterations $\mathcal{I}$. We can see that the mutual information $I^{[V]}_{\text{ap},j}$ increases in a wave-like way. Starting from the boundaries of the codeword, the mutual information converges towards~$1$ with an increasing number of iterations from the outside towards the inside until both waves meet and the whole codeword has been successfully decoded.

\subsection{Windowed Differential Decoding of SC-LDPC Codes}
By observing the wave-like decoding behavior~\cite{Kudekar0812} in Fig.~\ref{fig:wave}, we note that only parts of the protograph get updated when carrying out iterations. For instance, the reliability of the protograph variable node indices $110$ to $140$ stay at an almost constant value during the first $200$ iterations. Likewise, the protograph variable node indices $1$ to $30$ have already converged to $1$ after $110$ iterations and do not benefit anymore from additional decoding iterations. This wave-like behavior thus leads to an efficient \emph{windowed decoder}~\cite{Iyengar0711,Lentmaier_Window}, which follows the decoding wave and only carries out operations on the part of the protograph that benefits from further decoding iterations. The windowed decoder works in principle as an LDPC decoder, just with the difference that it operates on a fraction of the parity-check matrix. The windowed decoder is characterized by a window length $w$ and operates on a portion of the protomatrix containing $w$ vertically stacked copies of $\bm{B}_r$
\begin{align}
\widetilde{\bm{B}}^{[\text{conv}]} = \left(\begin{array}{cccccc}
\bB_{m_s} & \cdots & \bB_0 & & & \\
 & \bB_{m_s} & \cdots & \bB_0 & &  \\
 & & \ddots & \cdots & \ddots &  \\
  & & & \bB_{m_s} & \cdots & \bB_0 \end{array}\right)_{wm^\prime\times (w+m_s)n^\prime}\,.
\label{eq:protomatrix_window}
\end{align}
The decoder takes a block of $(w+m_s)n^\prime$ protograph variables (i.e., \mbox{$(w+m_s)Pn^\prime$} code bits), and carries out $\mathcal{I}_w$ decoding iterations on this block (using the conventional decoding scheme described in the Appendix). After having carried out $\mathcal{I}_w$ iterations, the window is shifted by $n^\prime$ protograph variables ($Pn^\prime$ code bits) and the left-most portion of $Pn^\prime$ code bits are considered to be decoded. Then the process starts again. At the beginning of the process, the decoding window 
is initialized with perfect knowledge of the boundary values. For details, we refer the interested reader to~\cite{Iyengar0711} and~\cite{Iyengar12}.

\subsection{Design of Protograph-Based SC-LDPC Codes for\\ Differential Coded Modulation}
We show by means of an example how to construct good protographs for SC-LDPC codes in the context of differential coded modulation~\cite{SchmalenOFC15}. For simplicity, we restrict ourselves to SC-LDPC codes with $m_s=2$ leading to a protomatrix given by
\[
\bm{B}_c \!=\! \left(\begin{array}{@{\:}c@{\ \ \:}c@{\ \ \:}c@{\ \ \:}c@{\:}}
\bm{B}_0 & \phantom{\ddots} & & \\
\bm{B}_1 & \bm{B}_0 & \phantom{\ddots} &  \\
\bm{B}_2 & \bm{B}_1 & \ddots &  \\
 & \bm{B}_2 & \ddots & \bm{B}_0 \\
 &          & \ddots & \bm{B}_1 \\
  & & & \vphantom{\ddots}\bm{B}_2\end{array}\right)\,.
\]
We select $m_s=2$, as this choice leads to windowed decoders with a relatively compact decoding window. Note that the minimum required decoding window size grows almost linearly with $m_s$. Another viable choice would be $m_s=1$, however, simulation results which are not shown here have indicated that a better performance can be expected with $m_s=2$.

We consider iterative differential decoding as described in the previous section using the modified slip-resilient BCJR algorithm based on the trellis of Fig.~\ref{fig:trellisV4slipsnew} and wish to design coding schemes of rate $r=0.8$ ($25$\% OH). We use protographs leading to regular SC-LDPC codes with variable degree $\ltwreg=3$ and check degree $\rtw=15$. The reason for using regular codes is that it has been shown in~\cite{Kudekar0211} that regular SC-LDPC are sufficient to achieve capacity and this particular code has a MAP threshold very close to capacity. Furthermore, due to the regularity, the implementation of the decoder can be simplified and the error floor is expected to be very low (increasing $\ltwreg$ further may even lead to lower error floors).

\begin{figure}[tbh!]
\centering\includegraphics[width=0.75\textwidth]{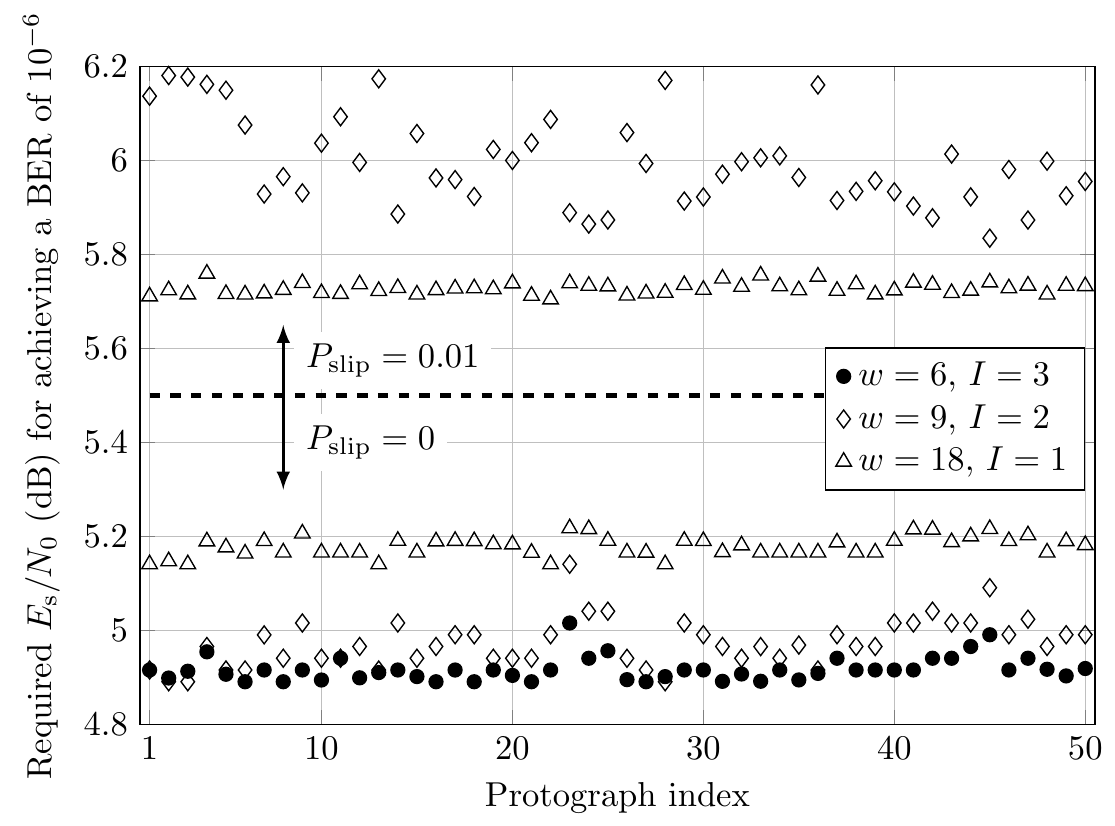}
\caption{Required $\Es/\No$ for 50 different protographs}
\label{fig:test_proto}
\end{figure}

Although $\ltwreg$, $\rtw$ and $m_s$ are fixed, we still need to find good protomatrices $\bB_0$, $\bB_1$ and $\bB_2$. In order to have a very simple structure, we fix $m^\prime=1$ and $n^\prime=5$, leading to the smallest possible protomatrices. We have constructed all $1837$ possible such combinations of protographs (unique up to column permutation) and computed decoding
thresholds for all these protographs using the above described method. We selected the protographs with the $50$ best thresholds and carried out Monte Carlo simulations with three different windowed decoders:
\begin{itemize}
\item[$\bullet$] The first setup uses a window size of $w=4$ and carries out $\mathcal{I}_w=3$ iterations per decoding step.
\item[$\diamond$] The second setup uses a window size of $w=7$ and carries out $\mathcal{I}_w=2$ iterations per decoding step.
\item[{\footnotesize $\bigtriangleup$}] The third setup uses a window size of $w=16$ and carries out a single iteration per decoding step, i.e., $\mathcal{I}_w=1$.
\end{itemize}
Note that with all three setups, every coded bit undergoes and equivalent number of $18$ iterations, leading to the same complexity of all decoders and the same complexity as the LDPC coded schemes presented in Sec.~\ref{sec:ldpc_diff_mod}. Further note that the number of differential decoder executions per coded bit amounts $w+m_s$ and thus depends on the setup. In the case of LDPC coded differential demodulation, we have executed the differential decoder for each iteration. The required $\Es/\No$ (in dB) to achieve a target BER of $10^{-6}$ for the $50$ selected protographs is shown in Fig.~\ref{fig:test_proto} for $L=100$ and $P_{\text{slip}}\in\{0, 0.01\}$. Based on the results of Fig.~\ref{fig:test_proto}, we select the protograph with index 3 which has the best performance compromise at $P_{\text{slip}}=0$ and $P_{\text{slip}}=0.01$, and which is given by $\bm{B}_0 = (1\ \ 1\ \ 1\ \ 1\ \ 1)$, $\bm{B}_1 = (1\ \ 0\ \ 0\ \ 0\ \ 0)$, and $\bm{B}_2=(1\ \ 2\ \ 2\ \ 2\ \ 2)$.

\begin{figure}[t!]
\centering\includegraphics[width=0.85\textwidth]{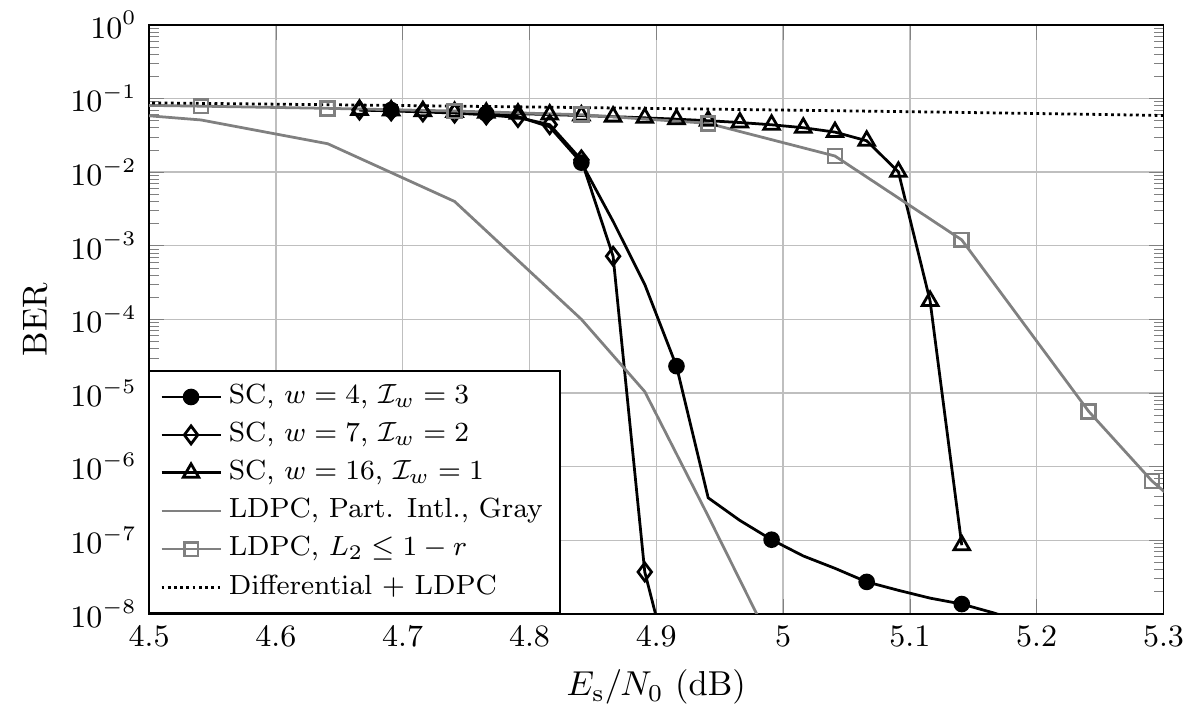}
\vspace*{-2ex}
\caption{Simulation results of proposed and reference schemes for QPSK with $\gamma = 0$}
\label{fig:simres_sc_codes_g0}
\end{figure}

\begin{figure}[t!]
\centering\includegraphics[width=0.85\textwidth]{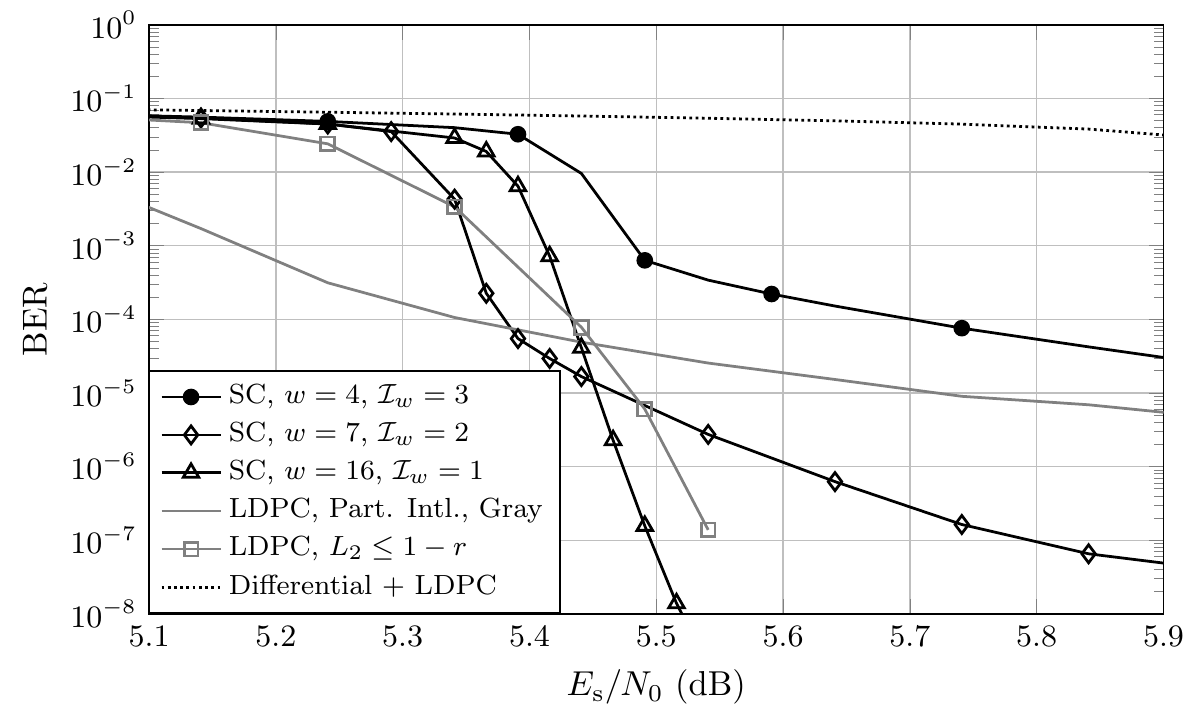}
\vspace*{-2ex}
\caption{Simulation results of proposed and reference schemes for QPSK with $\gamma = 0.2$}
\label{fig:simres_sc_codes_g1}
\end{figure}

Finally, we use this protograph to construct a code by first lifting the protographs $\bm{B}_i$ with $P=40$ and using the intermediate result in a second step to generate a \emph{quasi-cyclic} (QC) code with circulant permutation matrices of size $50\times 50$. The resulting parity-check submatrices $\bm{H}_i$ associated to $\bm{B}_i$ have size $\dim\bm{H}_i = 2000\times 10000$.  As reference, we use QPSK with Gray differential coding and pick the two best codes found in Sec.~\ref{sec:code_design}: partial interleaving with no restriction on $L_2$ and partial interleaving with $L_2 \leq 1-r$. As in Figs.~\ref{fig:simex_1} and~\ref{fig:simex_2}, we use $I=18$ decoding iterations in all cases. The results are shown in Fig.~\ref{fig:simres_sc_codes_g0} for $\gamma=0$ (i.e., no phase slips) and in Fig.~\ref{fig:simres_sc_codes_g1} for $\gamma=0.2$.

We can see from the results that for $\gamma=0$, the LDPC code with partial interleaving (no restriction on $L_2$) already yields a very good performance within 1\,dB of the theoretically minimum $\Es/\No$, while the code with the constraint on $L_2$  entails a performance loss of about $0.4$\,dB (see also Figs.~\ref{fig:simex_1} and~\ref{fig:simex_2}). The SC-LDPC code with the second decoder setup ($w=7$) outperforms the LDPC code for low BERs due to the steepness of the waterfall curve. If the phase slip probability is nonzero with $\gamma=0.2$, which may occur in a highly nonlinear DWDM transmission with OOK neighbors, we observe from Fig.~\ref{fig:simex_1} and Fig.~\ref{fig:simres_sc_codes_g1} that the LDPC code no longer performs well and has a severe error floor requiring strong outer coding. The error floor can be reduced by designing a code with $L_2 \leq 1-r$.  If $\gamma=0.2$, the SC-LDPC code with decoder setup 3 almost shows the same performance as the LDPC code with the constraint $L_2 \leq 1-r$ and outperforms it at low error rates (we did not observe any error floor in our simulations). 

The proposed SC-LDPC code has the important advantage of \emph{universality}~\cite{SchmalenSCC13}: A single SC code is powerful in all transmission scenarios requiring only a single \emph{channel agnostic transmitter} and a receiver selecting the best decoding setup depending on $\gamma$ and $P_{\text{slip}}$. The transmission scheme can thus be kept simple and only a \emph{single} code needs to be implemented. In the setup with conventional LDPC codes, two different codes have to be implemented, depending on the setup: one code for small $\gamma$ and another code, possibly with $L_2\leq 1-r$, for larger values of $\gamma$. The transmitter has to know the expected $\gamma$ on the channel and adapt the coding scheme accordingly, which requires an undesired feedback channel and changes in the network control plane. Another possibility is to deliberately design an LDPC code which shall perform well in both cases~\cite{PfluegerSCC13}: this leads however to a compromise in the construction and codes which are not able to compete with codes optimized for a specific scenario. With the SC-LDPC code, we can use a single code and use receiver processing to estimate $\gamma$ and setup the windowed decoder ($w$ and $\mathcal{I}_w$) accordingly. Note that for large $\gamma$, it is advantageous to use a long window $w$, which means that the number of differential decoding executions per bit shall be maximized. This is in contrast to the HTDD approach~\cite{BisplinghoffECOC2014}, which however uses a differential decoder not adapted to the channel. We conclude that executing the differential detector doesn't degrade the performance as long as it is well adapted to the channel model.

\section{Conclusions}
In this chapter, we have described some important aspects of soft-decision forward error correction with iterative decoding. We have shown that the phenomenon of phase slips, which occurs frequently in the case of coherent long-haul optical communications, can be combated effectively using iterative differential decoding. We have shown that the achievable information rate is not affected by iterative differential decoding and that
 differential decoding only leads to an unavoidable performance loss if not properly decoded. We have further shown how phase slips affect the achievable rate and how to design a trellis diagram describing the differential code affected by phase slips. In order to achieve the best possible performance, this differential decoder needs some well-adapted code design. We have proposed different design guidelines for LDPC codes and have shown that, depending on the channel quality, there is a different code design which may lead to the desired performance, especially if very low residual bit error rates are targeted. Finally, we have show that spatially coupled codes offer a more universal code design and lead to codes that are more agnostic to the channel and thus enable the implementation of a single code that performs equally well in all channel conditions and that even outperforms conventional LDPC codes.

\appendix
\section{LDPC Coded Differential Modulation -- Decoding Algorithms}

The decoders we consider (which can include simplified versions like binary message passing decoders~\cite{KramerBMP}) rely on the knowledge of the channel\footnote{In optical coherent receivers, either channel models like the AWGN channel are assumed together or histogram based methods may be employed~\cite[Sec. 6.2]{DjordjevicRyan}. Sometimes, the worst case channel for which the system is designed may be assumed.}. The communication channel, or an equivalent communication channel comprising the physical channel as well as various inner receiver and signal processing stages, can be characterized by its \emph{conditional probability} $P(z | y)$, i.e., the probability of observing $z$ at the receiver assuming a transmitted modulation symbol $y \in \mathcal{M}$. Note that we restrict ourselves to memoryless channels, which can be achieved in practice by sufficiently long interleaving of the channel input.
Notable examples are the \emph{binary symmetric channel} (BSC) with ($y=x$)
\begin{alignat*}{2}
P(z=0 | x=0) &= 1-\epsilon\quad & \quad P(z=1 | x=0) &= \epsilon \\
P(z=0 | x=1) &= \epsilon\quad & \quad P(z=1 | x=1) &= 1-\epsilon 
\end{alignat*}
which is frequently used to model the hard-decision channel. Another famous example is the real-valued \emph{additive white Gaussian noise} (AWGN) channel with \emph{binary phase shift keying} (BPSK) modulation with $y = (-1)^x$ and (real-valued) noise variance $\sigma_n^2 = N_0/2$, leading to
\[
p(z | y) = \frac{1}{\sqrt{\pi N_0}}\exp\left(\frac{-\left(z-y\right)^2}{N_0}\right)\,.
\]
As the computation with probabilities tends to be numerically unstable and hinders potential hardware implementations, frequently \emph{log-likelihood ratios} (LLRs)~\cite{Hagenauer0396} are employed. The LLR $L(z[t])$ for received symbol $z[t]$ at time instant $t$ is defined for BPSK as
\[
L(z[t]) = \log\frac{p(z[t] \big| x[t]=0)}{p(z[t] \big| x[t]=1)}\,,
\]
where $\log(\cdot)$ is the natural logarithm. It turns out that for the AWGN channel with BPSK modulation, we have
\[
L_{\text{AWGN}}(z[t]) = \frac{2}{\sigma_n^2}z[t] = 4\frac{\Es}{\No} z[t]\  = 4r\frac{\Eb}{\No} z[t]\ =: L_c\cdot z[t]\,.
\]
This last equation means that the LLR $L_{\text{AWGN}}(z[t])$ is obtained by multiplication of $z[t]$ with a constant $L_c:=\frac{2}{\sigma_n^2}$, which depends on the noise variance only. Usually, the noise variance is assumed to be constant and the constant $L_c$ is predetermined and set to a value suitable for implementation. The noise variance may also be estimated at the receiver~\cite{Mecklenbraeuker}.

If higher order modulation formats based on a constellation $\mathcal{M}$ are employed, we have to use a more involved computation rule. Starting from the mapping function $\phi(\bm{b})$, we first define the $i$th inverse mapping function 
\begin{multline*}
\phi_i^{-1}(b) =\\
 \left\{\tilde{y} = \phi(\tilde{b}_1,\ldots,\tilde{b}_{i-1},b,\tilde{b}_{i+1},\ldots,\tilde{b}_q)  : \tilde{y}\in\mathcal{M}, (\tilde{b}_1,\ldots,\tilde{b}_{i-1},\tilde{b}_{i+1},\ldots,\tilde{b}_q) \in \field_2^{q-1}\right\}
\end{multline*}
Thus, $\phi_i^{-1}(b)$ returns the set of all modulation symbols to which a bit pattern whereof the $i$th bit takes on the value $b$, is assigned. The LLR for the ubiquitous \emph{bit-wise} decoder~\cite{GuilleniFabregas} is then given by 
\begin{align}
L(b_{i}[t]) = \log\left(\frac{\sum_{\tilde{y}\in\phi^{-1}_i(0)}p(z[t]\big|\tilde{y})}{\sum_{\tilde{y}\in\phi^{-1}_i(1)}p(z[t]\big|\tilde{y})}\right)\,.\label{eq:bitwise}
\end{align}
Before we step ahead and describe the decoding algorithms, we first introduce the $\maxs$ operation, which simplifies the description of the BCJR decoder~\cite{Robertson0695}
\begin{definition}{The $\maxs$ operation.}
The $\maxs$ operations is defined as
\[
\maxso(\delta_1,\delta_2) := \max(\delta_1,\delta_2) + \log\left(1+e^{-|\delta_1-\delta_2|}\right) = \log\left(e^{\delta_1}+e^{\delta_2}\right)
\]
and can be conveniently approximated by
\[
\maxso(\delta_1,\delta_2) \approx \max(\delta_1,\delta_2).
\]
\end{definition}
The $\maxs$ operation has several properties, namely
\begin{align*}
\maxso(\delta_1,\delta_2) &= \maxso(\delta_2,\delta_1) \\
\lim_{\delta_1\to -\infty}\maxs(\delta_1,\delta_2) &= \delta_2 \\
\maxso(\delta_1,\delta_2,\delta_3) &= \maxso(\delta_1,\maxso(\delta_2,\delta_3))
\end{align*}
The latter property allows us to define
\begin{align*}
\maxso\limits_{j=1}^{\chi}\delta_j = \maxso(\delta_1,\delta_2,\ldots,\delta_{\chi}) = \maxso(\delta_1,\maxso(\delta_2,\cdots\maxso(\delta_{\chi-1},\delta_{\chi})\cdots ))
\end{align*}
and with the trivial case
\[
\maxso\limits_{j=1}^1 \delta_j = \delta_1\,.
\]

\subsubsection*{Differential Decoding}
The soft-input soft-output differential decoding is carried out using the BCJR algorithm~\cite{Bahl0374}. We just summarize the operations of the BCJR algorithm in the LLR domain, such that it can be immediately applied. We give the equations for the case $V=4$, which we have used in our simulations in this chapter. We use the trellis diagram of Fig.~\ref{fig:trellisV4slipsnew} to describe the BCJR algorithm. The algorithm consists of computing a forward and a backward recursion. In the forward recursion, the variables $\tilde{\alpha}_t(\stat_i)$ are updated for $i\in\{1,2,3,4\}$. The initialization, which describes the initial differential memory, is usually carried out as
\begin{align*}
\tilde{\alpha}_0(\stat_1) &= 0\\
\tilde{\alpha}_0(\stat_i) &= -\infty\qquad\text{for}\ i\in\{2,3,4\}\,.
\end{align*}
The recursive update is given by (for all $j\in\{1,2,3,4\}$)
\begin{align*}
\tilde{\alpha}_t(\stat_j) = \maxso\limits_{i=1}^4\maxso_{s=0}^{3}\left(\tilde{\alpha}_{t-1}(\stat_i)+\tilde{\gamma}_t(i,j,s)\right)
\end{align*}
Similarly, the backward recursion is carried out with the initialization $\tilde{\beta}_{\tilde{n}}(\stat_j) = 0$, for $j\in\{1,2,3,4\}$, where $\tilde{n}$ denotes the length of the sequence of modulation symbols to be decoded. We have
\begin{align*}
\tilde{\beta}_{t-1}(\stat_i) = \maxso\limits_{j=1}^4\maxso_{s=0}^{3}\left(\tilde{\beta}_{t}(\stat_j)+\tilde{\gamma}(i,j,s)\right)
\end{align*}
Before giving the equation to compute $\gamma(i,j,s)$, we first introduce the sets $\mathcal{M}_{\stat_i}\subset \mathcal{M}$ which contain all modulation symbols that are associated to state $\stat_i$ (that are within the region associated with state $\stat_i$). For the example of the QPSK constellation of Fig.~\ref{fig:qpsk_state}, $\mathcal{M}_{\stat_1} = \left\{\frac{1+\imath}{\sqrt{2}}\right\}$ and for the example of the 16-QAM constellation of Fig.~\ref{fig:qam16_diff}, we have 
\[
\mathcal{M}_{\stat_1} = \left\{\frac{1+\imath 1}{\sqrt{10}}, \frac{3+\imath 1}{\sqrt{10}},\frac{1+\imath 3}{\sqrt{10}},\frac{3+\imath 3}{\sqrt{10}}\right\}.
\]
The variable $\tilde{\gamma}_t(i,j,\zeta)$ describes the (logarithmic) probability of a state transition from state $\stat_i$ at time $t-1$ to state $\stat_j$ at time $t$ provided that the phase slip occurrence descriptor $s[t]$ takes on the value $\zeta$. We have
\begin{align*}
\tilde{\gamma}_t(i,j,\zeta) & =\frac{1}{\No}\sum_{\chi \in \mathcal{M}_{\stat_j}}\big|z[t]-\chi\big|^2   + \frac{1}{2}\sum_{\kappa=1}^{v}\left(1-2\check{f}_{\text{diff},\kappa}^{-1}(\stat_i,\stat_j,s)\right) L^{[\text{apriori},\Pi]}_{(t-1)v+\kappa} +\\
&\qquad +\log P(s = \zeta)\,. 
\end{align*}
Note that with the phase slip model introduced in Sec.~\ref{sec:channel_model}, we may
abbreviate $ \log P(s = \zeta) = |\zeta|\log\xi$. The function $\check{f}_{\text{diff},\kappa}^{-1}(\stat_i,\stat_j,s)$ returns the $\kappa$th bit $b_{\kappa}$ of the differential encoding map that causes a state transition from $\stat_i$ to $\stat_{j^\prime}$, where $j^\prime$ is an intermediate state leading to the final state $j=((j^\prime+s-1)\!\mod V)+1$ after taking into account the phase slip. $L^{[\text{apriori},\Pi]}_i$ contains the input LLR values $L^{[\text{apriori}]}_i$ that are provided by the LDPC decoder after full or partial interleaving. In the initial execution (first iteration) of the differential decoder, we may set $L^{[\text{apriori},\Pi]}_i=0$.

Finally, we obtain for each $t\in\{1,\ldots, \tilde{n}\}$ and $\kappa\in\{1,\ldots, v\}$
\begin{align*}
L_{(t-1)v+\kappa}^{[\text{diff},\Pi]} &= \maxso\limits_{\substack{(i,j,s) \\ \check{f}_{\text{diff},\kappa}^{-1}(\stat_i,\stat_j,s) = 0}}\left(\tilde{\alpha}_{t-1}(\stat_i) + \tilde{\gamma}_t(i,j,s) + \tilde{\beta}_t(\stat_j)\right) - \\
&\qquad - \maxso\limits_{\substack{(i,j,s) \\ \check{f}_{\text{diff},\kappa}^{-1}(\stat_i,\stat_j,s) = 1}}\left(\tilde{\alpha}_{t-1}(\stat_i) + \tilde{\gamma}_t(i,j,s) + \tilde{\beta}_t(\stat_j)\right) -L_{(t-1)v+\kappa}^{[\text{apriori},\Pi]} .
\end{align*}
After (partial or full) deinterleaving of $L_{i}^{[\text{diff},\Pi]}$, we obtain $L_{i}^{[\text{diff}]}$ which is used as input of the LDPC decoder.

\subsubsection*{LDPC decoding}
A vast collection of various decoding algorithms for LDPC codes exist and we refer to the broad body of literature for a good introduction (see, e.g.,~\cite{RyanBook,MoonBook}). Most of these decoders are message passing decoders, where the most prominent is probably the sum-product decoder~\cite{RyanBook}. In what follows, we describe the sum-product decoder and the closely related min-sum decoder, which can be interpreted as being an approximation of the sum-product decoder.

In order to describe the sum-product and min-sum decoders, we introduce the set $\mathcal{N}(m) := \{j : H_{m,j} \neq 0\}$ which contains the positions (columns) of non-zero entries at row $m$ of the parity-check matrix $\bm{H}$. For the matrix given in Example~\ref{ex:irregular_H}, we have $\mathcal{N}(1) = \{1;4;11;12;15;22;24;25;28;29;30;31\}$. Similary, the set $\mathcal{M}(n) := \{i : H_{i,n} \neq 0\}$ contains the positions (rows) of non-zero entries at column $n$ of the parity-check matrix $\bH$. Again, for the exemplary matrix of Example~\ref{ex:irregular_H}, we have $\mathcal{M}(1) = \{1;2\}$, $\mathcal{M}(2) = \{3;5\}$ and so on.

Within the sum-product decoder, messages are exchanged between the variable nodes and the check nodes, thus the name \emph{message passing decoder}. We denote the message that is passed from variable node $i$ towards check node $j$ by $L_{i,j}^{[\text{v}\rightarrow\text{c}]}$. Similarly, the message that is passed from check node $j$ towards variable node $i$ is denoted by  $L_{i,j}^{[\text{v}\leftarrow\text{c}]}$. Before first executing the LDPC decoder with a new frame of data, all messages are set to zero, i.e., $L_{i,j}^{[\text{v}\rightarrow\text{c}]} = L_{i,j}^{[\text{v}\leftarrow\text{c}]} = 0$ for all combinations of $(j,i)\in [1,\ldots,m]\times [1,\ldots, n]$ such that $H_{j,i} \neq 0$.

The sum-product LDPC coder computes for each of the $n$ variables, i.e., for each transmitted bits, the total sum
\[
L_i^{[\text{tot}]} = L_i^{[\text{diff}]} + \sum_{j\in\mathcal{M}(i)}L_{i,j}^{[\text{v}\leftarrow\text{c}]},\qquad\forall i\in\{1,\ldots, n\}
\]
Using this total sum, the variable-to-check messages may be computed as
\[
L_{i,j}^{[\text{v}\rightarrow\text{c}]} = L_i^{[\text{tot}]} - L_{i,j}^{[\text{v}\leftarrow\text{c}]}, \qquad \forall i \in\{1,\ldots, n\}, \forall j\in\mathcal{M}(i)
\]
In the second step, the check node update rule is carried out to compute new check-to-variable messages
\begin{align*}
L_{i,j}^{[\text{v}\leftarrow\text{c}]}=2\tanh^{-1}\left(\prod_{i^{\prime}\in\mathcal{N}(j)\backslash\{i\}}\!\!\!\!\tanh\left(\frac{L_{i^{\prime},j}^{[\text{v}\rightarrow\text{c}]}}{2}\right)\right),\forall j\in\{1,\ldots, m\}\  \forall i\in\mathcal{N}(j)
\end{align*}
where the inner product is taken over over all entries in $\mathcal{N}(j)$ except the one under consideration $i$. This is indicated by the notation $\mathcal{N}(m)\backslash\{i\}$. Usually, in practical implementations, simplified approximations to this update rule are implemented, for example the scaled min-sum rule~\cite{Chen0302}
\begin{align*}
L_{i,j}^{[\text{v}\leftarrow\text{c}]}=\nu\left(\prod_{i^{\prime}\in\mathcal{N}(j)\backslash\{i\}}\!\!\!\!\!\!\mathop{\text{sign}}\left(L_{i^{\prime},j}^{[\text{v}\rightarrow\text{c}]}\right)\right)\min_{i^{\prime}\in\mathcal{N}(j)\backslash\{i\}}\big|L_{i^{\prime},j}^{[\text{v}\rightarrow\text{c}]}\big|
\end{align*}
where $\nu$ is an appropriately chosen scaling factor. See~\cite{Zhao0405} for other simplified variants of the sum-product algorithm.

With the updated check-to-variable node messages, a new \emph{a priori} message that is transmitted to the differential decoder may be computed as
\[
L_i^{[\text{apriori}]} = \sum_{j\in\mathcal{M}(i)}L_{i,j}^{[\text{v}\leftarrow\text{c}]},\qquad\forall i\in\{1,\ldots, n\}\,.
\]
Note that the convergence of the sum-product decoder as described here can be considerable improved by using the so-called row-layered decoder~\cite{Hocevar2004}, which allows to roughly halve the number of decoding iterations. Many additional possible decoder variants, which may be better suited for an ASIC implementation than the message passing decoder described here, are discussed in~\cite{RyanBook}.

\bibliographystyle{ieeetr}

\end{document}